\documentclass[12pt]{article}
\usepackage{amsmath,amssymb, amsthm,graphicx,hyperref,booktabs,float,xcolor, tabularx, placeins, tcolorbox, mathtools, enumitem}
\usepackage{refcount}
\usepackage[left = 1in, right = 1in, top = 1in, bottom = 1in]{geometry}
\usepackage{subcaption}
\usepackage{rotating}
\usepackage{pdflscape}
\usepackage{placeins}
\newtheorem{proposition}{Proposition}
\DeclareMathOperator*{\argmax}{argmax}

\newtheorem{assumption}{Assumption}
\newtheorem{corollary}{Corollary}

\newtheorem{definition}{Definition}
\newtheorem{remark}{Remark}

\makeatletter
\renewenvironment{proof}[1][\proofname]{\par
  \pushQED{\qed}%
  \normalfont \topsep6\p@\@plus6\p@\relax
  \trivlist
  \item[\hskip\labelsep
        \itshape
    #1\@addpunct{.}]\ignorespaces
}{%
  \popQED\endtrivlist\@endpefalse
}
\makeatother
\newcolumntype{Y}{>{\centering\arraybackslash}X}
\usepackage{setspace} 

\usepackage[authordate,bibencoding=auto,backend=biber, natbib, doi=false, url=false, isbn=false]{biblatex-chicago}

\AtEveryBibitem{%
  \clearfield{url}%
  \clearfield{doi}%
  \clearfield{isbn}%
  \clearfield{eprint}%
}
 
\title{The Memorization Problem: \\ Can We Trust LLMs' Economic Forecasts?}
 \author{Alejandro Lopez-Lira, Yuehua Tang, Mingyin Zhu\thanks{We are grateful for the comments and feedback from Manuel Adelino, Yiwei Dou, Daniel Greene, Wei Jiang, Luka Vulicevic, Baozhong Yang, Mao Ye, and seminar and conference participants at the 2025 GSU-MS AI \& FinTech Conference, the Journal of Accounting, Auditing and Finance (JAAF) Symposium 2025, and the Applied Machine Learning, Economics, and Data Science (AMLED) Webinar. Contact information: Alejandro Lopez-Lira: alejandro.lopez-lira@warrington.ufl.edu;\phantom{ }Yuehua Tang: yuehua.tang@warrington.ufl.edu;\phantom{ } Mingyin Zhu: mingyin.zhu@warrington.ufl.edu. } \\University of Florida}
 \date{First version: April 15, 2025; This version: \today}
\begin{document}

\maketitle

\begin{abstract}
Large language models (LLMs) cannot be trusted for economic forecasts during periods covered by their training data. Counterfactual forecasting ability is non-identified when the model has seen the realized values: any observed output is consistent with both genuine skill and memorization. Any evidence of memorization represents only a lower bound on encoded knowledge. We demonstrate LLMs have memorized economic and financial data, recalling exact values before their knowledge cutoff. Instructions to respect historical boundaries fail to prevent recall-level accuracy, and masking fails as LLMs reconstruct entities and dates from minimal context. Post-cutoff, we observe no recall. Memorization extends to embeddings.

\end{abstract}

\small{
}

\newpage
\doublespacing
\section{Introduction}

A growing body of literature employs large language models (LLMs) to generate historical expectations, evaluate their forecasting accuracy, or backtest LLM-based investment strategies within periods covered by these models' training data. Most LLMs are trained on comprehensive internet-scale datasets up to a specific knowledge cutoff date, creating a fundamental challenge: when analyzing pre-cutoff data, we cannot distinguish whether a model demonstrates genuine forecasting ability or simply recalls memorized information.\footnote{Following initial training, models typically undergo reinforcement learning from human feedback (RLHF) to improve their usefulness and safety, but there is no evidence that this process extends their knowledge timeline.} For example, if LLMs have memorized historical S\&P 500 values, evaluating their ability to ``forecast" these values from any pre-cutoff information becomes unreliable. In this paper, we show that LLMs have memorized large amounts of economic and financial data, thus challenging the usual interpretation of LLMs' forecasting ability.

Theoretically, we formalize the memorization problem as a non-identification issue and prove that when a model has seen the realized values during training, its counterfactual forecasting ability cannot be recovered from its outputs. Any observed forecast is consistent with two contradictory explanations (genuine analytical skill or simple recall of memorized information), making inference impossible. Empirically, we show this problem applies in practice by providing systematic evidence that LLMs have memorized economic and financial data at scale. Together, these findings establish that pre-cutoff ``forecasting" studies face a fundamental methodological problem: they cannot distinguish genuine predictive ability from memorization. Constraining prompts (e.g., ``use only data before 2010") cannot resolve this because prompts cannot change what information is encoded in the model's parameters.

Our theoretical results prove that other common proposed solutions also fail. First, fine-tuning the model to ``forget" future information does not help: without examining what is actually removed from the model, we cannot tell whether the model actually forgot or merely learned to hide what it knows. Second, using small post-cutoff samples as ``robustness checks" is invalid: when these samples are small, genuine forecasting skill and undetected memorization produce statistically indistinguishable results. Finally, any evidence of memorization constitutes only a lower bound: the fact that one test fails to elicit memorized knowledge does not mean it is absent, since different prompts or contextual cues may trigger it.

\begin{figure}[htbp]
    \centering
    \caption{Recall of exact numerical levels of market indices.} \label{fig:indices}
    \caption*{\scriptsize This figure shows the LLM's estimated closing prices of the stock market indices compared to the actual values. Panels A, C, and E graph the actual values against the estimated values. Panels B, D, and F show the estimation error for the S\&P 500, Dow Jones Industrial Average, and Nasdaq Composite. Estimation error is calculated as \textit{(Estimated - Actual)/Actual} and is shown in percentage points (5 means 5\%). For the Nasdaq Composite panels, 10 outliers were removed for the ease of plotting. These values are still included in the evaluation metrics table. The post-cutoff period (10/2023 onward) is shaded gray.}
    \includegraphics[width=\linewidth]{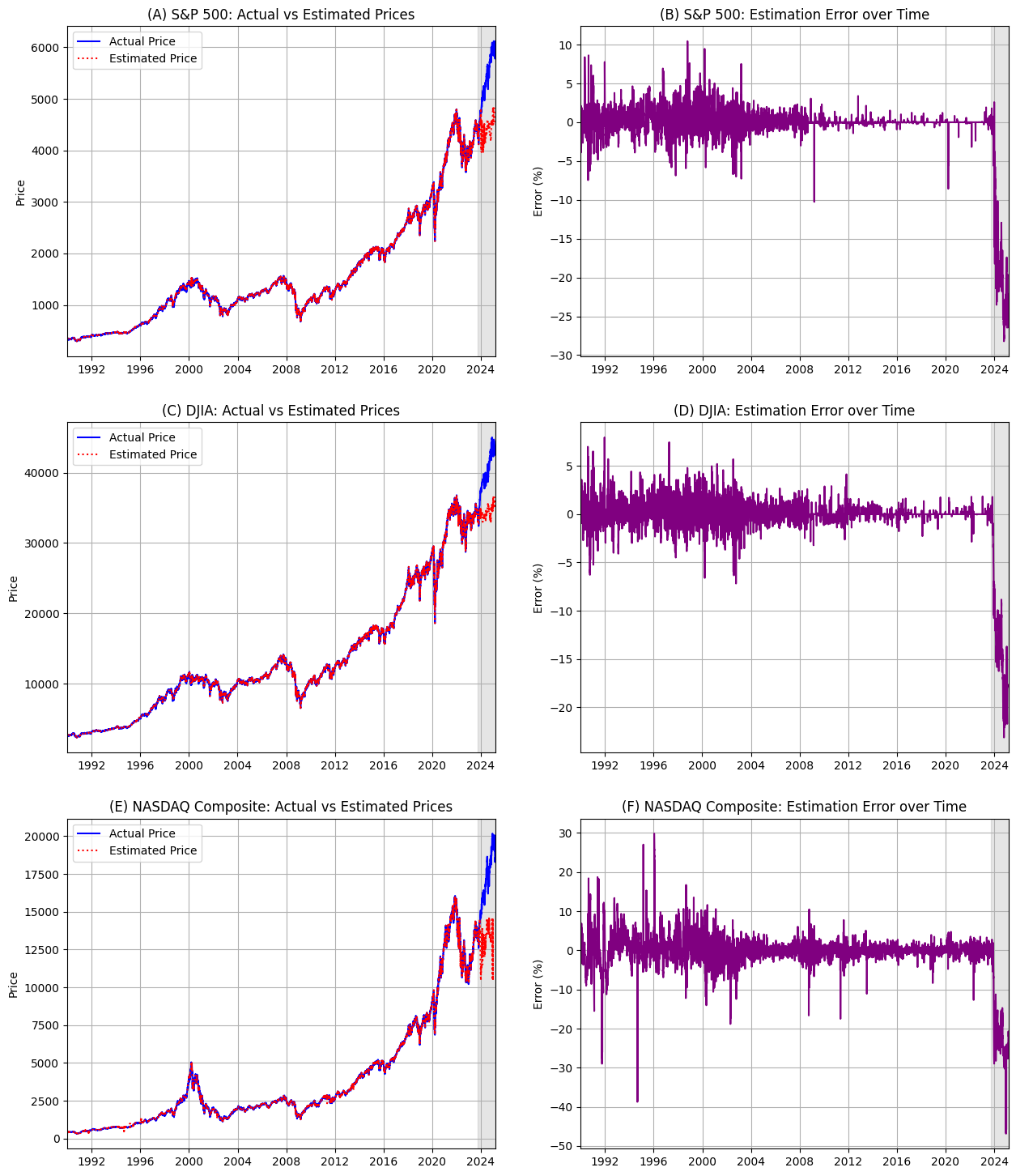}
\end{figure}

Using a novel testing framework, we show that LLMs can recall exact numerical values of economic data from their training. For example, before its knowledge cutoff date of October 2023, GPT-4o can recall specific S\&P 500 index levels with perfect precision on certain dates, unemployment rates accurate to a tenth of a percentage point, and precise quarterly GDP growth figures. Figure \ref{fig:indices} shows the model's memorized values for three stock market indices (S\&P 500, Dow Jones Industrial Average, and Nasdaq Composite) along with the actual values and the associated errors. In addition to U.S. and international macro indicators and market indices, we find clear evidence of memorization of individual security prices, which is more pronounced for prominent stocks and in recent periods. The model also shows near-perfect identification of the dates of major media front-page headlines. In short, we document pervasive memorization.   

Next, we demonstrate that standard techniques to constrain model knowledge fail to prevent access to memorized data. When we instruct GPT-4o to ignore any information after 2010 when forecasting quarterly GDP growth direction, the model achieves 97.6\% threshold accuracy before the artificial cutoff and 98.0\% after, nearly identical performance despite the explicit constraint.\footnote{Threshold accuracy measures the percentage of correct predictions for whether GDP growth exceeds 2.5\%, a threshold chosen to create a balanced binary classification (roughly 50\% of observations are above this threshold in the historical data).} In contrast, actual post-knowledge cutoff accuracy is only 40\%. These results suggest prompt-based constraints do not work in practice. Yet, even if constrained prompts produced lower accuracy, theory explains why this result would be equally uninformative: forecasting performance is not identified, consistent with the model having memorized the outcome and role-playing as a worse forecaster or truly being able to forget the underlying information.

To understand which applications are vulnerable, we must distinguish tasks where future knowledge affects the answer from those where it does not. The memorization problem applies to all tasks that are not \emph{future-invariant}: tasks where an analyst's answer would differ knowing future outcomes. Simple extraction tasks where the correct answer is identical regardless of future knowledge (such as extracting entity names or objective numerical facts) are typically unaffected. However, many tasks that appear to be pure extraction are actually judgment-laden: assessing whether something is ``relevant" or ``important," identifying ``risks," classifying ``sentiment," and generating ``expectations" all involve interpretations that can be influenced by knowledge of subsequent outcomes. For such tasks, the model's parameters encode future information despite pre-cutoff inputs: lookahead bias in the function rather than the data. Section~\ref{sec:formalization} formalizes this distinction and provides concrete guidance on common task categories. 

Masking techniques (anonymizing entity or company names or dates) present a more nuanced picture. In theory, valid masking requires two conditions: future-invariance (the masked task does not depend on post-cutoff information) and detectable skill (the model shows statistically significant performance). When both conditions hold, masking enables either task-preserving use (recovering the original target of estimation) or capability-demonstrating use (establishing the model's genuine capability on a different class of tasks). Without detectable skill, poor performance creates another identification problem: we cannot distinguish lack of capability from overly aggressive masking that removes necessary information.

In practice, attempts to prevent LLMs from accessing future information through masking face significant challenges. We show that LLMs can reconstruct original entities from seemingly minimal contextual clues in complex financial documents. For example, Figure \ref{fig:anonym_call} shows that when we present GPT-4o with an anonymized Ethan Allen (ETH) earnings call transcript, where company names, numbers, locations, and dates were all masked using the entity neutering approach proposed by \citet{engelbergEntityNeutering2025}, the model still correctly identified the company (ETH), quarter (Q1), and year (2018). The transcript contained only generic business language such as ``Our adjusted EPS of number\_e increased number\_f percent from the prior year," yet the model still recovers the precise corporate identity and reporting period.

\begin{figure}[htbp]
\input{Figures/conference_call_example}
\end{figure}

Systematically analyzing anonymized earnings call transcripts, we find GPT-4o correctly identifies the company in 100\% of Apple, Meta, and Microsoft calls and in above 85\% of the calls for all Magnificent Seven firms.\footnote{The Magnificent Seven refers to seven prominent technology companies: Apple (AAPL), Amazon (AMZN), Alphabet/Google (GOOGL), Meta (META), Microsoft (MSFT), NVIDIA (NVDA), and Tesla (TSLA).} For Apple, the model achieves 93.2\% accuracy in identifying the correct quarter and year. Successful reconstruction proves the masked task is not future-invariant---the model can still access contextual knowledge about those entities. Moreover, even when reconstruction fails, the model may access memorized knowledge through untested aggregate channels (e.g., industry patterns, business conditions, macroeconomic context). In short, reconstruction tests probe only a finite set of pathways; failure to reconstruct does not prove that future-invariance holds.


Further, memorization is not restricted to prompt-based tasks; even embedding vectors show signs of it.\footnote{Embeddings are high-dimensional vector representations that the model learns during training to place semantically similar text near each other in vector space. They are extracted from the model's internal representations (token or pooled sentence vectors) and may be further refined through contrastive learning to improve semantic similarity matching.} We test for memorization by encoding textual prompts that omit a specific numeric value for a given date (e.g., ``In Q4 2020, the earliest estimate of the US GDP growth rate was"), computing their embeddings, and regressing macroeconomic outcomes on those embeddings. Under a no-memorization null, these embeddings should have no predictive power; instead, we find evidence of predictive content for variables such as inflation and the unemployment rate.

Given these challenges with methods attempting to circumvent memorization, reliable evaluation of LLMs’ genuine forecasting abilities can only be conducted using data after their knowledge cutoff dates. One approach is to employ models explicitly designed with temporal cutoffs \citep[e.g.,][]{sarkarStoriesLMFamilyLanguage2024, rahimikiaReVisitingLargeLanguage2024, heChronologicallyConsistentLarge2025}. Another approach is to restrict the
analysis exclusively to the post-knowledge cutoff period \citep[e.g.,][]{lopez-liraCanChatGPTForecast2023, phamCanBaseChatGPT2024, bozman2025better}. Only by testing predictions for periods the models have not been exposed to during training can we confidently distinguish genuine forecasting ability from memorization. 



At a minimum, we recommend using our methodology to test whether the model has memorized the information in each research setting. Whenever an LLM's output would differ with the benefit of future knowledge, applying it to data within its training period is inherently risky.

\textbf{Related Literature and Contributions.} We contribute to the literature on the limits of LLMs in economics and finance through theoretical and empirical advances. Theoretically, we prove that when models have memorized outcomes, forecasting ability is non-identified. Empirically, we demonstrate that LLMs exhibit systematic memorization of economic and financial data at scale. Our key methodological contribution is a testing framework that reveals what indirect tests miss, providing unambiguous evidence of memorization.

Previous research has relied on clever, context-specific experiments to infer lookahead bias. \citet{sarkar_lookahead} use the appearance of COVID-19 in pre-pandemic firm risk factors as evidence that Llama models inadvertently include future information. While such natural experiments are valuable, they require specific historical events and cannot reveal the full extent of the problem. \citet{levyCautionAheadNumerical2024} find that GPT-4o performs poorly in numerical tasks and that perturbing financial statements causes LLMs' predictive accuracy to drop to random chance, and conjecture that LLMs are memorizing.\footnote{In other literature revealing LLMs' limitations, \citet{rossLLMEconomicusMapping2024} apply utility theory to evaluate economic biases in LLMs, revealing that the economic behavior of these models is neither fully rational nor entirely human-like. Furthermore, \citet{chenWhatDoesChatGPT2024} examines how LLMs forecast stock returns, finding that they exhibit human-like behavioral biases, such as over-extrapolation from recent performance, while being better calibrated in confidence intervals than humans.} Unlike these studies, our method is universally applicable: by directly eliciting numeric values using only variable names and dates, any researcher can test any economic series for memorization without needing creative designs. Moreover, our method also demonstrates how memorization affects embedding-based approaches.

Contemporaneous work by \citet{ludwigLargeLanguageModels2025} develops an econometric framework for using LLMs and identifies ``training leakage'' (whether specific texts in the researcher's sample were in the model's training data) as a threat to valid inference. We address a more fundamental problem: functional lookahead bias, where the model's parameters encode post-$t$ information learned from the aggregate training corpus, thereby contaminating the decision rule even when the specific input text was never seen in training. We prove the target estimand is non-identified due to observational equivalence (Proposition~\ref{prop:nonid}, Corollary~\ref{cor:sharp-nonid}) and common remedies provably fail (Corollaries~\ref{cor:finetuning}, Proposition~\ref{prop:lowpower}). Empirically, we provide systematic evidence that models have memorized outcomes and fundamentals at scale for economic and financial data and develop a direct elicitation framework to test what information is encoded in model parameters.

Research has also focused on potential solutions. \citet{sarkarStoriesLMFamilyLanguage2024}, \citet{rahimikiaReVisitingLargeLanguage2024}, and \citet{heChronologicallyConsistentLarge2025} train chronologically consistent language models that avoid entirely the lookahead bias by training different checkpoints on a dataset that is temporally ordered. \citet{engelbergEntityNeutering2025} proposes ``entity neutering", using LLMs to remove identifying information from text, and finds that masked text maintains similar sentiment and return predictability as unmasked text. Relatedly, \citet{glassermanAssessingLookAheadBias2023} find that forecasting with anonymized headlines outperforms originals within the training window, suggesting that the distraction effect from general company knowledge (in the case of anonymized headlines) appears to outweigh lookahead bias. Finally, other researchers have restricted themselves exclusively to the post-knowledge cutoff period to avoid lookahead bias \citep[e.g.,][]{lopez-liraCanChatGPTForecast2023, phamCanBaseChatGPT2024, bozman2025better}, exploiting the fact that the older GPT-3.5 and GPT-4 versions have a knowledge cutoff date of September 2021. 

ChatGPT and other LLMs have been recently used in forecasting or eliciting expectations of diverse economic series that include LLMs' training period by querying the model \citep[e.g.,][]{chenChatGPTDeepseekCan2023, bond2024large, tanLargeLanguageModels2024, jhaChatGPTCorporatePolicies2025, degenLargeLanguageModels2024}. Our findings suggest that caution is warranted when interpreting some of these results, as apparent forecasting accuracy may reflect the model's memorization of training data rather than genuine predictive capability. Moreover, studies that find inaccuracies or biases in LLM predictions during their training period may not be measuring actual forecasting limitations but instances where the model attempts to provide helpful responses by pretending not to know information it has memorized. 

With the growing number of applications of LLMs in economics and finance research \citep[e.g.,][] {jhaHarnessingGenerativeAI2024,caoCanGenerativeAI2025,vanbinsbergenTextualAnalysisShortseller2022, bai2023executives, chenExpectedReturnsLarge2022, kim2024financial, beckmannUnusualFinancialCommunication2024, breitungGlobalBusinessNetworks2025, bybeeGhostMachineGenerating2023, hortonLargeLanguageModels2023, hansenSimulatingSurveyProfessional2024, manningAutomatedSocialScience2024,clayton2025geoeconomic}, greater work is needed to evaluate the extent to which LLM memorization may affect specific applications such as similarity matching, sentiment scoring, survey generation, and expectation extraction.

\section{A Formalization of the Memorization Problem}
\label{sec:formalization}

Econometrically, the problem is an information-set and parameter mismatch. The researcher wishes to evaluate decisions using only what was known at time $t$, but the model's parameters already encode information revealed after $t$. This section provides intuition with simplified notation; formal definitions, assumptions, and proofs are in Appendix~\ref{sec:formalization-proof}.

\subsection{Lookahead Bias in LLM Forecasting}
\paragraph{Setup.}
Let $I_t$ denote information available at $t$ and $I_{>t}$ information revealed later; $I_{\text{all}}=I_t\cup I_{>t}$. Let $Q_t$ be a task posed ``as of $t$'' (comprising an instruction and any input data drawn from $I_t$). Let $Y_{\text{true}}$ denote the realized outcome (revealed after time $t$). Let $\theta=\theta(I_{\text{all}})$ be the \textbf{factual parameters} of the fully trained model (trained on all data) and $S_\phi(\cdot)$ the model's internal scoring function, where $\phi$ denotes generic model parameters. We define $\theta_t=\theta(I_t)$ as the \textbf{counterfactual parameters} that \textit{would have been obtained} by applying the same training procedure to the restricted information set $I_t$. We model the LLM's output as the optimal decision $y$ from a set of possible answers $\mathcal{Y}$, selected by the decision rule $\delta_\phi(Q,P) \coloneqq \argmax_{y\in\mathcal{Y}} S_\phi(y;Q,P)$, where $Q$ is the task and $P$ represents additional instructions that modify the task (with $P=\varnothing$ denoting no additional instructions beyond the task itself).\footnote{We focus on deterministic (greedy) decoding for clarity. The results generalize to stochastic decoding (temperature $> 0$); see Appendix~\ref{sec:formalization-proof}.}

\paragraph{Ideal (target estimand).}
The counterfactual decision the researcher wants is

\begin{equation}\label{eq:greedy-argmax}
Y_t^\star \;\coloneqq\; \delta_{\theta_t}(Q_t, \varnothing) \;=\; \argmax_{y\in\mathcal{Y}} S_{\theta_t}\!\big(y; Q_t, \varnothing\big).
\end{equation}

\paragraph{Observed (implemented estimand).}
In practice, we observe the decision of the fully trained model,
\begin{equation}\label{eq:llm-argmax}
\begin{aligned}
Y_{\mathrm{LLM}} &= \delta_\theta(Q_t, \varnothing) \;=\; \argmax_{y\in\mathcal{Y}} S_{\theta}\!\big(y; Q_t, \varnothing\big),\\
\theta &= \theta(I_{\text{all}}).
\end{aligned}
\end{equation}

\begin{definition}[Task invariance and lookahead bias]\label{def:lookahead-bias}
A task $Q_t$ is either:
\begin{itemize}
    \item \textbf{Future-variant:} The task is \emph{future-variant} (equivalently, exhibits \emph{lookahead bias}) if the LLM's decision depends on post-$t$ information embedded in its parameters:
    \begin{equation}\label{eq:future-variant}
    \delta_{\theta_t}(Q_t, \varnothing) \;\neq\; \delta_{\theta}(Q_t, \varnothing),
    \end{equation}
    that is, the counterfactual answer (what the model would have predicted if trained only on pre-$t$ information) differs from the answer produced by the fully trained model. This definition is model-relative: it assumes the training procedure is deterministic conditional on the data, such that the only difference between $\theta$ and $\theta_t$ is the inclusion of future information, $I_{>t}$.

    \item \textbf{Future-invariant:} The task is \emph{future-invariant} if the identity of the optimal answer is invariant to post-$t$ information:
    \begin{equation}\label{eq:future-invariant}
    \delta_{\theta_t}(Q_t, \varnothing) \;=\; \delta_{\theta}(Q_t, \varnothing),
    \end{equation}
    that is, training on the full information set $I_{\text{all}}$ vs. the restricted set $I_t$ yields the same answer.
\end{itemize}
\end{definition}

Since the model's parameters encode information from the full training data, other uses of these parameters (e.g., embedding vectors) may also reflect memorized information.

Lookahead bias is pervasive, affecting tasks requiring judgment, selection, or prediction. It contaminates any task that is not future-invariant. For example, classifying the sentiment of a 2007 housing-related article: under $\theta_t$, \textsc{``Optimistic''} may be the answer that maximizes the score; under $\theta$ (which includes the 2008 crisis), \textsc{``Negative''} may instead be optimal. This differs from standard data lookahead bias in econometrics (e.g., accidentally including Q4 data in a Q3 predictive regression): the input $Q_t$ here is valid and contains only information available at time $t$, but the function $\delta_\theta$ processing it is contaminated by future information embedded in the model's parameters (i.e., functional-form lookahead bias).

\paragraph{Practical guidance for researchers.}
To apply this distinction in practice, researchers can assess whether a task is future-invariant by asking: Would an analyst's answer differ if they knew what happened after time $t$? If the answer is yes, or even plausibly yes, the task is not future-invariant. When recovering answers about specific firms, events, or time periods, using LLM outputs on pre-cutoff data is problematic. 

Common categories of tasks:

\textit{Likely future-invariant (safe for pre-cutoff use):}
\begin{itemize}[noitemsep]
\item Factual extraction: entity names, dates, reported numbers
\item Structural parsing: word counts, sentence boundaries, document structure
\item Objective classification: language detection, part-of-speech tagging
\end{itemize}

\textit{Not future-invariant:}
\begin{itemize}[noitemsep]
\item Judgment-based assessment: identifying ``important'' or ``relevant'' information
\item Sentiment and tone analysis: evaluating whether text is positive or negative
\item Risk and uncertainty: extracting ``risks,'' ``concerns,'' or ``uncertainties''
\item Forecasting and expectations: generating predictions or eliciting expectations
\item Similarity and comparison: determining which documents are ``similar'' in economically meaningful ways
\end{itemize}

For the future-variant tasks in the second category, pre-cutoff use is invalid. Regardless of task type, researchers should verify what the model has memorized using our direct elicitation procedure (see Section~\ref{sec:methodology}). Even seemingly safe factual extraction tasks may be contaminated if the model has memorized specific facts relevant to the analysis. For instance, a task like ``Extract all risks mentioned in this 2007 earnings call" is \emph{not} future-invariant, even though it appears to be a simple extraction. The concept of ``risk" is judgment-laden---what the model identifies as salient risks is informed by knowledge of which concerns actually materialized into problems.

\begin{remark}[Lower bound property of memorization tests]\label{rem:lower-bound}
Any memorization test establishes only a \textbf{lower bound} on what is encoded in $\theta$: positive evidence is conclusive, but negative evidence is not.
\end{remark}

Different query-prompt pairs may access different subsets of encoded information through different mechanisms, so failure to elicit knowledge in a particular test does not prove its absence. Consequently, conditioning on cases where tests fail (e.g., restricting to observations where the model cannot identify the firm) does not resolve non-identification---the model may still access memorized information through alternative contextual cues.

\subsection{Non-identification of Constrained Forecasts}

Researchers sometimes add constraining prompts $P$ (e.g., ``use only data before 2010''), hoping to emulate $Y_t^\star$ without retraining the model. For any such prompt, define the constrained output as

\begin{equation}\label{eq:constrained-argmax}
\begin{aligned}
Y_{\mathrm{constrained}}(P)
&\coloneqq \delta_{\theta}(Q_t, P) \;=\; \argmax_{y\in\mathcal{Y}} S_{\theta}\!\big(y;\,Q_t, P\big),\\
\theta &= \theta(I_{\text{all}}).
\end{aligned}
\end{equation}

The researcher's goal is to use the observable $Y_{\mathrm{constrained}}(P)$ as a proxy for the unobservable, ideal estimand $Y_t^\star$. The central methodological question is therefore whether $Y_t^\star$ can be uniquely recovered from the constrained outputs. This is a formal identification question. The following proposition states that this is not possible.

\begin{proposition}[Non-identification]\label{prop:nonid}
Given observations $Y_{\mathrm{constrained}}(P)$ for any set of constraining prompts $\{P_k\}$ designed to restrict the model to information available at time $t$, the ideal estimand $Y_t^\star$ cannot be uniquely identified. Any observed constrained output is consistent with multiple, contradictory values of the true counterfactual forecast $Y_t^\star$.
\end{proposition}

The problem is observational equivalence: any observed output is consistent with both the constraint working (the model genuinely forecasts using only pre-$t$ information) and the constraint failing (the model retrieves a memorized answer). Without additional assumptions, these alternatives cannot be distinguished. See Appendix~\ref{sec:formalization-proof} for the formal construction.

\begin{corollary}[Sharp non-identification]\label{cor:sharp-nonid}
Under the same conditions as Proposition~\ref{prop:nonid}, for any observed constrained output, the identified set for $Y_t^\star$ equals the entire label set $\mathcal{Y}$. That is, observing constrained outputs provides zero information about the ideal estimand.
\end{corollary}

\noindent(See Appendix~\ref{sec:formalization-proof} for the proof.)

To illustrate the practical implications, consider the realized outcome that was eventually revealed after time $t$. Let $Y_{\text{true}} \in \mathcal{Y}$ denote this true outcome. Counterintuitively, both good and bad forecasts are uninterpretable due to exact observational equivalence. A ``good'' forecast ($Y_{\mathrm{constrained}} = Y_{\text{true}}$) could indicate the constraint worked and the model demonstrated genuine skill, or that the constraint failed and the model simply recalled the memorized answer. A ``bad'' forecast ($Y_{\mathrm{constrained}}\neq Y_{\text{true}}$) is equally uninformative: the constraint may have worked, yielding a genuine but incorrect forecast ($Y_t^\star = Y_{\mathrm{constrained}}$), or the model may be using post-$t$ information in unintended ways (for instance, drawing on memorized knowledge of what historical forecasts or beliefs looked like to simulate a period-appropriate answer). As our empirical tests show, implausibly high accuracy across many items suggests constraints typically fail. Consequently, constrained prompts render forecasts fundamentally uninformative about the target estimand.

\begin{corollary}[Black-box fine-tuning]\label{cor:finetuning}
The same non-identification result holds for black-box fine-tuning to ``forget'' post-$t$ information: without white-box verification, $Y_t^\star$ cannot be identified from fine-tuned outputs.
\end{corollary}

The problem is observational equivalence between genuine forgetting (parameters changed to approximate $\theta_t$) and behavioral suppression (post-$t$ information still embedded, but the model learned to hide it). See Appendix~\ref{sec:formalization-proof} for the formal construction.

\subsection{Masking as a Potential Solution}

Masking procedures attempt to remove identifying information (firm names, dates, industry labels) from tasks, hoping to prevent the model from accessing memorized information while preserving the ability to perform useful analysis.

\begin{definition}[Valid masking]\label{def:valid-masking}
Let $M: Q_t \mapsto Q_t^{\text{mask}}$ be a masking procedure. Masking is \textbf{valid} if:
\begin{enumerate}[label=(\roman*)]
    \item \textbf{Future-invariance:} $\delta_{\theta_t}(Q_t^{\text{mask}}, \varnothing) = \delta_{\theta}(Q_t^{\text{mask}}, \varnothing)$.
    \item \textbf{Detectable skill:} The model's accuracy on pre-cutoff masked data exceeds baseline ($p^{\text{mask}} > p_0$).
\end{enumerate}
\end{definition}

Condition (ii) is necessary for identification: without it, poor performance could reflect either lack of capability or over-aggressive masking---observationally equivalent alternatives analogous to Proposition~\ref{prop:nonid}. Valid masking enables two research uses: \textbf{task-preserving} (if $\delta_{\theta}(Q_t^{\text{mask}}, \varnothing) = Y_t^\star$, recovering the original estimand) or \textbf{capability-demonstrating} (if the estimand changes, establishing capability on the masked task class).

\begin{remark}[Reconstruction tests and the lower bound property]\label{rem:masking-verification}
By Remark~\ref{rem:lower-bound}, reconstruction tests provide asymmetric evidence: successful reconstruction proves condition (i) fails, but failure to reconstruct does not prove it holds.
\end{remark}

Verifying condition (i) requires both empirical evidence (reconstruction fails) and theoretical arguments (no information pathway exists), analogous to establishing instrument variable exogeneity in econometrics. When masking preserves future-invariance, it enables task-preserving use (recovering the original estimand); when it changes the estimand but maintains future-invariance, it enables capability-demonstrating use. See Appendix~\ref{sec:formalization-proof} for details.

\subsection{Post-Cutoff Data and the Identification Problem}

Post-cutoff data provide a potential solution to Proposition~\ref{prop:nonid}: since the model has not been trained on these data, post-cutoff accuracy directly measures forecasting ability. However, a common practice is to report pre-cutoff accuracy as the primary result and use small post-cutoff samples as ``robustness checks,'' arguing that similar accuracy implies no memorization. The next proposition shows why this practice is unjustified.

\begin{proposition}[Statistical indistinguishability in pre- vs. post-cutoff comparisons]\label{prop:lowpower}
When post-cutoff sample size is small relative to pre-cutoff sample size, genuine forecasting skill and undetected memorization cannot be reliably distinguished by hypothesis tests. Economically substantial memorization gaps remain statistically undetectable due to low power.
\end{proposition}

Consequently, if post-cutoff data are sufficient to be informative, they should be the primary analysis, not a robustness check; if they are insufficient, comparing pre- and post-cutoff accuracy fails to distinguish skill from memorization. See Appendix~\ref{sec:formalization-proof} for the formal power analysis.

\section{Methodology} \label{sec:methodology}

To evaluate LLMs' memorization of economic and financial data, we develop a testing framework that isolates recall abilities from forecasting. Our approach formalizes the information environment by providing a context set $x_t$ and requesting a prediction about $y_{t+1}$, where $t$ represents a specific point in time. The query structure explicitly references periods, asking the model to provide economic or financial data for particular dates. For instance, we might ask ``What was the level of the S\&P 500 on May 2nd, 2020?'' 

We vary the information set $x_t$ to isolate different memory access mechanisms. In the baseline case, we provide no context, testing the model's pure recall ability. We then augment this with two progressively richer information environments: (1) historical context containing the recent history of $y_t$ up to time $t$, and (2) news context including headlines from major financial publications from the period leading up to $t$.

Our tests of LLMs' recall capabilities span several categories of economic variables. First, we test macroeconomic indicators by querying precise values (e.g., unemployment rates) and directional trends.  Second, we examine stock market indices through questions about exact numerical levels, directional changes, percentage movements, and relative performance. For example, we ask for the S\&P 500 closing value on specific dates or whether the NASDAQ increased or decreased on particular days. Third, we test LLMs' ability to identify news headlines' dates by presenting sets of \textit{The Wall Street Journal} front-page headlines without dates and asking it to identify when these headlines appeared and to predict the corresponding S\&P 500 level on the next trading day. Fourth, we assess their ability to recall individual securities information, including specific stock price levels and directional movements. Fifth, we examine the effectiveness of requesting LLMs to impose artificial knowledge cutoff dates. Finally, we test the LLM's ability to recover company identities from anonymized firm-specific text, including earnings calls and news headlines. 

\subsection{Period Selection and Supplementary Analysis}

Our experimental design strategically spans three temporal zones relative to an LLM's training cutoff date. First, we include periods before the cutoff where we expect high recall accuracy if memorization occurs. Second, we examine the periods within 10 years preceding the cutoff to assess potential recency effects in memorization patterns. Finally, we include post-cutoff periods as a control condition, where memorization is not expected.

\subsection{Prompt}

We implement a standardized prompt template across all model queries to ensure consistency and minimize experimental variation. Each prompt includes an optional context section, a specific question about economic data, and explicit instructions for response formatting. The general template is:

\begin{singlespace}
\begin{quote}
\texttt{[Context: \{context\_information\}]\\
\\
\{question\_about\_economic\_data\}\\
\\
Provide a precise answer based on your knowledge. Indicate your level of confidence. Format as a JSON object with the following fields:
\begin{itemize}
    \item \texttt{answer}: The precise answer to the question.
    \item \texttt{confidence}: A number between 0 and 100 indicating the model's confidence in its answer.
\end{itemize}
}
\end{quote}
\end{singlespace}

The context information represents the information set $x_t$ for our experimental conditions, which may be empty (when testing pure recall), contain historical data points, include relevant news headlines, or provide general knowledge about the period. More detailed information is provided in Section A of the Appendix. 

For example, in a prompt testing recall with historical context, we might provide: ``Context: The S\&P 500 closed at 2,834.40 on March 14, 2019, and at 2,808.48 on March 13, 2019. What was the S\&P 500 closing value on March 15, 2019?'' This standardized approach allows us to systematically vary the information provided while controlling for confounding factors in question phrasing or response expectations.

\subsection{LLMs}

Given its wide usage in research, we conduct our main analysis using ChatGPT-4o with the specific version of ``gpt-4o-2024-08-06.'' This model is a snapshot of GPT-4o, ensuring consistent performance and behavior. It will not receive updates, and its training data ends in October 2023.\footnote{More information about this model is available here: \url{https://platform.openai.com/docs/models/gpt-4o}.} Importantly, the temperature of the model is set to 0 for all of the analyses to maximize the reproducibility of the results.\footnote{Temperature is a parameter of ChatGPT models that governs the randomness and the creativity of the responses. A temperature of 0 essentially means that the model will always select the highest probability word conditional on the text, which will eliminate the effect of randomness in the responses and maximize the reproducibility of the results.} In additional analyses, we also test for the recall capabilities of open-source models such as Llama-3.1-70b-Instruct.\footnote{Llama-3.1-70B-Instruct is Meta’s 70B-parameter instruction-tuned Llama-3.1 model, released July 23, 2024. More information is available here: \url{https://huggingface.co/meta-llama/Llama-3.1-70B-Instruct}.} 


\section{Data}

Most of the tests use data from January 1990 to September 2023, which precedes GPT-4o’s knowledge cutoff date of October 2023. To test the LLM's memorization, we use three categories of datasets: (1) stock index and individual stock prices, (2) macroeconomic indicators, and (3) textual data on WSJ front-page news, earnings calls, and firm-specific news headlines. 

We ask the LLM to give us the closing value of the stock indices and a sample of individual stocks. We use the daily closing values of the S\&P 500, the Dow Jones Industrial Average, and the Nasdaq Composite from Yahoo Finance to evaluate the LLM's answers. We use Center for Research in Security Prices (CRSP) data to obtain daily stock market data for individual stock closing prices. The sample of individual stocks includes the Magnificent 7 (AAPL, AMZN, GOOGL, META, MSFT, NVDA, TSLA) and randomly selected subsamples of stocks in different size categories, representing a total of over 4,200 unique ticker symbols. 

We also ask the LLM to give us estimates of various macroeconomic indicators. The indicators we test are (i) US GDP growth, (ii) inflation, (iii) unemployment rate, (iv) 10-year Treasury Yield, (v) VIX, (vi) housing starts, and (vii) change in nonfarm payrolls. We obtain the actual unemployment rate and 10-year Treasury Yield values from Federal Reserve Economic Data (FRED). We obtain the VIX levels from Yahoo Finance. For GDP growth, inflation, housing starts, and change in nonfarm payrolls, we use the Philadelphia Federal Reserve Real-Time Data Set to get the first vintage and ask the LLM to give us the earliest estimate of these indicators. 

The textual data we use include \textit{The Wall Street Journal} (WSJ) front-page headlines obtained from Factiva, earnings conference call transcripts from Capital IQ, firm-specific headlines, and the RavenPack news database. The WSJ front-page news dataset comprises 90,123 headlines from January 1990 to February 2025. There are, on average, approximately 9 headlines for each date. Given each set of headlines, we ask the LLM to provide the date and S\&P 500 level on the next trading day. The conference call dataset starts in January 2006 and ends in May 2024. We extract the opening statement delivered by the CEO, anonymize the text using an entity neutering approach as proposed by \citet{engelbergEntityNeutering2025}, and ask the LLM to provide the firm, quarter, and year of the conference call. 

We implement a similar test for the firm-specific headlines from January 2000 to June 2024. We follow the filtering procedures based on the RavenPack news database as in \citet{lopez-liraCanChatGPTForecast2023}, including keeping only headlines with a relevance score of 100, keeping complete articles and press releases only and excluding headlines categorized as ``stock-gain'' and ``stock-loss'', and avoiding repeated news through requiring the “event similarity days” to exceed 90 and removing duplicate headlines. We arrive at a sample of 1,358,737 headlines covering 4,391 unique firms from January 2000 to June 2024. 
About 70\% of headlines are from overnight articles released either before 9 a.m. or after 4 p.m., with the remaining 30\% from intraday articles.

\section{Results}

In this section, we present a comprehensive evaluation of GPT-4o's memorization of economic and financial data, spanning macro indicators, market indices, individual stocks, headlines, and attempts to mitigate memorization through fake knowledge cutoffs and masking. Across these domains, we assess the model's ability to recall precise values, identify contextual details, and adhere to constraints. 
Our goal is to study the extent and selectivity of memorization, highlighting its implications for using LLMs in economic forecasting and the challenges of isolating genuine predictive ability. Each subsection examines a specific data type or mitigation strategy, building a cohesive picture of how memorization manifests and persists. 

First, we establish that memorization exists through direct elicitation tests on pre-cutoff data, where we query the model for specific values using only variable names and dates. High accuracy on these tests provides unambiguous evidence of memorization. Second, we examine post-cutoff performance as descriptive evidence regarding data contamination. When post-cutoff samples are small, hypothesis tests have low power to distinguish genuine forecasting skill from undetected memorization. Therefore, we interpret post-cutoff results cautiously: poor post-cutoff performance is descriptively consistent with absence of data contamination, but our post-cutoff samples are often small and do not constitute definitive statistical tests.

\subsection{Macroeconomic Data Recall}

To assess GPT-4o's memorization of macroeconomic indicators, we tested its ability to recall monthly values across various variables (quarterly for GDP). In the baseline analysis, we use data from January 1990 to September 2023, all within the model's training cutoff of October 2023. Importantly, for comparison, we also examine the post-cutoff period from October 2023 to February 2025. The indicators were divided into two groups: rates (GDP Growth, Inflation, Unemployment Rate, and 10-year Treasury Yield) and levels (Housing Starts, VIX, and Nonfarm Payrolls). For rates, we requested percentage values, evaluating accuracy through Mean Error, Mean Absolute Error, Threshold Accuracy (correctly identifying whether the rate was above a threshold value), and Directional Accuracy (correct direction of change from the previous period). For levels, we requested raw values, with errors calculated relative to actual levels through Mean Percent Error, Mean Absolute Percent Error, Threshold Accuracy, and Directional Accuracy. We also examined levels over the 10-year period preceding the cutoff to explore potential recency effects. We measure performance against the actual values of macro indicators and report the findings in Table \ref{tab:macro}.\footnote{Table \ref{tab:summary} in the Appendix reports summary statistics for the data to serve as benchmark accuracy rates for comparing the model's ability to recall values against the true data.}

\begin{figure}[htbp]
    \centering
    \caption{Recall of exact numerical values of macro indicators.} \label{fig:macro}
    \caption*{\scriptsize This figure shows the LLM's estimated values of macro indicators including \textit{Inflation}, \textit{10-yr Treasury Yield}, \textit{GDP Growth}, and \textit{Unemployment Rate} compared to the actual values. Panels A, C, E, and G graph the actual values against the estimated values. Panels B, D, F, and H show the estimation error. Estimation error is calculated as \textit{(Estimated - Actual)/Actual} and is shown in percentages (5 means 5\%). The post-cutoff period (10/2023 onward) is shaded gray.}
    \includegraphics[width=0.9\linewidth]{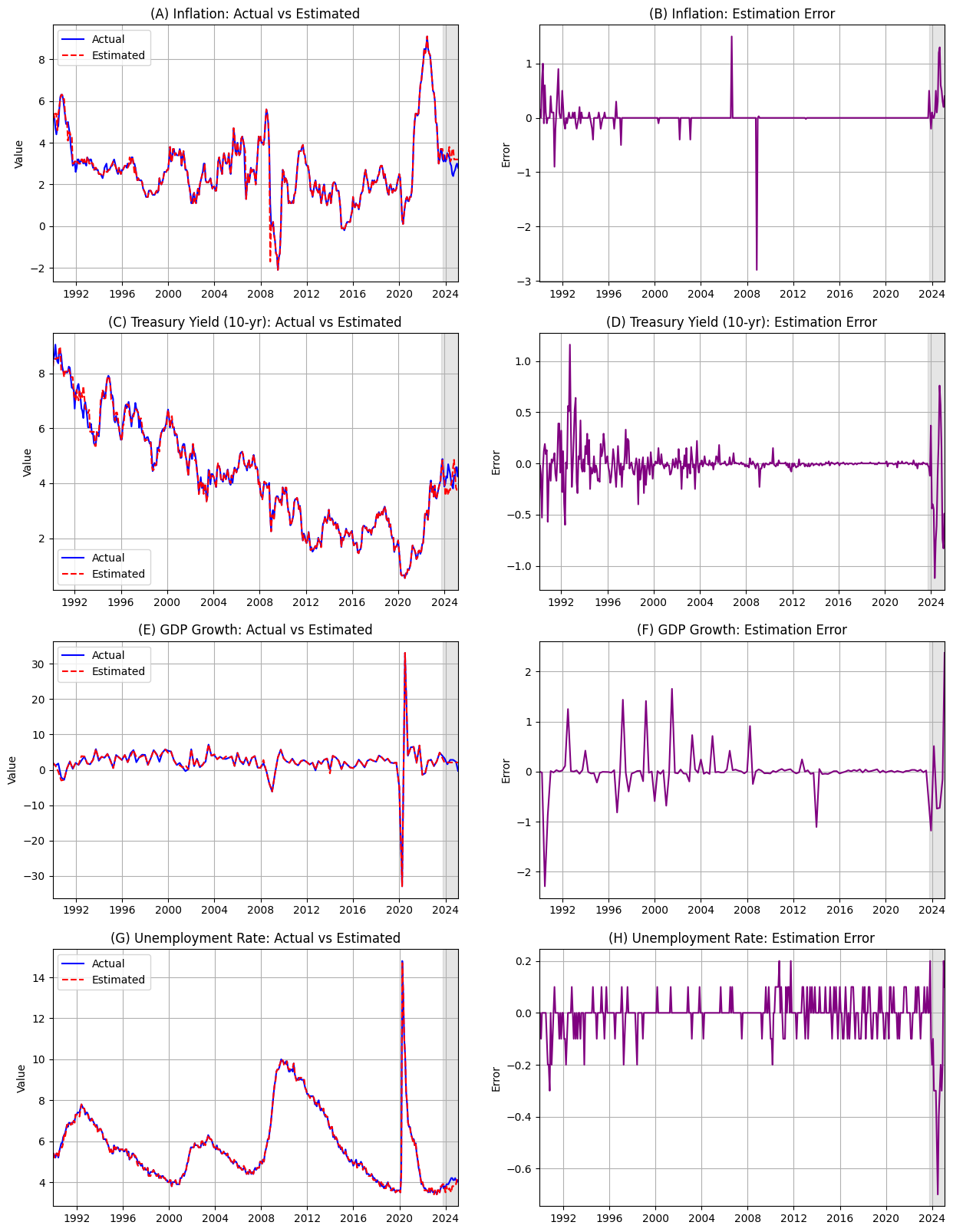}
\end{figure}

The results reveal a strong ability of GPT-4o to recall macroeconomic data. Figure \ref{fig:macro} plots the recalled values against actual values as well as the estimation error for GDP Growth, Inflation, Unemployment Rate, and 10-year Treasury Yield. The series for recalled and actual values are almost identical, particularly in more recent periods. For rates, the model demonstrates near-perfect recall, with Mean Absolute Errors ranging from 0.03\% (Unemployment Rate) to 0.15\% (GDP Growth) and Threshold Accuracy exceeding 96\% across all indicators, reaching 98\% for 10-year Treasury Yield and 99\% for Unemployment Rate. This set of results suggests that GPT-4o has memorized these percentage-based indicators with a high degree of accuracy.

For levels, the recall remains high, with Threshold Accuracies between 94\% and 100\% for all indicators during the whole pre-training sample. Moreover, when focusing on the most recent 10-year period in the pre-training sample, performance improves dramatically--Mean Absolute Percent Errors fall to 1.06\% for Housing Starts, 0.34\% for VIX, and 0.00\% for Nonfarm Payrolls, with Threshold Accuracy rising to 95\%–100\%. This recency effect indicates stronger memorization for more recent data, likely due to denser representation in the training corpus.

In contrast, the evaluation metrics in the post-cutoff period do not suggest recall. For rates, the Mean Absolute Errors increase significantly, ranging between 0.26\% (Unemployment Rate) to 0.95\% (GDP Growth). Threshold Accuracy falls to between 29.41\% (10-Yr Treasury Yield) to 52.94\% (Unemployment Rate). For levels, we see the same breakdown in recall. The Mean Absolute Percent Error jumps dramatically when compared to the recall in the more recent pre-cutoff period. For example, the Mean Absolute Percent Error rises from 0\% to 97.4\% for Nonfarm Payrolls. Threshold Accuracy drops from above 94\% for all indicators to 35.3\% (Housing Starts), 50.0\% (VIX), or 58.8\% (Nonfarm Payrolls). 

The high recall accuracy for rates and recent levels underscores the memorization problem when evaluating LLMs' forecasting capabilities during the training period. The model's ability to reproduce precise macroeconomic values, especially for percentage-based indicators and recent periods, suggests that apparent forecasting success for pre-cutoff data may stem from retrieving memorized information rather than genuine economic analysis. The weaker performance for levels over the full period, particularly for volatile indicators like Nonfarm Payrolls, hints at selective memorization, where certain data types or time frames are less reliably retained. On the other hand, we do not observe post-cutoff recall. 

\clearpage
\begin{landscape}
\begin{table}[htbp]
\centering
\caption{Evaluation Metrics for Macro Indicators}\label{tab:macro}
\caption*{\scriptsize This table reports a set of evaluation metrics for various macroeconomic indicators grouped into two panels: Rates and Levels. We ask the LLM to recall monthly values (quarterly for GDP, specific end of month date for 10-Year Treasury Yield and VIX) for each indicator. The indicators in the \textit{Rates} panel include GDP Growth, Inflation, Unemployment Rate, and the 10-Year Treasury Yield. For these indicators, we ask the LLM to give us a percentage. The \textit{Levels} panel includes Housing Starts, VIX, and Nonfarm Payrolls, evaluated over the full sample period. In each panel, we examine the periods before and after the knowledge cutoff date (October 2023) separately. The \textit{Levels, Recent Pre-cutoff Period: Past 10 years} panel evaluates these same indicators over a more recent, shorter period. \textit{Mean Error (ME)}, \textit{Mean Absolute Error (MAE)}, \textit{Mean Percent Error (MPE)}, \textit{Mean Absolute Percent Error (MAPE)}, \textit{Threshold Accuracy},and \textit{Directional Accuracy} are reported in percentage points (0.01 means 0.01\%). For \textit{Rates}, the \textit{ME} is the difference $Estimated Rate  - Actual Rate$. \textit{MAE} is calculated by taking the average of the absolute value of the \textit{ME}. \textit{Threshold Accuracy} is the proportion of predictions that correctly identify whether the rate or level is above a threshold value (2.5\% for GDP Growth, 3\% for Inflation, 4\% for Unemployment Rate, 4\% for the 10-Year Treasury Yield, 16 for VIX, 1400 for Housing Starts, and 200 for Nonfarm Payrolls). For \textit{Levels}, the \textit{MPE} is calculated by taking the average of the percent error $(Estimated Level  - Actual Level)/Actual Level$. \textit{MAPE} is calculated by taking the average of the absolute value of the percent error. \textit{Directional Accuracy} is the proportion of predictions that correctly identify the direction of change (up or down) relative to the previous month. \textit{Confidence Calibration} is the correlation between the LLM's confidence level (on a scale from 0 to 100) and the MAPE.  \textit{Num Obs} is the number of observations used in the evaluation. \textit{Refusals} are the number of instances in which the model withheld a prediction by either answering ''null" or 0. Refusal count also includes instances of missing data.}
\scriptsize

\begin{tabularx}{\linewidth}{lYYYYYYY}
  \toprule
\textit{Panel A: Rates} & ME (\%) & MAE (\%) & Threshold Accuracy (\%) & Directional Accuracy (\%) & Confidence Calibration & Num Obs & Refusals \\
  \midrule
  \multicolumn{8}{l}{\textit{Pre-cutoff, 01/1990 to 09/2023}} \\
  \midrule
  GDP Growth & 0.01 & 0.15 & 96.27 & 96.99 & -0.27 & 134 & 1 \\
  Inflation & 0.00 & 0.04 & 98.02 & 93.07 & -0.11 & 405 & 0 \\
  Unemployment Rate & -0.00 & 0.03 & 99.26 & 83.42 & 0.09 & 405 & 0 \\
  10-Yr Treasury Yield & -0.00 & 0.06 & 98.52 & 88.12 & -0.40 & 405 & 0 \\ 
  \midrule
  \multicolumn{8}{l}{\textit{Post-cutoff, 10/2023 to 02/2025}} \\
  \midrule
  GDP Growth & 0.01 & 0.95 & 40.00 & 100.00 & - & 6 & 0 \\
  Inflation & 0.35 & 0.38 & 47.06 & 56.25 & 0.70 & 17 & 0 \\
  Unemployment Rate & -0.20 & 0.26 & 52.94 & 31.25 & 0.47 & 17 & 0 \\
  10-Yr Treasury Yield & -0.24 & 0.48 & 29.41 & 50.00 & 0.16 & 17 & 0 \\
  \bottomrule
\end{tabularx}

\vspace{5pt}

\begin{tabularx}{\linewidth}{lYYYYYYY}
  \toprule
  \textit{Panel B: Levels} & MPE (\%) & MAPE (\%) & Threshold Accuracy (\%) & Directional Accuracy (\%) & Confidence Calibration & Num Obs & Refusals \\
  \midrule
  \multicolumn{8}{l}{\textit{Pre-cutoff, 01/1990 to 09/2023}} \\
  \midrule
  VIX & 3.25 & 6.74 & 94.62 & 87.40 & -0.42  & 390 &  15 \\
  Housing Starts & -2.38 & 3.92 & 100.00 & 81.91 & -0.24 & 399 & 6 \\
  Nonfarm Payrolls & -7.65 & 66.30 & 95.56 & 94.06 & -0.11 & 405 & 0 \\
  \midrule
  \multicolumn{8}{l}{\textit{Recent Pre-cutoff Period, 10/2014 to 09/2023}} \\
  \midrule
  VIX & 0.04 & 0.34 & 100.00 & 98.13 & -0.20 & 108 & 0 \\
  Housing Starts & -0.22 & 1.06 & 95.28 & 98.10 & -0.14 & 106 & 2 \\ 
  Nonfarm Payrolls & -0.00 & 0.00 & 100.00 & 100.00 & -0.14 & 108 & 0 \\
  \midrule
  \multicolumn{8}{l}{\textit{Post-cutoff, 10/2023 to 02/2025}} \\
  \midrule
  VIX & 16.87 & 21.14 & 50.00 & 61.54 & - & 14 & 3 \\ 
  Housing Starts & 2.25 & 8.54 & 35.29 & 56.25 & - & 17 & 0 \\
  Nonfarm Payrolls & 69.51 & 97.44 & 58.82 & 68.75 & -0.21 & 17 & 0 \\
  \bottomrule
\end{tabularx}
\end{table}
\end{landscape}

\subsection{Market Index Recall}

We next evaluate GPT-4o's memorization of market index data by testing its ability to recall daily and monthly values for the S\&P 500, Dow Jones Industrial Average (DJIA), and Nasdaq Composite, using data from January 1990 to February 2025. For numerical recall tests, we requested exact closing values at daily frequency, both without context and with the previous two days' levels provided, as well as monthly returns. Additionally, we assessed directional changes (up or down) and relative performance between index pairs at the monthly frequency. Performance metrics include Mean Percent Error, Mean Absolute Percent Error, and Directional Accuracy (proportion of predictions correctly identifying the direction of change relative to the previous period) for numerical predictions and accuracy for directional and relative performance tasks, all compared against actual values. 

\begin{table}[htbp]
\centering
\caption{Evaluation Metrics for Stock Market Indices}\label{tab:index}
\caption*{\scriptsize This table reports a set of evaluation metrics assessing the LLM’s ability to recall market index levels and their changes over time. These tests are done at the daily or monthly frequency. We ask the LLM to recall the closing value of the index each trading day. Panel A provides metrics for predictions of \textit{Daily Levels} and \textit{Daily Levels with context} (where the previous two days' index levels are provided). We ask the LLM to provide monthly returns for these indices as well. Metrics include \textit{Mean Percent Error (MPE)}, \textit{Mean Absolute Percent Error (MAPE)}, and \textit{Directional Accuracy}, all reported in percentage points (0.10 means 0.10\%). \textit{MPE} is calculated by averaging the percent error \((Estimated Level - Actual Level) / Actual Level\). \textit{MAPE} takes the average absolute value of the percent errors. \textit{Directional Accuracy} measures the proportion of predictions of the market index levels correctly following the direction of change (up or down) relative to the previous day. \textit{Confidence Calibration} reports the correlation between the LLM's confidence level (on a scale from 0 to 100) and mean absolute percent error. Panel B presents accuracy metrics related to predicting \textit{Directional Changes} and \textit{Relative Performance} between indices. \textit{Directional Changes} asks the LLM directly for an up or down answer for each month. \textit{Relative Performance} asks the LLM to answer which index of the index pair performed better during the month. \textit{Accuracy} reports the proportion of predictions correctly identifying either the direction of change or relative performance in percentage points. \textit{Confidence Calibration} in this panel reflects the correlation between the LLM’s confidence and the MAPE. Results are separately provided for the S\&P 500 (SP500), Dow Jones Industrial Average (DJIA), and Nasdaq Composite indices.}
\scriptsize
\begin{tabularx}{\linewidth}{lYYYYYY}
  \toprule
  \multicolumn{7}{l}{\textit{Panel A: Numerical Tests}} \\
  \midrule
     & MPE (\%) & MAPE (\%) & Directional Accuracy (\%) & Confidence Calibration & Num Obs & Refusals \\
  \midrule
  \multicolumn{7}{l}{\textit{Daily Levels: Pre-cutoff, 01/02/1990 to 09/29/2023}} \\
  \midrule
   SP500 & 0.12 & 0.61 & 80.58 & -0.14 & 8,488 & 0 \\
   DJIA & 0.11 & 0.53 & 80.66 & -0.36 & 8,488 & 0 \\
   Nasdaq Composite & 0.18 & 1.80 & 69.38 & -0.12 & 8,488 & 0 \\
  \midrule
   \multicolumn{7}{l}{\textit{Daily Levels: Post-cutoff, 10/02/2023 to 02/28/2025}} \\
  \midrule
   
   SP500 & -16.78 & 16.87  & 45.70 & -0.10 & 292 &  62  \\ 
   DJIA & -12.96 & 13.01  & 49.26 & -0.14 & 271 &  83  \\
   Nasdaq Composite & -20.40 & 20.48  & 44.03 & -0.25 & 294 &  60  \\
  \midrule
   \multicolumn{7}{l}{\textit{Daily Levels: Pre-cutoff with context, 01/02/1990 to 09/29/2023}}\\
  \midrule
   SP500 & 0.13 & 0.50 & 80.78 & -0.16 & 8,488 & 0 \\
   DJIA & 0.06 & 0.46 & 80.06 & -0.17 & 8,488 & 0 \\
   Nasdaq Composite & 0.00 & 1.06 & 68.62 & -0.18 & 8,488 & 0 \\
  \midrule
   \multicolumn{7}{l}{\textit{Daily Levels: Post-cutoff with context, 10/02/2023 to 02/28/2025}} \\
  \midrule
   SP500 & 0.00 & 0.64 & 54.40 & -0.07 & 319 &  35  \\
   DJIA & -0.04 & 0.54 & 53.17 & 0.02  & 332 &  22  \\ 
   Nasdaq Composite & -0.03 & 0.92 & 51.85 & -0.09  & 298 &  56 \\
  \midrule
   \multicolumn{7}{l}{\textit{Monthly Returns: Pre-cutoff, 01/1990 to 09/2023}} \\
  \midrule
   SP500 & -0.70 & 3.36 & 85.19 & 0.27 & 405 & 0 \\
   DJIA & -0.70 & 3.28 & 80.49 & 0.26 & 405 & 0 \\
   Nasdaq Composite & -1.03 & 4.83 & 79.51 & 0.21 & 405 & 0 \\
    \midrule
   \multicolumn{7}{l}{\textit{Monthly Returns: Post-cutoff, 10/2023 to 02/2025}} \\
   \midrule
   SP500            & -1.16 & 3.20 & 41.67 & -0.28 & 12 & 5 \\
   DJIA             & -0.80 & 3.18 & 30.77 & -0.06 & 13 & 4 \\
   Nasdaq Composite & -2.46 & 4.07 & 41.67 & -0.39 & 12 & 5 \\
  \bottomrule
\end{tabularx}

\vspace{10pt}

\begin{tabularx}{\linewidth}{lYY}
  \toprule
  \multicolumn{3}{l}{\textit{Panel B: Other Tests}} \\
  \midrule
    & Accuracy (\%)  & Confidence Calibration \\
  \midrule
  \multicolumn{3}{l}{\textit{Monthly Directional Changes: Pre-cutoff}} \\
    \midrule
   SP500 & 82.80 & 0.33 \\
   DJIA & 80.63  & 0.29 \\
    Nasdaq Composite & 79.36  & 0.29 \\
   \midrule
  \multicolumn{3}{l}{\textit{Monthly Relative Performance: Pre-cutoff}}\\
  \midrule
SP500, DJIA & 87.10  & 0.23 \\
  SP500, NDAQ & 82.86  & 0.49 \\
  NDAQ, DJIA & 83.87  & 0.42 \\
   \bottomrule
\end{tabularx}

\end{table}

Results are reported in Table \ref{tab:index}, distinguishing pre-cutoff (before October 2023) and post-cutoff (after October 2023) periods to isolate memorization effects. For pre-cutoff daily exact numerical levels, GPT-4o exhibits strong recall, with Mean Absolute Percent Errors of 0.61\% for S\&P 500, 0.53\% for DJIA, and 1.80\% for Nasdaq Composite, and Directional Accuracy ranging from 69.4\% (Nasdaq) to 80.7\% (DJIA). Providing context improves slightly accuracy for S\&P 500 (0.50\%) and Nasdaq (1.06\%). Prompting directly for returns, which we test monthly, yields higher Directional Accuracy (79.5\%–85.2\%), reflecting robust memorization of directional trends. Other tests further confirm memorization: prompting directly for directional performance (``up'' or ``own'') exceeds 79\% accuracy across indices, and relative performance accuracy ranges from 82.9\% (S\&P 500 vs. Nasdaq) to 87.1\% (S\&P 500 vs. DJIA). 

In contrast, post-cutoff performance collapses. For instance, without context, the Mean Absolute Percent Errors balloon to 13.01\%–20.48\% for exact levels and Directional Accuracy dropping to near-random levels (44.0\%–49.3\%). We find similar evidence for recall with context in terms of Directional Accuracy which ranges from 51.9\% to 54.4\% in the post-cutoff period, much lower than the Directional Accuracy pre-cutoff which ranges from 68.6\% to 80.8\%. These findings do not suggest data leakage beyond the training cutoff date.

The sharp pre-cutoff accuracy, particularly for exact levels and relative performance, highlights GPT-4o's extensive memorization of historical index data, posing challenges for forecasting studies in economics and finance. The model's ability to recall precise closing values and correctly identify directional trends within its training period suggests that any apparent predictive success may reflect memorized data rather than analytical capability. The negligible improvement from context and the complete performance drop post-cutoff reinforce that these results stem from training data exposure. These findings caution against using LLMs for historical market analysis without ensuring data is outside their training scope, as their outputs risk being artifacts of memorization rather than genuine economic foresight. 


\subsection{Headline Date Identification}

In this section, we assess LLMs’ ability to extract date information from front-page news headlines and, in turn, predict the aggregate stock market index. To do so, we present GPT-4o with sets of \emph{The Wall Street Journal} front-page headlines (approximately 9 headlines per day) from our dataset of 90,123 headlines spanning January 1990 to February 2025 without revealing their publication dates. We asked the model to identify when these headlines appeared and, in a separate test variant, predict the S\&P 500 level on the next trading day. Performance is evaluated using multiple accuracy metrics: year accuracy, month-and-year accuracy, exact date accuracy, mean absolute days difference, and confidence calibration. By comparing results between pre-cutoff headlines (where memorization could occur) and post-cutoff headlines (where memorization is impossible), we can clearly distinguish between the model's inferential abilities and its capacity to recall memorized chronological information.  

\begin{table}[htbp]
\centering
\caption{Evaluation Metrics for News Headlines}\label{tab:headlines}
\caption*{\scriptsize This table reports a set of evaluation metrics assessing the LLM’s ability to recall dates associated with historical headlines, along with corresponding levels of the S\&P 500 index. Metrics are separated into two panels: \textit{Headline Dates}, focusing solely on the accuracy of predicted dates, and \textit{Headline Dates and Levels}, evaluating the accuracy of results when we prompt the LLM to give both the dates and S\&P 500 levels on the next trading day. \textit{Mean Days Difference} is the average signed difference (in days) between predicted and actual dates, while \textit{Mean Absolute Days Difference} reports the average absolute difference. \textit{Year Accuracy}, \textit{Month and Year Accuracy}, and \textit{Exact Date Accuracy} measure the percentage of predictions correctly recalling the year, the month and year, and the exact date, respectively. \textit{Confidence Calibration} indicates the correlation between the LLM’s confidence level (on a scale from 0 to 100) and the accuracy of date predictions. \textit{Mean Percent Error S\&P 500} and \textit{Mean Absolute Percent Error S\&P 500} measure the accuracy of the LLM’s predicted index levels, calculated as the average and average absolute values of the percent error, respectively, and reported in percentage points (-0.01 means -0.01\%). Results are provided separately for headlines from the \textit{Pre-training Period} and the \textit{Post-training Period}.}
\scriptsize
\begin{tabularx}{\linewidth}{lYYYYYYYY}
  \toprule
   & Mean Days Difference & Mean Absolute Days Difference & Year Accuracy (\%) & Month and Year Accuracy (\%) & Exact Date Accuracy (\%) & Confidence Calibration & MPE S\&P 500 (\%) & MAPE S\&P 500 (\%)\\\\
  \midrule
  \multicolumn{9}{l}{\textit{Headline Dates}} \\
   \midrule
   Pre-training Period & -0.77 & 9.52 & 98.45 & 90.38 & 47.03 & -0.10 &  -  &  -\\
   Post-training Period & 413.46 & 414.54 & 28.81 & 20.71 & 7.86 & -0.12 & -  & - \\
   \midrule 
   \multicolumn{9}{l}{\textit{Headline Dates and Levels}}\\
   \midrule
   Pre-training Period & -1.30 & 9.63 & 98.50 & 90.31 & 39.31 & -0.10 & 0.00 & 0.01 \\
   Post-training Period & 456.84 & 457.13 & 26.20 & 19.47 & 5.53 & 0.29 & -0.21 & 0.22 \\
   \bottomrule
\end{tabularx}

\end{table}

We present the results in Table \ref{tab:headlines}. GPT-4o demonstrates remarkable memorization of headline chronology within its training period. For pre-cutoff headlines, it achieves 98.5\% accuracy in determining the correct year and 90.4\% in identifying the correct month and year. Even for exact date identification, the model achieves 47.0\% accuracy—significantly above chance levels. When incorrect, the model's estimates remain close to the actual date, with a mean absolute difference of 9.5 days.

In stark contrast, for headlines published after the model's training cutoff date, performance deteriorates dramatically across all metrics. Year accuracy drops to 28.8\%, month-and-year accuracy falls to 20.7\%, and exact date accuracy declines to just 7.9\%. The mean absolute difference increases to 414.5 days, indicating essentially random guessing.

We observed a similar pattern when we extend our test to ask the model to provide both the headline date and the corresponding S\&P 500 level on the next trading day. For the pre-training period, the model achieves high temporal accuracy while maintaining near-perfect recall of index values (mean absolute percent error of just 0.01\%). For post-training headlines, both date identification and index level predictions became significantly less accurate.

The sharp performance discontinuity at the training cutoff date provides compelling evidence that the model's apparent ``knowledge'' of financial chronology stems primarily from memorization rather than inference or reasoning. This finding raises significant concerns about using LLMs to analyze historical relationships between news events and market movements within their training period, as their responses may reflect memorized associations rather than genuine analytical insights.

\subsection{Individual Stock Price Recall}

To complement our earlier findings on market indices, we examine memorization at the individual-security level in this section. In particular, we test the model's ability to recall end-of-month closing prices for prominent stocks such as Magnificent 7 and individual stocks in the cross section. 

\subsubsection{Magnificent 7 stocks}
\begin{figure}[htbp]
    \centering
    \caption{Recall of closing prices for AMZN, GOOGL, META, and TSLA without context.} \label{fig:mag7_1}
    \caption*{\scriptsize This figure shows the LLM's estimated closing prices for AMZN, GOOGL, META, and TSLA compared to the actual values. Panels A, C, E, and G graph the actual values against the estimated values. Panels B, D, F, and H show the estimation error. Estimation error is calculated as \textit{(Estimated - Actual)/Actual} and is shown in percentages (5 means 5\%). The post-cutoff period (10/2023 onward) is shaded gray.}
    \includegraphics[width=\linewidth]{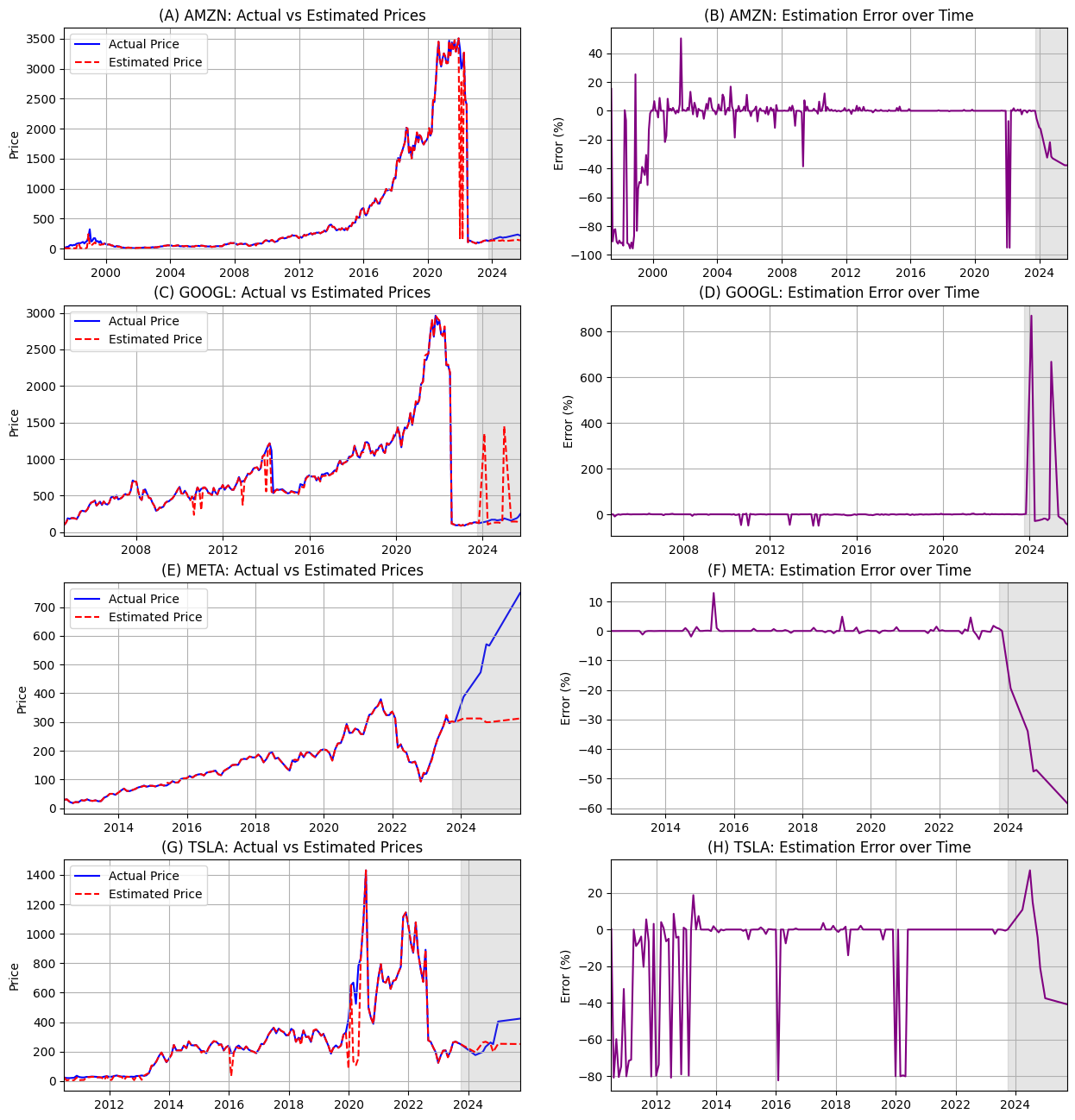}
\end{figure}

\begin{figure}[htbp]
    \centering
    \caption{Recall of closing prices for AAPL, MSFT, and NVDA without context.} \label{fig:mag7_2}
    \caption*{\scriptsize This figure shows the LLM's estimated closing prices for AAPL, MSFT, and NVDA compared to the actual values. Panels A, C, and E graph the actual values against the estimated values. Panels B, D, and F show the estimation error. Estimation error is calculated as \textit{(Estimated - Actual)/Actual} and is shown in percentages (5 means 5\%). The post-cutoff period (10/2023 onward) is shaded gray.}
    \includegraphics[width=0.9\linewidth]{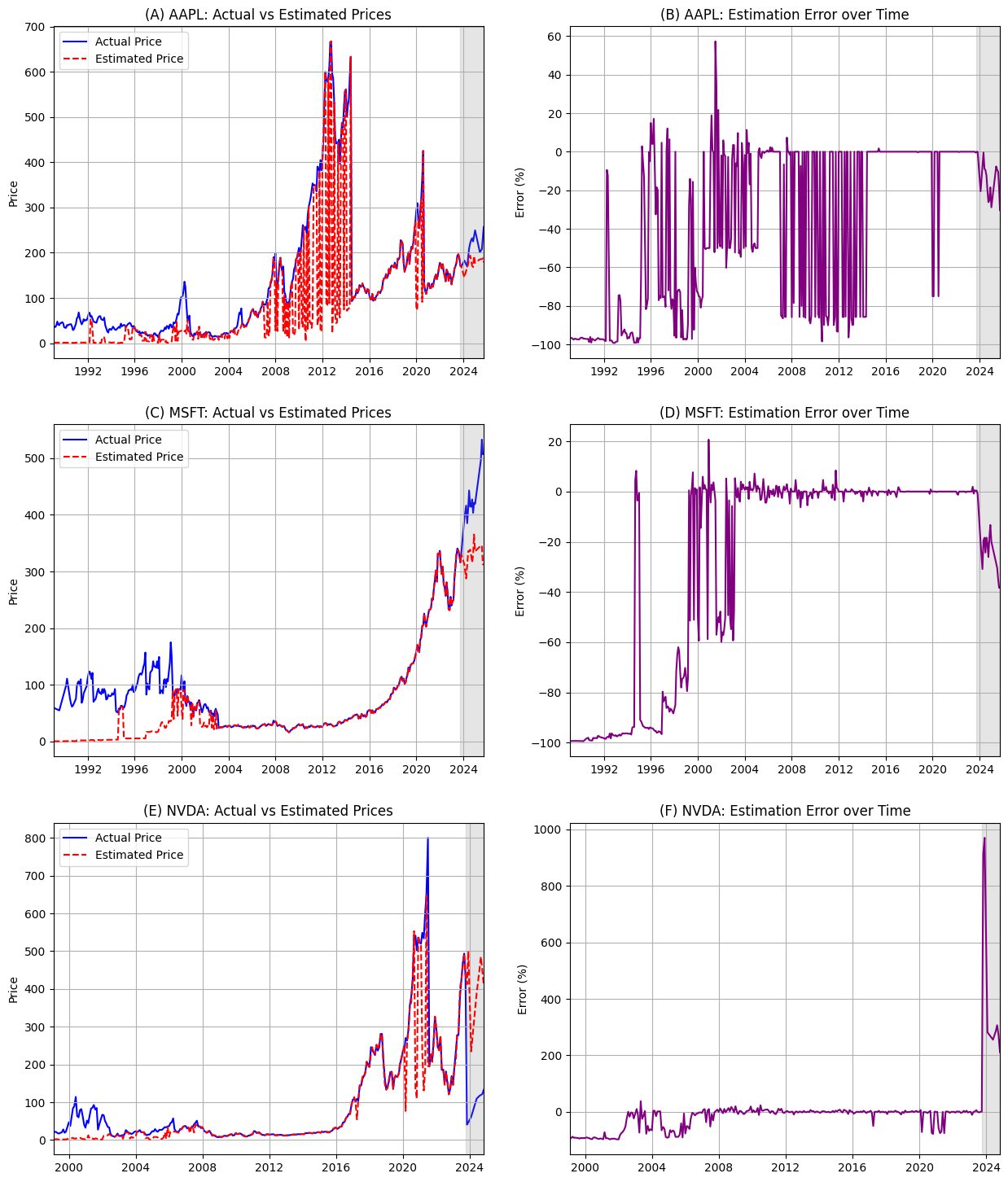}
\end{figure}

We begin our analysis with the Magnificent 7 stocks (META, GOOGL, AMZN, TSLA, NVDA, MSFT, AAPL) from January 1990 to September 2023, which precedes the model's training cutoff of October 2023. We queried prices both without context and with the previous two months' closing prices provided, using data from the Center for Research in Security Prices (CRSP). Performance was evaluated using Mean Percent Error, Mean Absolute Percent Error, and Directional Accuracy (correctly identifying the direction of change relative to the previous month), with results compared against actual closing prices. We report these metrics in Table \ref{tab:mag7} and plot the actual vs estimated values without context in Figures \ref{fig:mag7_1} and \ref{fig:mag7_2}. 

\begin{table}[htbp]
\centering
\caption{Evaluation Metrics for Magnificent 7 Stocks}\label{tab:mag7}
\caption*{\scriptsize This table reports a set of evaluation metrics for the Magnificent 7 stocks which includes META, GOOGL, AMZN, TSLA, NVDA, MSFT, and AAPL. We ask the LLM to recall closing prices at the end of each month. \textit{Mean Percent Error (MPE)}, \textit{Mean Absolute Percent Error (MAPE)}, and \textit{Directional Accuracy } are reported in percentage points (0.18 means 0.18\%). \textit{MPE} is calculated by taking the average of the percent error $(Predicted Price - Actual Price) / Actual Price$. \textit{MAPE} is calculated by taking the average of the absolute value of the percent error. \textit{Directional Accuracy} is the proportion of predictions that went in the correct direction (up or down) with respect to the previous month. \textit{Confidence Calibration} is the correlation between the LLM's confidence level (on a scale of 0 to 100) and the MAPE. \textit{Num Obs} is the number of observations used in the evaluation, \textit{Start Date} and \textit{End Date} indicate the period over which the metrics were computed. \textit{Refusals} are the number of instances in which the model withheld a prediction by either answering "null" or 0. Results are provided for a prompt that contains an empty context in panel A and a prompt that provides the previous two month's closing prices as context in panel B.}
\scriptsize
\begin{tabularx}{\linewidth}{lYYYYYYYY}
  \toprule
  \multicolumn{9}{l}{\textit{Panel A: No Context}} \\
  \midrule
   & MPE (\%) & MAPE (\%) & Directional Accuracy (\%) & Confidence Calibration & Num Obs & Start Date & End Date & Refusals \\
  \midrule
  \multicolumn{8}{l}{\textit{Pre-cutoff, 01/1990 to 09/2023}} \\
  \midrule
    META  & 0.18   & 0.37  & 99.26 & -0.08 & 137 & 05/2012 & 09/2023 & 0 \\
    GOOGL & -1.41  & 1.79  & 93.42 & -0.19 & 229 & 08/2004 & 09/2023 & 1 \\
    AMZN  & -5.87  & 7.98  & 91.77 & -0.12 & 317 & 05/1997 & 09/2023 & 0 \\
    TSLA  & -9.21  & 9.99  & 92.45 & -0.13 & 160 & 06/2010 & 09/2023 & 0 \\
    NVDA  & -20.60 & 23.92 & 77.05 & -0.53 & 293 & 01/1999 & 09/2023 & 4 \\
    MSFT  & -25.72 & 26.62 & 76.75 & -0.65 & 401 & 01/1990 & 09/2023 & 14 \\
    AAPL  & -35.14 & 36.44 & 72.61 & -0.54 & 399 & 01/1990 & 09/2023 & 6 \\
    \midrule
  \multicolumn{8}{l}{\textit{Post-cutoff, 10/2023 to 02/2025}} \\
  \midrule
    META    & -29.59 & 29.60  & 50.00 & 0.48 & 5 &  10/2023 & 02/2025  & 12 \\
    GOOGL   & 138.06 & 170.09  & 55.56 & -0.98  & 10 & 10/2023  & 02/2025  & 7 \\
    AMZN    & -22.06 & 22.06 & 42.86 & 0.48 & 8 &  10/2023 & 02/2025  & 7 \\
    TSLA    &   -0.81 & 20.15   & 100.00 & 0.61 & 6 & 10/2023  & 02/2025  & 11 \\
    NVDA    &  436.43 & 436.43  & 100.00 & -0.64  & 8 &  10/2023 & 02/2025  & 9  \\
    MSFT    & -19.25 & 19.25  & 45.45 & -0.48 & 12 & 10/2023  & 02/2025  & 5  \\
    AAPL    & -15.46 & 15.46  & 30.00 & 0.36 & 11 &  10/2023 & 02/2025  & 6  \\
\bottomrule
\end{tabularx}

\vspace{5pt}

\begin{tabularx}{\linewidth}{lYYYYYYYY}
  \toprule
  \multicolumn{9}{l}{\textit{Panel B: With Context}} \\
  \midrule
    & MPE (\%) & MAPE (\%) & Directional Accuracy (\%) & Confidence Calibration & Num Obs & Start Date & End Date & Refusals \\
  \midrule
      \multicolumn{8}{l}{\textit{Pre-cutoff, 01/1990 to 09/2023}} \\
  \midrule
    META  & -0.18 & 0.40 & 98.52 & -0.10 & 136 & 05/2012 & 09/2023 & 1 \\ 
    GOOGL & -0.27 & 0.84 & 95.52 & 0.10 & 224  & 08/2004 & 09/2023 & 6 \\
    AMZN  & -0.42 & 3.12 & 93.35 & 0.02 & 317  & 05/1997 & 09/2023 & 0 \\
    TSLA  & -0.87 & 2.34 & 94.87 & -0.17 & 157 & 06/2010 & 09/2023 & 3 \\
    NVDA  & 1.01  & 8.80 & 80.68 & -0.17 & 296 & 01/1999 & 09/2023 & 1 \\
    MSFT  & 1.24  & 4.57 & 84.71 & -0.21 & 400 & 01/1990 & 09/2023 & 5 \\
    AAPL  & -1.37 & 5.89 & 83.91 & -0.25 & 405 & 01/1990 & 09/2023 & 0 \\
    \midrule
    \multicolumn{8}{l}{\textit{Post-cutoff, 10/2023 to 02/2025}} \\
  \midrule
    META  & -5.04 & 5.29  & 50.00 & 0.14 & 7 &  10/2023 & 02/2025  & 10 \\ 
    GOOGL & 0.88 & 6.26  & 50.00 & -0.36 & 9 &  10/2023 & 02/2025  & 8 \\
    AMZN  & -2.72 & 6.06 &54.55 &  0.54 & 12 &  10/2023 & 02/2025  & 5 \\ 
    TSLA  & -5.67 & 10.75  & 60.00 & 0.02 & 11 &  10/2023 & 02/2025  & 6 \\ 
    NVDA  & 85.89 & 99.39  & 66.67 & 0.45  & 10 &  10/2023 & 02/2025  & 7 \\ 
    MSFT  & -0.80 & 2.75  & 50.00 & -0.14 & 9 &  10/2023 & 02/2025  & 8 \\ 
    AAPL  & -2.90 & 4.12 & 55.56 & -0.10 & 10 &  10/2023 & 02/2025  & 7 \\ 
   \bottomrule
\end{tabularx}
\end{table}

The results reveal varying recall accuracy across stocks, with notable improvements when context is provided. Without context (Panel A), GPT-4o performs best on more recently listed stocks, such as META, with a Mean Absolute Percent Error of 0.37\% and Directional Accuracy of 99.26\%, but struggles with stocks with a longer history, such as AAPL (36.44\% error, 72.61\% accuracy) and MSFT (26.62\% error, 76.75\% accuracy). Errors are also high for NVDA (23.92\%) and TSLA (9.99\%), suggesting selective memorization tied to the length of the price series or data prominence. With context, accuracy improves significantly as shown in Panel B: Mean Absolute Percent Errors drop to 0.40\% for META, 0.84\% for GOOGL, and 5.89\% for AAPL, with Directional Accuracy rising to 98.52\%, 95.52\%, and 83.91\%, respectively. This context-driven enhancement mirrors the slight improvements seen for market indices, indicating that recent price cues help the model anchor its recall more precisely, particularly for stocks with longer price histories.

\begin{figure}[htbp]
    \centering
    \caption{Recall of closing prices for AAPL, MSFT, and NVDA with context.} \label{fig:mag7_2_context}
    \caption*{\scriptsize This figure shows the LLM's estimated closing prices for AAPL, MSFT, and NVDA compared to the actual values. We give the LLM two previous end of the month closing prices given as context. Panels A, C, and E graph the actual values against the estimated values. Panels B, D, and F show the estimation error. Estimation error is calculated as \textit{(Estimated - Actual)/Actual} and is shown in percentages (5 means 5\%). The post-cutoff period (10/2023 onward) is shaded gray.}
    \includegraphics[width=0.9\linewidth]{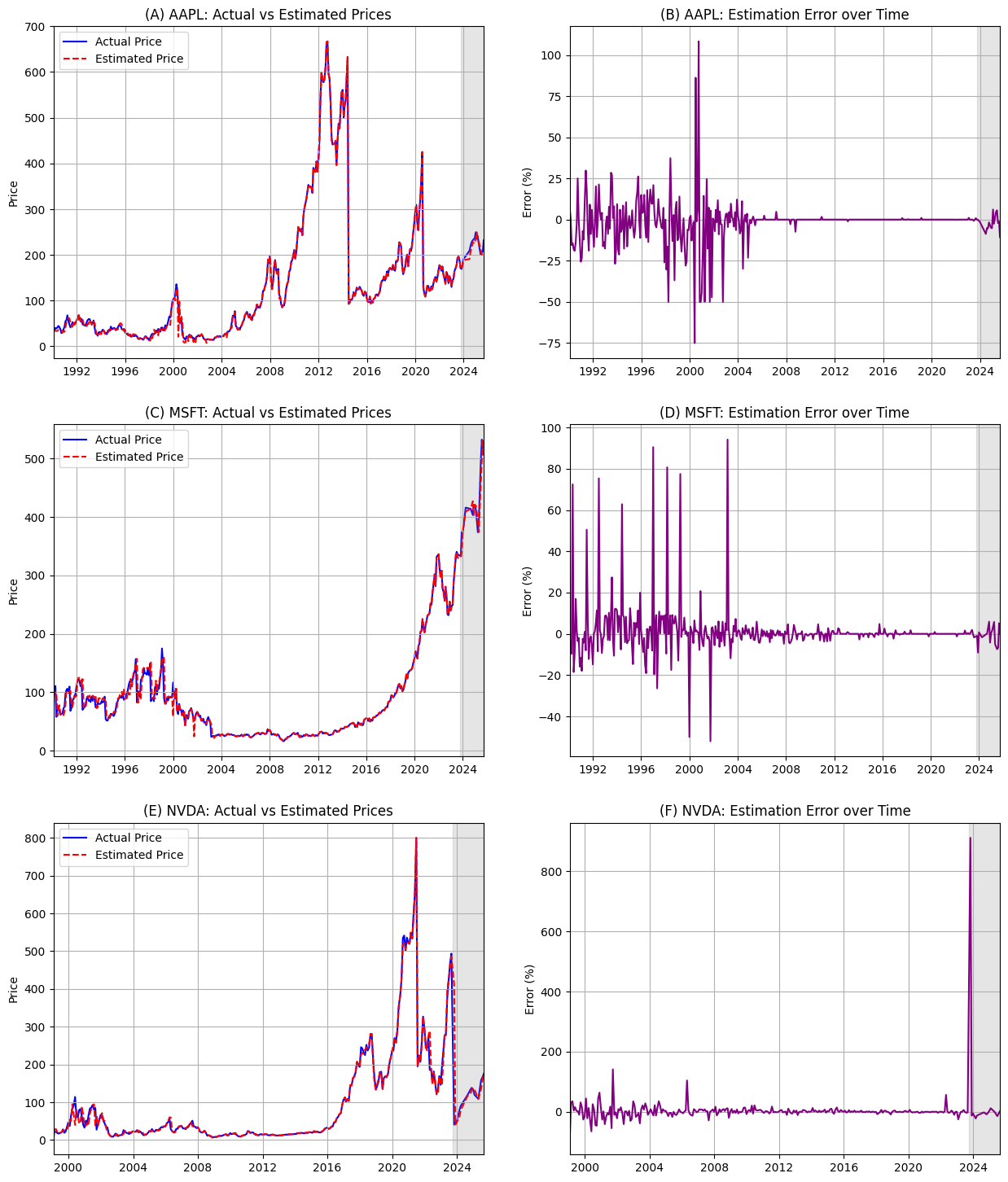}
\end{figure}

We plot the actual vs estimated values with context for NVDA, AAPL, and MSFT in Figure \ref{fig:mag7_2_context}. Compared to Figure \ref{fig:mag7_2}, this figure shows why any memorization performance only provides a lower bound. The errors substantially decrease when we give ChatGPT the prices for the previous two months. This situation is relevant as when forecasting, researchers typically provide contextual information.

Overall, these findings extend our market index results, highlighting that GPT-4o's memorization is not uniform across securities and is sensitive to contextual cues and stock-specific factors. The high accuracy for META and GOOGL, especially with context, parallels the model's strong recall of recent macroeconomic levels and index values, suggesting robust memorization of prominent, frequently referenced data. Conversely, larger errors for long-listed stocks like AAPL and MSFT, even with context, align with the weaker recall of long-horizon macroeconomic levels, pointing to the potential dilution of older data in the training corpus. This selective memorization reinforces the challenge for financial research: apparent forecasting success for individual stocks within the training period may reflect memorized prices rather than predictive ability, necessitating evaluations with post-cutoff data to ensure methodological rigor.

\subsubsection{Individual Stocks in the Cross Section}
\begin{figure}[H]
    \centering
    \caption{Recall of exact numerical levels of closing prices by size quintiles.} \label{fig:stocks}
    \caption*{\scriptsize This figure shows the LLM's estimated closing prices for randomly selected stocks compared to the actual closing prices. Panels A, C, E, G, and I graph the actual values against the estimated values. Panels B, D, F, H, and J show the estimation error. Estimation error is calculated as \textit{(Estimated - Actual)/Actual} and is shown in percentages (5 means 5\%). The price plotted is the equal-weighted average price of the five quintile market cap groups, Q1 (smallest) through Q5 (largest). Each size quintile includes 50 randomly sampled stocks, resampled annually. The market cap breakpoints were calculated using NYSE stocks. The post-cutoff period (10/2023 onward) is shaded gray.}
    \includegraphics[width=\linewidth]{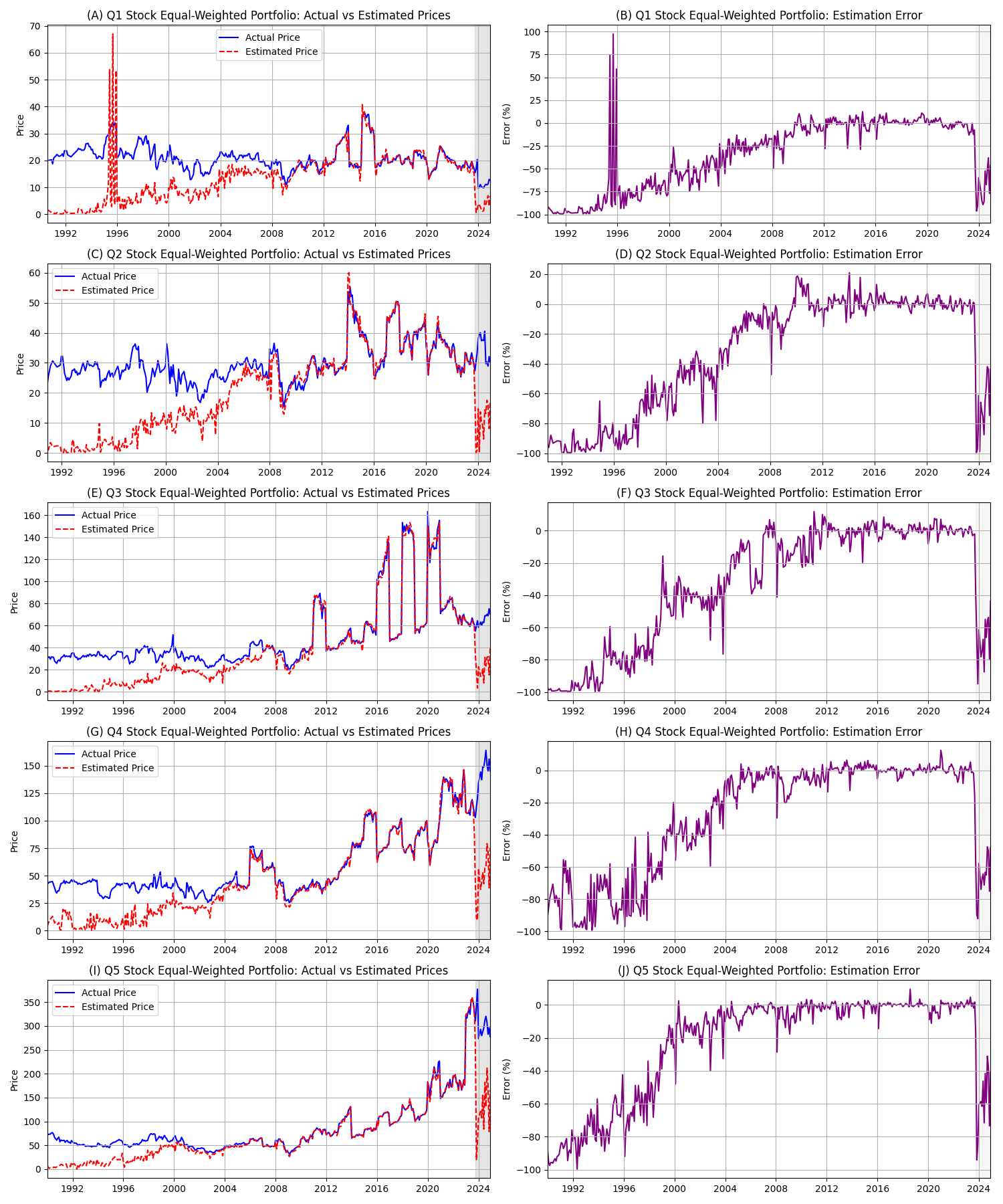}
\end{figure}

To broaden our analysis of GPT-4o's memorization, we also examine its recall of end-of-month closing prices for individual stocks grouped by market capitalization covering January 1990 to December 2024. Stocks are divided into market capitalization quintiles with size breakpoints calculated using NYSE stocks only. We randomly sample 50 stocks from each quintile and resampled annually to account for size changes, using stock price data from CRSP. We test recall without context, with the previous two months' prices provided, and over the recent 10-year pre-cutoff period to assess recency effects. Performance metrics, including Mean Percent Error, Mean Absolute Percent Error, and Directional Accuracy (correct direction of change relative to the prior month), are reported in Table \ref{tab:stocks}. We also plot the actual vs estimated values in Figure \ref{fig:stocks}.

\begin{table}[H]
\centering
\caption{Evaluation Metrics for Individual Stocks by Size Quintile}\label{tab:stocks}
\caption*{\scriptsize This table reports a set of evaluation metrics for portfolios of stocks grouped by size quintiles. We divide stocks into quintiles by market cap (using the NYSE stocks only to calculate cutoffs) and sample 50 stocks from each quintile, resampling each year. We then ask the LLM to recall end of month closing prices for each stock. \textit{Mean Percent Error (MPE)}, \textit{Mean Absolute Percent Error (MAPE)}, and \textit{Directional Accuracy} are reported in percentage points (0.78 means 0.78\%). \textit{MPE} is calculated by taking the average of the percent error $(Predicted Price - Actual Price)/Actual Price$. \textit{MAPE} is calculated by taking the average of the absolute value of the percent error. \textit{Directional Accuracy} is the proportion of predictions that went in the correct direction (up or down) relative to the previous month. \textit{Confidence Calibration} is the correlation between the LLM's confidence level (on a scale from 0 to 100) and the mean absolute percent error. \textit{Num Obs} is the number of observations used in the evaluation. The panel labels and subheadings indicate the period over which the metrics were computed. \textit{Refusals} represent the number of instances in which the model withheld a prediction by either answering "null" or 0. Results are provided for the full sample period, a recent period covering the past 10 years, and \textit{With Context}, in which the previous two months' closing prices are provided to the model.}
\scriptsize

\begin{tabularx}{\linewidth}{lYYYYYY}
  \toprule
  \multicolumn{7}{l}{\textit{Panel A: No Context}}\\
\midrule
Quintile & MPE (\%) & MAPE (\%) & Directional Accuracy (\%) & Confidence Calibration & Num Obs & Refusals \\
\midrule
  \multicolumn{7}{l}{\textit{Pre-cutoff, 01/1990 to 09/2023}} \\
\midrule
1    & 1.20 & 24.70 & 42.66 & -0.10 & 11,925 & 6828 \\
2    & -0.55 & 19.17 & 48.78 & -0.13 & 12,262 & 6394 \\
3    & -3.04 & 17.06 & 51.02 & -0.32 & 13,077 & 5469 \\
4    & -3.36 & 14.18 & 53.64 & -0.22 & 13,593 & 4908 \\
5    & -4.55 & 11.29 & 61.11 & -0.26 & 14,816 & 3576 \\
\midrule
  \multicolumn{7}{l}{\textit{Recent Period, 01/2014 to 09/2023}} \\
\midrule
1  & 4.24 & 15.24 & 48.62 & -0.03 & 5,415 &  96 \\ 
2  & 2.71 & 10.18 & 58.59 & -0.04 & 5,448 &  26 \\
3  & 1.53 & 6.45  & 63.61 & -0.02 & 5,403 &  20 \\
4  & 0.54 & 4.79  & 69.21 & 0.01  & 5,309 &  10 \\
5  & 0.21 & 3.16  & 79.54 & -0.01 & 5,444 &   7 \\
\midrule
\multicolumn{7}{l}{\textit{Post-cutoff Period, 10/2023 to 12/2024}} \\
\midrule
1  & 53.36 & 97.83 & 29.71 & 0.01 & 238 & 479 \\
2  & 27.24 & 46.06 & 39.47 & 0.03 & 302 & 426 \\
3  & 12.63 & 39.18 & 38.79 & 0.04 & 290 & 440 \\
4  & -6.04 & 23.52 & 37.14 & -0.03 & 322 & 413 \\ 
5  & 0.23  & 27.32 & 46.36 & 0.03 & 390 & 348 \\
\bottomrule
\end{tabularx}

\vspace{10pt}

\begin{tabularx}{\linewidth}{lYYYYYY}
  \toprule
  \multicolumn{7}{l}{\textit{Panel B: With Context}}\\
\midrule
Quintile & MPE (\%) & MAPE (\%) & Directional Accuracy (\%) & Confidence Calibration & Num Obs & Refusals \\
\midrule
  \multicolumn{7}{l}{\textit{Pre-cutoff, 01/1990 to 09/2023}} \\
\midrule
1  & 1.35 & 10.80 & 74.46 & 0.01  & 18187 & 565 \\
2  & 1.47 & 10.14 & 74.35 & -0.01 & 18029 & 627 \\
3  & 0.65 & 8.47  & 74.01 & -0.03 & 17839 & 707 \\
4  & 1.15 & 7.77  & 74.45 & -0.01 & 17767 & 734 \\ 
5  & 0.72 & 5.73  & 75.58 & -0.06 & 17565 & 827 \\
\midrule
  \multicolumn{7}{l}{\textit{Recent Period, 01/2014 to 09/2023}} \\
\midrule
1  & 1.86 & 10.69 & 73.35 & -0.01 & 5376 & 135 \\
2  & 0.97 & 8.55  & 73.62 & -0.04 & 5311 & 163 \\
3  & 0.45 & 6.68  & 72.88 & -0.10 & 5192 & 231 \\
4  & 0.47 & 5.48  & 74.52 & -0.10 & 5083 & 236 \\
5  & 0.35 & 3.29  & 80.56 & -0.11 & 5217 & 234 \\ 
\midrule
\multicolumn{7}{l}{\textit{Post-cutoff Period, 10/2023 to 12/2024}} \\
\midrule
1  & 10.55 & 25.56 & 73.50 & 0.03  & 533 & 184 \\ 
2  & 1.32  & 11.49 & 76.40 & 0.04  & 525 & 203 \\
3  & 0.95  & 9.93  & 72.38 & -0.11 & 456 & 274 \\
4  & 0.90  & 8.74  & 72.18 & -0.10 & 455 & 280 \\
5  & 1.44  & 9.06  & 76.02 & 0.01  & 438 & 300 \\
\bottomrule
\end{tabularx}
\end{table}

The results show weaker recall compared to the Magnificent 7, with accuracy improving for larger stocks and recent periods. Without context in the pre-cutoff period, Mean Absolute Percent Errors are high: ranging from 11.29\% for Quintile 5 containing the largest stocks to 24.70\% for Quintile 1 containing the smallest stocks. Directional Accuracy is low, ranging from 42.66\% to 61.11\%. The high refusal rates suggest uncertainty for less prominent stocks. Over the recent 10-year period, errors decrease significantly ranging from 3.16\% for Quintile 5 to 15.24\% for Quintile 1. Directional Accuracy rises to 48.6\%-79.5\% and near-zero refusal rates, echoing the recency effect seen in macroeconomic levels. In the post-cutoff period, the error rates jump with Mean Absolute Percent Errors ranging from 23.52\% to 97.83\%. Directional Accuracy for all quintiles falls below 46.4\%, with Quintile 1 having only 29.7\% accuracy. With context, errors further improve for Directional Accuracy, with all quintiles reaching at least 74\% in the pre-cutoff period. Larger stocks consistently show better recall, likely due to greater data prominence, aligning with the Magnificent 7's stronger performance for high-profile securities.

These findings complement our Magnificent 7 results, revealing that GPT-4o's memorization weakens for less prominent stocks, particularly smaller ones, and is strongest for recent, larger-cap data. The high errors and refusals for small stocks contrast with the precision for META or GOOGL, suggesting memorization is skewed toward widely covered securities, similar to the robust recall of market indices and macroeconomic rates. The recency and context effect parallel improvements seen in prior tests, but the lower directional accuracy with context indicates limits in capturing trends for diverse portfolios. Again, this selective memorization underscores the risk of relying on LLMs for historical stock analysis, as apparent forecasting accuracy may stem from memorized prices of prominent stocks rather than broad predictive insight, reinforcing the need for post-cutoff evaluations.

\subsection{Can LLMs Follow ``Fake'' Knowledge Cutoff Prompts?}

To assess whether GPT-4o can adhere to instructions not to use information beyond an artificially imposed cutoff date (i.e., fake cutoff), we test its performance in predicting U.S. real GDP growth rates over the period from March 1990 to June 2023. 

We design three prompting conditions using an artificial cutoff date of \textit{December 2010}: (i) one where both system and user messages explicitly restrict knowledge to pre-2010 data, (ii) one where only the system prompt imposes this fake cutoff, and (iii) one where only the user prompt imposes this fake cutoff. The task requires the model to predict quarterly year-over-year GDP growth, with data split into pre-fake-cutoff (1990–2010) and post-fake-cutoff (2011 onward) periods to evaluate compliance with the constraint. Performance metrics include Mean Percent Error, Mean Absolute Percent Error, Directional Accuracy (correctly identifying up/down changes), and refusal counts, and are reported in Table \ref{tab:cutoff}. 

When both system and user prompts enforce the pre-2010 cutoff, GPT-4o performs well preceding this artificial cutoff, with a Mean Absolute Percent Error of 0.27\% and Directional Accuracy of 95.12\%, but struggles post-fake-cutoff, showing a higher error (0.59\%) and lower accuracy (69.23\%), with 38 refusals out of 51 post-fake-cutoff observations. By Proposition~\ref{prop:nonid}, this behavior is observationally equivalent to multiple contradictory explanations: the constraint genuinely working (the model cannot access post-2010 information), behavioral suppression (the model retains memorized information but learned to hide it), or strategic compliance (the model role-plays what it believes a restricted model would say). We also find that the results are similar when only the user prompt specifies the 2010 cutoff. In contrast, when only the system prompt specifies the cutoff, pre-fake-cutoff accuracy improves slightly (0.11\% error, 97.59\% accuracy). Still, post-fake-cutoff performance remains implausibly strong (0.05\% error, 98.00\% accuracy) with only one refusal across 50 observations. This minimal drop in accuracy post-fake-cutoff indicates that without reinforced user instructions, the model likely accesses memorized data, undermining the cutoff constraint. \footnote{We also test alternative pre-2005 and pre-2015 fake knowledge cutoffs for GDP growth in Appendix Tables \ref{tab:cutoff_05} and \ref{tab:cutoff_15}, and find similar patterns.}

In an additional set of analyses reported in Table \ref{tab:cutoff_roll}, we conduct an analogous experiment for S\&P 500 index levels at the daily frequency. In the Rolling Fake Cutoff test, we modify the user prompt to include instructions ``Do not use any knowledge after time $\{t-1\}$'' where $\{t-1\}$ is the date before the current date we ask the model to recall. We implement this procedure in the pre-cutoff period and the post-cutoff period and find the results to match the original recall procedure that does not include the fake knowledge cutoff instructions since the accuracy in the pre-cutoff period is still very high compared to the post-cutoff period. For the original benchmark, the model recalls pre-cutoff index levels with small errors (Mean Average Percent Error of 0.61\% and Directional Accuracy of 80.6\%), but performs poorly after the true cutoff (Mean Average Percent Error of 16.9\% and Directional Accuracy of 45.7\%). Under the Rolling Fake Cutoff instructions, the pre-cutoff performance remains strong (Mean Average Percent Error of 0.81\% and Directional Accuracy of 69.6\%), while the post-cutoff performance is weak (Mean Average Percent Error of 17.2\% and Directional Accuracy of 42.2\%). We also conduct two subsample tests for the periods 1990 to 2008 and 2009 to 2023 with two individual cutoff dates, December 31, 1999 and December 31, 2015, respectively. The results before and after the cutoff dates should be similar in both periods since all dates fall in the model's pre-cutoff period. We do find this to be the case. For the 1990 to 2008 subsample, the Mean Average Percent Error actually falls from 1.39\% pre-fake-cutoff to 0.67\% post-fake-cutoff and Directional Accuracy rises from 59.3\% to 74.4\%. Similarly, in the 2009 to 2023 subsample, we find that the Mean Average Percent Error slightly falls from 0.05\% to 0.03\% and Directional Accuracy slightly rises from 97.3\% to 98.1\%.

These results align well with our earlier findings on macroeconomic indicators and market indices, where high pre-cutoff accuracy reflects memorization. The strong post-fake-cutoff performance without user prompt reinforcement (96\% accuracy) demonstrates that prompt-based constraints empirically fail to prevent access to memorized data. Moreover, even if fake cutoffs had successfully reduced accuracy, Proposition~\ref{prop:nonid} establishes we could not verify whether the constraint genuinely worked: reduced accuracy could reflect either successful restriction to historical information or the model learning to suppress known answers. These observationally equivalent outcomes make the target estimand fundamentally non-identified. The high refusal rate with dual prompts indicates partial compliance but not complete isolation from memorized information, illustrating the practical difficulty of constraining model outputs even when the theoretical impossibility of verification remains.

\begin{landscape}
\begin{table}
\centering
\caption{Fake Knowledge Cutoff}\label{tab:cutoff}
\caption*{\scriptsize This table reports GPT-4o’s performance on U.S. real GDP growth predictions evaluated on data from the Philadelphia Fed’s Real-Time Data Set. We evaluate model accuracy under different artificially-imposed knowledge cutoff constraints: one where both system and user prompts reinforce the fake knowledge cutoff (pre-2010 only), another where only the system prompt specifies the cutoff, and one where only the user prompt specifies the cutoff. The task involves predicting the quarterly year-over-year GDP growth rate, with test data split into pre-fake-cutoff (1990–2010) and post-fake-cutoff (2011 onward) periods to assess whether the model respects the stated cutoff. Metrics include Mean Error (ME), Mean Absolute Error (MAE), Threshold Accuracy (percentage of guesses correctly above a threshold of 2.5\%), Directional Accuracy (percentage of correct up/down changes), Confidence Calibration (correlation between the LLM's confidence level and the MAPE), total observations, and refusal counts. The results indicate that explicitly instructing the model not to use post-2010 data yields higher refusal rates and weaker post-fake-cutoff performance, consistent with adherence to the knowledge constraint.}
\scriptsize
\begin{tabularx}{\linewidth}{lYYYYYYYYY}
\toprule
 & ME (\%) & MAE (\%) & Threshold Accuracy (\%) & Directional Accuracy (\%) & Confidence Calibration & Start Date & End Date & Num Obs & Refusals  \\
  \midrule
  \multicolumn{10}{l}{\textit{GDP Growth: Our prompt with both system and user message knowledge cutoff}}  \\
  \midrule
  Pre fake-cutoff & 0.07 & 0.27 & 97.59 & 95.12 & -0.02 & 03/01/1990 & 12/01/2010 &  83 &   0  \\
  Post fake-cutoff & 0.58 & 0.59 & 69.23 & 75.00 & 0.51 & 03/01/2011 & 09/01/2023 &  13 &  38  \\
  \midrule
  \multicolumn{10}{l}{\textit{GDP Growth: Our prompt with system but no user message knowledge cutoff}} \\
  \midrule
  Pre fake-cutoff & 0.02 & 0.11 & 97.59 & 97.56 & -0.12 & 03/01/1990 & 12/01/2010 &  83 &   0 \\
  Post fake-cutoff & -0.01 & 0.05 & 98.00 & 100.00 & -0.24 & 03/01/2011 & 09/01/2023 &  50 &   1 \\
  \midrule
  \multicolumn{10}{l}{\textit{GDP Growth: Our prompt with user but no system message knowledge cutoff}} \\
  \midrule
  Pre fake-cutoff & 0.08 & 0.22 & 98.80 & 96.34 & -0.18 & 03/01/1990 & 12/01/2010 &  83 &   0  \\
  Post fake-cutoff & 0.60 & 0.61 & 66.67 & 70.59 & -0.15 & 03/01/2011 & 09/01/2023 &  18 &  32  \\
  \bottomrule
\end{tabularx}
\end{table}
\end{landscape}

   
  

\subsection{Testing Masking Effectiveness}

To evaluate whether masking can prevent GPT-4o from accessing memorized information, we examine its ability to identify the firm, year, and quarter of anonymized earnings conference call transcripts from January 2006 to September 2023, all within the pre-cutoff period before the model's training knowledge cutoff of October 2023. Transcripts were anonymized using the entity neutering approach of Engelberg et al.\ (2025), removing identifying details like company names and dates. We focus on the Magnificent 7 stocks (AAPL, META, MSFT, GOOGL, NVDA, TSLA, AMZN) and portfolios grouped by market capitalization terciles (Large, Medium, Small), measuring performance through Mean Years Difference, Mean Absolute Years Difference, and accuracy in identifying the exact year, quarter and year, and firm. Results are reported in Table \ref{tab:anon}. 

\begin{table}[H]
\centering
\caption{Anonymized Conference Calls}\label{tab:anon}
\caption*{\scriptsize This table reports GPT-4o's performance in identifying the correct firm, year, and quarter of anonymized earnings conference call transcripts obtained from Capital IQ during the pre-cutoff period from January 2006 to September 2023. For each firm, we report the mean and mean absolute difference between the model's recollection of the transcript years and the actual transcript years, as well as the percentage of calls for which the model correctly identified the exact year, the exact quarter and year, and the correct firm. Results are grouped by firm and by market capitalization terciles (Large, Mid, Small). Tercile breakpoints are formed using NYSE stocks.}
\scriptsize
\begin{tabularx}{\linewidth}{lYYYYYY}
\toprule
 \multicolumn{7}{l}{\textit{Panel A: Magnificent Seven Firms}} \\
  \midrule
  Ticker & Mean Years Difference & Mean Absolute Years Difference & Year Accuracy (\%) & Quarter and Year Accuracy (\%) & Firm Accuracy (\%) & Num Obs \\
  \midrule
      \multicolumn{7}{l}{\textit{Pre-cutoff, 01/2006 to 09/2023}} \\
    \midrule
AAPL & -0.03 & 0.05 & 95.95 & 93.24 & 100.00 & 74 \\
META & 0.22 & 0.27 & 73.47 & 2.04 & 100.00 & 49 \\
MSFT & -0.67 & 0.78 & 61.64 & 1.37 & 100.00 & 73 \\
GOOGL & -1.05 & 1.59 & 59.46 & 6.76 & 91.89 & 74 \\
NVDA & -1.34 & 1.66 & 50.00 & 1.61 & 93.55 & 62 \\
TSLA & -3.27 & 3.45 & 39.29 & 3.57 & 85.71 & 56 \\
AMZN & -1.87 & 2.11 & 30.67 & 4.00 & 85.33 & 75 \\
  \bottomrule
\end{tabularx}

\vspace{5pt}

\begin{tabularx}{\linewidth}{lYYYYYY}
  \toprule
\multicolumn{7}{l}{\textit{Panel B: Firms by Market Cap}} \\
\midrule
  Size & Mean Years Difference & Mean Absolute Years Difference & Year Accuracy (\%) & Quarter and Year Accuracy (\%) & Firm Accuracy (\%) & Num Obs \\
  \midrule

  
    \multicolumn{7}{l}{\textit{Pre-cutoff, 01/2006 to 09/2023}} \\
    \midrule
    Large & -1.54 & 1.92 & 44.69 & 8.09 & 65.77 & 593 \\
    Medium & -1.87 & 2.37 & 39.93 & 7.34 & 40.78 & 586 \\
    Small & -1.52 & 2.26 & 33.26 & 5.79 & 30.90 & 466 \\
  \bottomrule
\end{tabularx}
\end{table}

GPT-4o demonstrates a remarkable ability to deanonymize transcripts, particularly for prominent firms. For AAPL, the model achieves 100\% firm accuracy, 95.95\% year accuracy, and 93.24\% quarter-and-year accuracy, with a Mean Absolute Years Difference of just 0.05 years. META and MSFT also show perfect firm identification (100\%), though year accuracy drops to 73.47\% and 61.64\%, respectively, and quarter-and-year accuracy is low (2.04\% and 1.37\%). Performance is slightly weaker for GOOGL (91.89\% firm accuracy), NVDA (93.55\%), TSLA (85.71\%), and AMZN (85.33\%), with year accuracy ranging from 30.67\% to 59.46\% and Mean Absolute Years Differences rising to 1.59--3.45 years. Across market cap terciles, firm accuracy declines from 65.77\% for Large to 30.90\% for Small stocks, with year accuracy dropping from 44.69\% for Large to 33.26\% for Small stocks, indicating stronger memorization for larger, more prominent firms, consistent with our portfolio stock findings.

These results demonstrate that masking empirically fails to prevent deanonymization. However, even when deanonymization tests fail, Proposition~\ref{prop:nonid} and Remark~\ref{rem:lower-bound} establish we cannot conclude the constraint succeeded: failed reconstruction may indicate genuine anonymization or simply inaccessible memorized knowledge via that particular test. The high firm and year accuracy for Magnificent 7 stocks (100\% for AAPL/META/MSFT) represents a lower bound on what the model has memorized and the actual knowledge is likely substantially greater. Weaker performance for smaller stocks mirrors the higher errors for small-cap portfolios, reinforcing that memorization favors well-represented entities. The ability to identify firms and years from anonymized texts implies that LLMs can leverage subtle cues to access memorized data, undermining masking as a safeguard against forecasting contamination. This finding underscores the need for post-cutoff data to evaluate true predictive ability, as masked historical analyses may still reflect memorized outcomes rather than genuine insight.

In the next analysis, we further evaluate whether masking can prevent GPT-4o from accessing memorized information by repeating the test for firm-specific news headlines from January 2000 to September 2023 (the pre-cutoff period). Additionally, we have data from October 2023 to June 2024 in the post-cutoff period. In contrast with the earnings call transcripts, the firm headlines are shorter texts and are more uniform in wording. After the entity neutering process, routine news such as executives trading shares of the firm or announcements of earnings are almost indistinguishable between firms and across time. For these reasons, we grouped firm headlines by month in this test rather than giving GPT-4o the headlines one at a time. 

We show in Table \ref{tab:firm_headlines} that this masking approach could reasonably obscure the dates of the news, but GPT-4o could guess the Magnificent 7 firms with 70\% accuracy, the top decile of firms by market capitalization with 55.64\% accuracy, and the top quintile of firms by market capitalization with 46.71\% accuracy. Even for the dates, while the LLM is not extremely accurate, it could guess the year during the pre-cutoff period almost 20\% of the time, with accuracy rising to 50\% in the more recent period near the cutoff date of October 2023.

While the masking approach was more effective for firm headlines, this approach to mitigate memorization still fails to completely prevent GPT-4o from reconstructing various identifying information. However, it is possible that if the masked text is short enough, GPT-4o cannot effectively identify firm identity. To test this conjecture, we further examine the effectiveness of masking firm headlines by sampling 50 stocks for each market capitalization decile (resampling annually), and testing identification of firms by GTP-4o from single headlines. In contrast to the previous test, we give the masked headlines one at a time to be identified by GPT-4o rather than giving it a group of headlines. Since the single headlines tend to be short and generic, especially after masking, there is not much differentiating information for the LLM to use in identifying firms. For this reason, the recall rates for the pre-cutoff period are generally lower, as shown in Panel A of Table \ref{tab:firm_headlines_sampled}. However, the recall rates are higher than random chances in all deciles. For instance, the firm accuracy for Decile 10 containing the largest firms is 7.53\%, more than five times higher than randomly guessing one of the firms in the decile (1.41\%).  In contrast, the results from the post-cutoff period show little recall capabilities. As shown in Panel B, firm accuracy is close to or lower than the random chances in all size deciles. 

\begin{table}[H]
\centering
\caption{Anonymized Firm Headlines}\label{tab:firm_headlines}
\caption*{\scriptsize This table reports a set of evaluation metrics assessing the LLM’s ability to recall dates and stock tickers associated with historical headlines for specific firms. The LLM was prompted with a group of anonymized headlines from RavenPack about a firm during a month and asked to state what month, year, and company the headlines were referencing. We only keep the groups that had at least 10 headlines. Panel A reports the results for identifying the anonymized headlines. \textit{Mean Months Difference} is the average signed difference (in months) between predicted and actual dates, while \textit{Mean Absolute Days Difference} reports the average absolute difference. \textit{Year Accuracy} measures the percentage of predictions correctly recalling the year. \textit{Firm Accuracy} measures the percentage of predictions correctly identifying the ticker associated with the headlines. Results are provided separately for firm headlines from the \textit{Pre-training Period}, a more recent time frame \textit{Pre-training Period Recent} from January 2014 to September 2023, the \textit{Pre-training Period Near Cutoff} period from January 2022 to September 2023 and the \textit{Post-training Period}. Results are also shown by various firm sizes including the Magnificent Seven stocks, the top decile and quintile by market cap, and large stocks versus the rest categorized by NYSE market cap cutoffs. In Panel B, we break down the results further and provide the statistics by market capitalization deciles. In Panel C, we provide a benchmark for date identification accuracy using the deanonymized headlines. }
\scriptsize
\begin{tabularx}{\linewidth}{lYYYYY}
  \toprule
     \multicolumn{6}{l}{\textit{Panel A: Date and Firm Recall With Anonymized Headlines}} \\
  \midrule
 & Mean Months Difference & Mean Absolute Months Difference & Year Accuracy (\%) & Firm Accuracy (\%) & Num Obs \\
  \midrule
   \multicolumn{6}{l}{\textit{Various sample periods}} \\
   \midrule
 Pre-training Period & 37.76 & 55.17 & 19.65 & 21.13 & 8714 \\
  Pre-training Period Recent & 26.34 & 42.44 & 23.44 & 23.21 & 2133 \\
  Pre-training Period Near Cutoff & -3.65 & 13.01 & 50.47 & 27.96 & 422 \\
  Post-training Period & -15.26 & 15.26 & 15.79 & 28.57 & 133 \\ 
    \midrule
   \multicolumn{6}{l}{\textit{Various firm sizes}} \\
   \midrule
  Magnificent Seven & 13.39 & 26.10 & 33.33 & 70.00 & 390 \\
  Top Decile &  18.68    &  32.17    &  30.25    &   55.64   &  638\\
  Top Quintile &  23.14    &   37.46   &  25.71     &  46.71    &  1276\\
  Large &  25.99     &  44.36   & 22.59   &  32.68    &  5233\\
  Small and Medium &  43.55     &  62.98   &  15.11  & 8.12     & 1145\\
   \bottomrule
\end{tabularx}

\vspace{5pt}

\begin{tabularx}{\linewidth}{lYYYYYYYYY}
\toprule
 \multicolumn{10}{l}{\textit{Panel B: Firm Size Deciles}} \\
 \midrule
Decile & Mean Months Diff & Mean Abs Months Diff & Year Accuracy (\%) & Firm Accuracy (\%) & Num Month Firms & Num Unique Firms & Min Cap (\$B) & Max Cap (\$B) & Median Cap (\$B) \\

\midrule
1  & 48.00 & 67.49 & 14.29 &  8.79 & 637 & 263 &   0.25 &   2.28 &   1.04 \\
2  & 40.76 & 60.54 & 14.73 &  9.09 & 638 & 255 &   2.28 &   5.65 &   3.88 \\
3  & 31.48 & 51.50 & 22.10 & 12.54 & 638 & 226 &   5.65 &  12.01 &   8.57 \\
4  & 27.58 & 48.54 & 17.43 & 23.86 & 637 & 173 &  12.01 &  21.32 &  16.33 \\
5  & 25.75 & 47.96 & 20.06 & 25.24 & 638 & 147 &  21.32 &  36.98 &  27.93 \\
6  & 20.47 & 42.65 & 22.41 & 36.21 & 638 & 106 &  36.98 &  59.98 &  47.59 \\
7  & 29.21 & 48.53 & 21.51 & 36.73 & 637 &  79 &  59.98 &  99.72 &  78.37 \\
8  & 22.02 & 34.97 & 28.53 & 36.83 & 638 &  52 &  99.72 & 157.61 & 127.24 \\
9  & 27.59 & 42.74 & 21.16 & 37.77 & 638 &  42 & 157.61 & 241.22 & 191.06 \\
10  & 18.71 & 32.21 & 30.14 & 55.57 & 637 &  32 & 241.22 &2831.10 & 355.71 \\
\bottomrule
\end{tabularx}

\vspace{5pt}

\begin{tabularx}{\linewidth}{lYYYY}
  \toprule
   \multicolumn{5}{l}{\textit{Panel C: Date Recall With Deanonymized Headlines}} \\
  \midrule
 & Mean Months Difference & Mean Absolute Months Difference & Month Year Accuracy (\%) & Num Obs \\ 
  \midrule
  Pre-Training Period & 1.53 & 3.94 & 76.71 & 8733 \\ 
  Pre-Training Period Recent & 2.85 & 3.52 & 77.81 & 2181\\ 
  Post-Training Period & -1.20 & 1.47 & 57.89 & 152 \\ 
   \bottomrule
\end{tabularx}

\end{table}


\begin{table}[H]
\centering
\caption{Anonymized Firm Headlines - Single Headline Ticker Accuracy}\label{tab:firm_headlines_sampled}
\caption*{\scriptsize This table reports a set of evaluation metrics assessing the LLM’s ability to recall stock tickers associated with historical headlines for specific firms. The LLM was prompted with an anonymized headline from RavenPack about a firm asked to state what month, year, and company the headline was referencing. The sample covers Janaury 2000 to June 2024. Panel A reports the Firm Accuracy for identifying the anonymized headlines. \textit{Firm Accuracy} measures the percentage of predictions correctly identifying the ticker associated with the headlines. \textit{Most News} indicates the accuracy achieved by always predicting the firm with the highest total number of headlines in the sample. \textit{Random} is the inverse of Num Unique Firms, the accuracy of a uniform random guess across all unique firms in the decile. \textit{AAPL} is the accuracy obtained by always predicting AAPL. Results are shown for the market capitalization deciles. We form decile breaks using NYSE stocks for each year and randomly sample 50 stocks from each decile to test, resampled annually.}
\scriptsize
\begin{tabularx}{\linewidth}{lYYYYYYYYY}
\toprule
Decile & Firm Accuracy (\%) & Random (\%) & Most News (\%)  & AAPL (\%) & Num Headlines & Num Unique Firms & Median Market Cap (\$B) & Min Market Cap (\$B) & Max Market Cap (\$B) \\
\midrule
\multicolumn{6}{l}{\textit{Panel A: Pre-cutoff period, 01/2000 to 09/2023}} \\
\midrule
1  & 0.57 & 0.20 & 0.00 & 0.00 & 11500 & 501 & 0.45 & 0.25 & 1.09 \\
2  & 0.70 & 0.25 & 0.00 & 0.00 & 11449 & 408 & 1.08 & 0.47 & 2.15 \\
3  & 0.78 & 0.26 & 0.00 & 0.00 & 11642 & 384 & 1.82 & 0.76 & 3.49 \\
4  & 1.00 & 0.29 & 0.00 & 0.00 & 11489 & 345 & 2.94 & 1.49 & 6.07 \\
5  & 1.01 & 0.30 & 0.00 & 0.00 & 11682 & 332 & 4.82 & 2.23 & 10.96 \\
6  & 1.26 & 0.35 & 0.00 & 0.49 & 11352 & 284 & 8.01 & 3.13 & 17.71 \\
7  & 1.80 & 0.43 & 0.00 & 0.00 & 11615 & 232 & 13.87 & 5.49 & 46.48 \\
8  & 2.99 & 0.54 & 0.00 & 0.00 & 11262 & 184 & 24.37 & 9.22 & 92.19 \\
9  & 3.89 & 0.73 & 0.00 & 1.13 & 11263 & 137 & 51.52 & 17.10 & 239.51 \\
10 & 7.53 & 1.41 & 7.01 & 3.11 & 12012 & 71  & 174.06 & 42.13 & 2434.21 \\
\midrule
\multicolumn{6}{l}{\textit{Panel B: Post-cutoff period, 10/2023 to 06/2024}} \\
\midrule
1  & 0.92 & 1.59 & 0.00 & 0.00 & 433 & 63 & 0.26 & 0.00 & 0.54 \\
2  & 0.00 & 1.96 & 0.00 & 0.00 & 378 & 51 & 0.87 & 0.38 & 1.42 \\
3  & 0.00 & 1.96 & 0.00 & 0.00 & 425 & 51 & 1.79 & 1.30 & 2.43 \\
4  & 0.50 & 1.85 & 0.00 & 0.00 & 404 & 54 & 2.82 & 2.17 & 3.95 \\
5  & 0.00 & 2.00 & 0.00 & 0.00 & 396 & 50 & 4.98 & 3.22 & 7.40 \\
6  & 1.95 & 2.27 & 0.00 & 0.00 & 411 & 44 & 8.85 & 5.81 & 14.40 \\
7  & 0.72 & 2.50 & 0.00 & 0.00 & 415 & 40 & 15.59 & 8.94 & 27.22 \\
8  & 2.84 & 2.44 & 0.00 & 0.00 & 423 & 41 & 38.44 & 17.34 & 71.06 \\
9  & 3.79 & 3.13 & 0.00 & 0.00 & 396 & 32 & 97.38 & 33.65 & 428.85 \\
10 & 6.94 & 7.14 & 42.82 & 10.05 & 418 & 14 & 497.48 & 83.83 & 2434.21 \\
\bottomrule
\end{tabularx}
\end{table}

These masking results reveal a fundamental trade-off. For an LLM to generate useful forecasts, text must contain specific contextual information. Yet this specificity enables entity identification. Consider a masked headline: ``Firm\_x reported weak demand for product\_y in region\_z.'' If details are generic enough to prevent identification (``a company reported weak demand''), the signal loses predictive value. If details are specific enough to predict (``a smartphone maker reported weak China demand in Q4 2018''), they reveal the identity (Apple). Achieving both effective anonymization and meaningful forecasting power simultaneously is highly challenging. Combined with Remark~\ref{rem:lower-bound}, which establishes that failed identification does not prove successful anonymization, this trade-off suggests masking cannot reliably solve the memorization problem for forecasting tasks.

\subsection{Addressing Lookahead Bias with Economic Logic}

In previous sections, we have shown that the effectiveness of mitigating lookahead bias using artificially imposed knowledge cutoff dates and anonymization through masking entities is limited. A natural question is whether further abstraction by removing the original wording while retaining economic meaning can weaken the link between the text and GPT-4o's internal memory. In this section, we test whether transforming firm headlines into anonymized ``economic logic'' reduces the rate of memorization while still allowing the model to make forecasts. If abstraction successfully removes identifying information and sentence structure while retaining the underlying economic logic, the model should be able to forecast stock returns but should no longer be able to recover the firm or date of the headline. Valid masking requires both future-invariance (the task does not depend on information revealed after time t) and detectable skill (above-baseline accuracy). We find that while this procedure yields detectable skill, reconstruction tests reveal multiple pathways through which the model can access memorized information, indicating the masking does not achieve future-invariance.

We begin with a set of firm-specific headlines. For each headline, we ask the LLM to describe the effect this news may have on the firm omitting any specifics, outputting the economic logic only using one sentence. Additionally, we ask the LLM to anonymize this economic logic using the masking technique. Using the set of economic logic constructed in the first step, we then ask the LLM to predict whether the stock will be up or down by giving it the anonymized economic logic. An example is shown in Figure \ref{fig:econ_logic_examples_one_sentence}. 

\begin{figure}[htbp]
    \centering
    \begin{tcolorbox}[
        colback=gray!5,
        colframe=gray!75!black,
        width=0.95\textwidth,
        arc=1mm,
        boxrule=0.5pt
    ]
    \small
    \textbf{Economic Logic Example (One Sentence):}
    \vspace{0.2cm}
    \begin{quote}
     \textbf{Headline:} Amazon Launches Kindle in Mexico
    
     \textbf{Economic Logic:} The firm could experience increased revenue and market expansion opportunities due to tapping into a new consumer base in Mexico.
     
     \textbf{Anonymized Economic Logic:} The firm could experience increased revenue and market\_x expansion opportunities due to tapping into a new consumer\_base\_x in location\_x.
     
    \end{quote}
    \end{tcolorbox}
    \caption{An example of a headline, the one sentence economic logic generated from GPT-4o, and the anonymized economic logic masked using the entity neutering approach \citep{engelbergEntityNeutering2025} using GPT-4o mini. }
    \label{fig:econ_logic_examples_one_sentence}
\end{figure}

    
     
     

Table \ref{tab:econ_logic_one} reports the results of testing this procedure on the Magnificent 7 stocks using GPT-4o on a daily frequency. For each day where there was news either before 9 a.m. on the current day or after 4 p.m. the previous day, we ask the LLM to predict whether the stock will be up or down from open to close based on the anonymized economic logic. For the pre-cutoff period, directional accuracies for all seven stocks exceed 50\%, ranging from 50.74\% to 58.19\%, satisfying the detectable skill requirement for valid masking. 

To test the future-invariance condition, we attempt to reconstruct identifying information from the anonymized economic logic by asking GPT-4o to identify the firm and the date of the news from the anonymized text. The results show that for prominent stocks (AAPL, META, MSFT, TSLA, AMZN), GPT-4o can identify the firm from anonymized economic logic with accuracy rates ranging from 11.9\% to 46.3\%. For GOOGL and NVDA, firm identification rates are lower (3.3\% and 0.6\%, respectively). Date reconstruction largely fails, with the LLM essentially unable to guess the exact date and only identifying the year between 3-13\% of the time, depending on the stock. For prominent stocks, successful firm reconstruction directly demonstrates that future-invariance fails: the model can access information that links the masked task to memorized knowledge about what happened after the headline date.

\begin{table}[H]
\centering
\caption{Anonymized Economic Logic}\label{tab:econ_logic_one}
\caption*{\scriptsize This table reports GPT-4o's performance in using the underlying economic logic of headlines to directionally forecast stock price movement on a daily frequency from January 2000 to June 2024. We use RavenPack news headlines for the Magnificent Seven stocks. For each headline, we ask GPT-4o to describe how the firm will be impacted by the news using economic logic only without specifics. We then anonymize the economic logic as another layer of abstraction away from the original text. Panel A reports the percentage accuracies of the forecasts for each firm reported in percentage points (-0.01 means -0.01\%). Panel B reports the success rate of identifying the correct firm, year, and the exact date of the anonymized economic logic. }
\scriptsize

\begin{tabularx}{\linewidth}{lYYYYYY}
\toprule
 \multicolumn{5}{l}{\textit{Panel A: Forecast Accuracy using Anonymized Economic Logic}} \\
  \midrule
  Ticker & Forecast Accuracy Excl Unknown (\%) & Forecast Accuracy Incl Unknown (\%) & Num Obs & Num Headlines & Num Headlines (before 9am or after 4pm)  & Avg Num Headlines Per News Day \\
    \midrule
  \multicolumn{5}{l}{\textit{Pre-cutoff period, 01/2000 to 09/2023}} \\
  \midrule
  AAPL  & 54.89 & 49.51 & 919  & 2857& 2008  &2.04 \\
  META  & 50.77 & 41.38 & 319  & 933    & 660	 &1.62 \\
  MSFT  & 50.74 & 46.06 & 1333 & 3671   & 2385 & 1.77\\
  GOOG & 56.61 & 48.74 & 835  & 2009   & 1284  & 1.47\\
  NVDA  & 55.29 & 44.53 & 411  & 1070  & 775	 & 1.73\\
  TSLA  & 58.19 & 49.43 & 352  & 899    & 621   & 1.59\\
  AMZN  & 52.85 & 41.14 & 722  & 2025   & 1342	 & 1.72\\
    \midrule
   \multicolumn{5}{l}{\textit{Post-cutoff period, 10/2023 to 06/2024}} \\
  \midrule
  AAPL  & 68.18 & 51.72 & 29 & 107 & 62 & 1.95\\
  META  & 50.00 & 43.75 & 16 & 44  & 32 & 1.76\\
  MSFT  & 55.88 & 52.78 & 36 & 126 & 64 & 1.62\\
  GOOG & 74.07 & 60.61 & 33 & 97  & 62 & 2.16\\
  NVDA  & 53.33 & 41.11 & 19 & 68  & 46 & 1.94\\
  TSLA  & 71.43 & 62.50 & 16 & 62  & 24 & 1.72\\
  AMZN  & 53.33 & 51.61 & 31 & 88  & 59 & 1.69\\
  \bottomrule
\end{tabularx}

\vspace{5pt}

\begin{tabularx}{\linewidth}{lYYY}
\toprule
 \multicolumn{4}{l}{\textit{Panel B: Identification of Anonymized Economic Logic}} \\
Ticker & Firm Accuracy (\%) & Year Accuracy (\%) & Date Accuracy (\%) \\
\midrule
  \multicolumn{4}{l}{\textit{Pre-cutoff period, 01/2000 to 09/2023}} \\
  \midrule
AAPL  & 46.3 &  3.5 & 0.0 \\
META  & 11.9 &  7.2 & 0.4 \\
MSFT  & 21.6 &  3.9 & 0.1 \\
GOOG &  3.3 &  7.3 & 0.0 \\
NVDA  &  0.6 &  7.5 & 0.0 \\
TSLA  & 27.5 & 12.2 & 0.0 \\
AMZN  & 19.2 &  5.4 & 0.0 \\
    \midrule
   \multicolumn{4}{l}{\textit{Post-cutoff period, 10/2023 to 06/2024}} \\
  \midrule
AAPL  & 39.2  & 34.0  & 0.0  \\
META  & 11.4  & 25.0  &  4.5 \\
MSFT  & 27.0  & 26.2  & 0.0  \\
GOOG & 1.9  &  27.1 &  0.0 \\
NVDA  & 5.9  & 30.9  & 0.0  \\
TSLA  & 27.4  & 21.0  & 0.0  \\
AMZN  & 22.7  &  27.3 &  0.0 \\
\bottomrule
\end{tabularx}
\end{table}




For a broader cross-section of stocks, this procedure yields similar directional accuracy rates to short-text masking, as shown in Table \ref{tab:econ_logic_one_sampled}. Firm identification rates from the anonymized economic logic are above random (randomly picking a unique firm from the decile) for deciles 5 to 10 in the pre-cutoff period, though lower than for the Magnificent 7 stocks. For the post-cutoff period, firm identification rates fall below the random benchmark, consistent with the model not having seen these data during training.

\begin{table}[H]
\centering
\caption{Anonymized Economic Logic - Sampled Stocks}\label{tab:econ_logic_one_sampled}
\caption*{\scriptsize  This table reports GPT-4o's performance in using the underlying economic logic of headlines to directionally forecast stock price movement on a daily frequency from January 2000 to June 2024. We use RavenPack news headlines for a sample of stocks, the same sample used in Table \ref{tab:firm_headlines_sampled}. For each headline, we ask GPT-4o to describe how the firm will be impacted by the news using economic logic only without specifics. We then anonymize the economic logic as another layer of abstraction away from the original text. Panel A reports the Firm Accuracy for identifying the anonymized economic logic. \textit{Firm Accuracy} measures the percentage of predictions correctly identifying the ticker associated with the anonymized economic logic. \textit{Most News} indicates the accuracy achieved by always predicting the firm with the highest total number of headlines in the sample. \textit{Random} is the accuracy of a uniform random guess across all firms in the decile. \textit{AAPL} is the accuracy obtained by always predicting AAPL. Results are shown for the market capitalization deciles. We form decile breaks using NYSE stocks for each year and randomly sample 50 stocks from each decile to test, resampled annually.}
\scriptsize
\begin{tabularx}{\linewidth}{lYYYYYYYYY}
\toprule
Decile & Firm Accuracy (\%) & Random (\%) & Most News (\%)  & AAPL (\%) & Num Headlines & Num Unique Firms & Median Market Cap (\$B) & Min Market Cap (\$B) & Max Market Cap (\$B) \\
\midrule
\multicolumn{10}{l}{\textit{Panel A: Pre-cutoff period, 01/2000 to 09/2023}} \\
\midrule
1  & 0.03 & 0.20 & 0.00 & 0.00 & 11500 & 501 & 0.45  & 0.25  & 1.09 \\
2  & 0.01 & 0.25 & 0.00 & 0.00 & 11449 & 408 & 1.08  & 0.47  & 2.15 \\
3  & 0.11 & 0.26 & 0.00 & 0.00 & 11642 & 384 & 1.82  & 0.76  & 3.49 \\
4  & 0.02 & 0.29 & 0.00 & 0.00 & 11489 & 345 & 2.94  & 1.49  & 6.07 \\
5  & 0.34 & 0.30 & 0.00 & 0.00 & 11682 & 332 & 4.82  & 2.23  & 10.96 \\
6  & 0.42 & 0.35 & 0.00 & 0.49 & 11352 & 284 & 8.01  & 3.13  & 17.71 \\
7  & 0.41 & 0.43 & 0.00 & 0.00 & 11615 & 232 & 13.87 & 5.49  & 46.48 \\
8  & 1.16 & 0.54 & 0.00 & 0.00 & 11262 & 184 & 24.37 & 9.22  & 92.19 \\
9  & 1.86 & 0.73 & 0.00 & 1.13 & 11263 & 137 & 51.52 & 17.10 & 239.51 \\
10 & 6.15 & 1.41 & 7.01 & 3.11 & 12012 & 71  & 174.06 & 42.13 & 2434.21 \\
\midrule
\multicolumn{10}{l}{\textit{Panel B: Post-cutoff period, 10/2023 to 06/2024}} \\
\midrule
1  & 0.23 & 1.59 & 0.00 & 0.00 & 433 & 63 & 0.26  & 0.00  & 0.54 \\
2  & 0.00 & 1.96 & 0.00 & 0.00 & 378 & 51 & 0.87  & 0.38  & 1.42 \\
3  & 0.00 & 1.96 & 0.00 & 0.00 & 425 & 51 & 1.79  & 1.30  & 2.43 \\
4  & 0.00 & 1.85 & 0.00 & 0.00 & 404 & 54 & 2.82  & 2.17  & 3.95 \\
5  & 0.00 & 2.00 & 0.00 & 0.00 & 396 & 50 & 4.98  & 3.22  & 7.40 \\
6  & 0.49 & 2.27 & 0.00 & 0.00 & 411 & 44 & 8.85  & 5.81  & 14.40 \\
7  & 0.00 & 2.50 & 0.00 & 0.00 & 415 & 40 & 15.59 & 8.94  & 27.22 \\
8  & 0.47 & 2.44 & 0.00 & 0.00 & 423 & 41 & 38.44 & 17.34 & 71.06 \\
9  & 1.52 & 3.12 & 0.00 & 0.00 & 396 & 32 & 97.38 & 33.65 & 428.85 \\
10 & 4.55 & 7.14 & 42.82 & 10.05 & 418 & 14 & 497.48 & 83.83 & 2434.21 \\
\bottomrule
\end{tabularx}
\end{table}

Additionally, Table \ref{tab:econ_logic_industry} shows that GPT-4o can identify the correct Fama–French industry classification at rates far above chance even from anonymized economic logic. Under the five industry classification, the model correctly assigns the industry 62.4\% of the time in the pre-cutoff period and 59.0\% in the post-cutoff period for the Consumer industry. Even in the more granular ten industry grouping, GPT-4o exhibits non-trivial classification performance, correctly identifying industries such as HiTec, Manuf, and NoDur at roughly 20–33\% accuracy.

This industry reconstruction result is critical for evaluating future-invariance. Even when the model cannot directly identify the specific firm, it can reconstruct industry membership at rates far exceeding chance. By Remark~\ref{rem:lower-bound}, reconstruction tests provide only lower bounds on accessible information: failure to reconstruct firm identity does not prove the model cannot access other memorized contextual information (industry patterns, business conditions, macroeconomic context) that links the masked task to information revealed after time t. The successful industry reconstruction demonstrates that alternative information pathways remain available even when direct firm identification fails. Combined with the direct firm identification results for prominent stocks, these findings indicate that the economic logic procedure does not satisfy the future-invariance requirement for valid masking.

\begin{table}[H]
\centering
\caption{Industry Identification Accuracy}\label{tab:econ_logic_industry}
\caption*{\scriptsize
This table reports GPT-4o's accuracy in identifying a firm's industry from anonymized economic logic. Panel A summarizes accuracy under the Fama--French 5-industry grouping; Panel B summarizes accuracy under the Fama--French 10-industry grouping. \textit{Industry Accuracy} is \textit{Correct} divided by \textit{Total}, expressed in percentage points. \textit{Random} is a benchmark representing the accuracy of a classifier that guesses uniformly at random across industries within the grouping (20\% for the 5-industry grouping and 10\% for the 10-industry grouping). The \textit{Pre-cutoff period} uses observations dated before 2023-10-01, and the \textit{Post-cutoff period} uses observations dated on or after 2023-10-01.}
\scriptsize

\begin{tabularx}{\linewidth}{lYYYY}
  \toprule
  \multicolumn{5}{l}{\textit{Panel A: Fama--French 5-Industry Grouping}} \\
  \midrule
  \multicolumn{5}{l}{\textit{Pre-cutoff period, 01/2000 to 09/2023}} \\
  \midrule
  Industry (FF5) & Total & Correct & Industry Accuracy (\%) & Random (\%) \\
  \midrule
  Cnsmer & 18,175 & 11,337 & 62.4 & 20.0 \\
  HiTec  & 14,055 & 2,588  & 18.4 & 20.0 \\
  Hlth   & 5,203  & 308    & 5.9  & 20.0 \\
  Manuf  & 18,144 & 2,722  & 15.0 & 20.0 \\
  Other  & 60,640 & 12,532 & 20.7 & 20.0 \\
  \midrule
  \multicolumn{5}{l}{\textit{Post-cutoff period, 10/2023 to 06/2024}} \\
  \midrule
  Industry (FF5) & Total & Correct & Industry Accuracy (\%) & Random (\%) \\
  \midrule
  Cnsmer & 363  & 214 & 59.0 & 20.0 \\
  HiTec  & 345  & 42  & 12.2 & 20.0 \\
  Hlth   & 161  & 6   & 3.7  & 20.0 \\
  Manuf  & 429  & 52  & 12.1 & 20.0 \\
  Other  & 2,685 & 609 & 22.7 & 20.0 \\
  \bottomrule
\end{tabularx}

\vspace{10pt}

\begin{tabularx}{\linewidth}{lYYYY}
  \toprule
  \multicolumn{5}{l}{\textit{Panel B: Fama--French 10-Industry Grouping}} \\
  \midrule
  \multicolumn{5}{l}{\textit{Pre-cutoff period, 01/2000 to 09/2023}} \\
  \midrule
  Industry (FF10) & Total & Correct & Industry Accuracy (\%) & Random (\%) \\
  \midrule
  Durbl & 3,225  & 533    & 16.5 & 10.0 \\
  Enrgy & 2,867  & 164    & 5.7  & 10.0 \\
  HiTec & 11,603 & 2,493  & 21.5 & 10.0 \\
  Hlth  & 5,200  & 308    & 5.9  & 10.0 \\
  Manuf & 12,042 & 2,482  & 20.6 & 10.0 \\
  NoDur & 4,332  & 1,347  & 31.1 & 10.0 \\
  Other & 60,628 & 11,467 & 18.9 & 10.0 \\
  Shops & 10,613 & 431    & 4.1  & 10.0 \\
  Telcm & 2,451  & 168    & 6.9  & 10.0 \\
  Utils & 3,229  & 25     & 0.8  & 10.0 \\
  \midrule
  \multicolumn{5}{l}{\textit{Post-cutoff period, 10/2023 to 06/2024}} \\
  \midrule
  Industry (FF10) & Total & Correct & Industry Accuracy (\%) & Random (\%) \\
  \midrule
  Durbl & 48  & 6   & 12.5 & 10.0 \\
  Enrgy & 66  & 3   & 4.5  & 10.0 \\
  HiTec & 312 & 62  & 19.9 & 10.0 \\
  Hlth  & 161 & 6   & 3.7  & 10.0 \\
  Manuf & 271 & 50  & 18.4 & 10.0 \\
  NoDur & 133 & 44  & 33.1 & 10.0 \\
  Other & 2,685 & 544 & 20.3 & 10.0 \\
  Shops & 182 & 4   & 2.2  & 10.0 \\
  Telcm & 33  & 0   & 0.0  & 10.0 \\
  Utils & 92  & 0   & 0.0  & 10.0 \\
  \bottomrule
\end{tabularx}
\end{table}



\subsection{Memorization in Embeddings?}

Another question to ask is whether the embeddings in LLMs show signs of memorization. Recent work using LLM embeddings for financial applications, such as \citet{chenExpectedReturnsLarge2022}, result in substantially better stock-return predictions than traditional indicators or simpler NLP methods.  Outside of finance, embeddings have been shown to leak a similar amount of sensitive information as text \citep{morris2023text}. Embeddings are numerical vector representations of text generated by the LLM. These vectors contain the semantic content of the text input, including contextual information drawn from patterns the model encountered during pre-training. The ability to recover the historically reported figures from embeddings of prompts that omit the numeric token would constitute strong evidence of memorization in the LLM embeddings.

\begin{figure}[htbp]
    \centering
    \caption{Recall through Embeddings} \label{fig:embeddings}
    \caption*{\scriptsize This figure shows the comparison of actual values and predicted values for GDP growth, inflation, 10-year treasury rate, and unemployment rate using Ridge regression with regularization parameter of 0.01 trained on embeddings of textual prompts such as ``In Q4 2020, the earliest estimate of the US GDP growth rate was" and the corresponding economic data, either rates or levels. We use a 5 year rolling window to train the Ridge regression each period on the most recent five years of data to predict the next value. The solid blue lines show the actual values and dashed red lines show the predicted values. }




  \begin{subfigure}[b]{0.58\textwidth}
    \includegraphics[width=\textwidth]{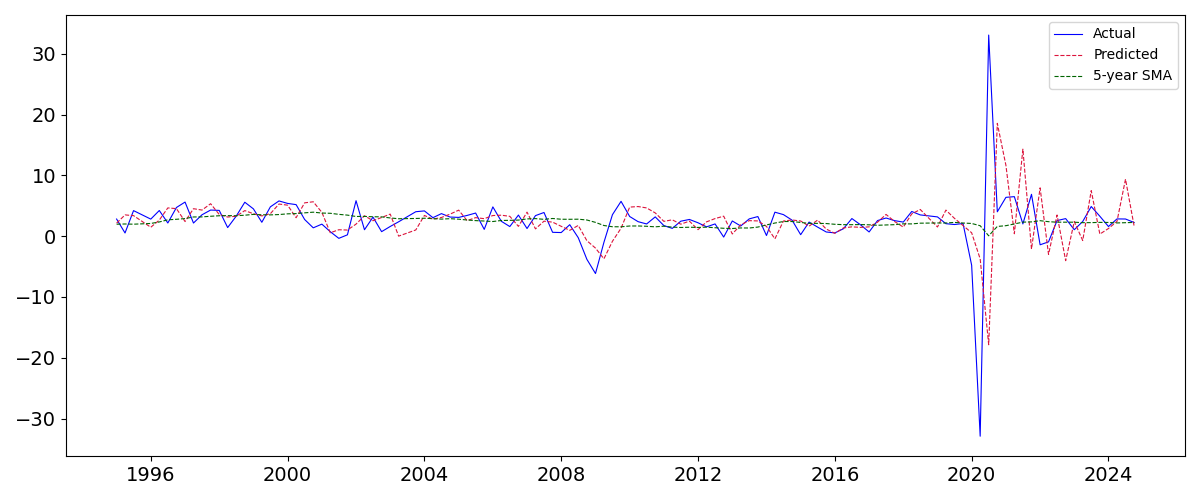}
    \caption{GDP Growth (Rolling Window)}
    \label{fig:gdp_rolling}
  \end{subfigure}
  \hfill
  \begin{subfigure}[b]{0.58\textwidth}
    \includegraphics[width=\textwidth]{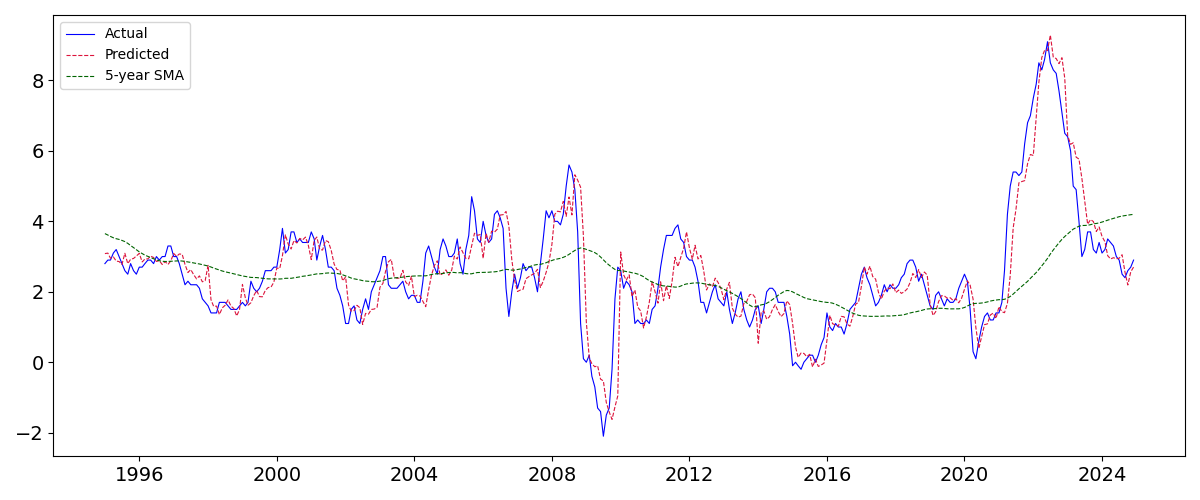}
    \caption{Inflation (Rolling Window)}
    \label{fig:inflation_rolling}
  \end{subfigure}

  \vspace{1em}

  \begin{subfigure}[b]{0.58\textwidth}
    \includegraphics[width=\textwidth]{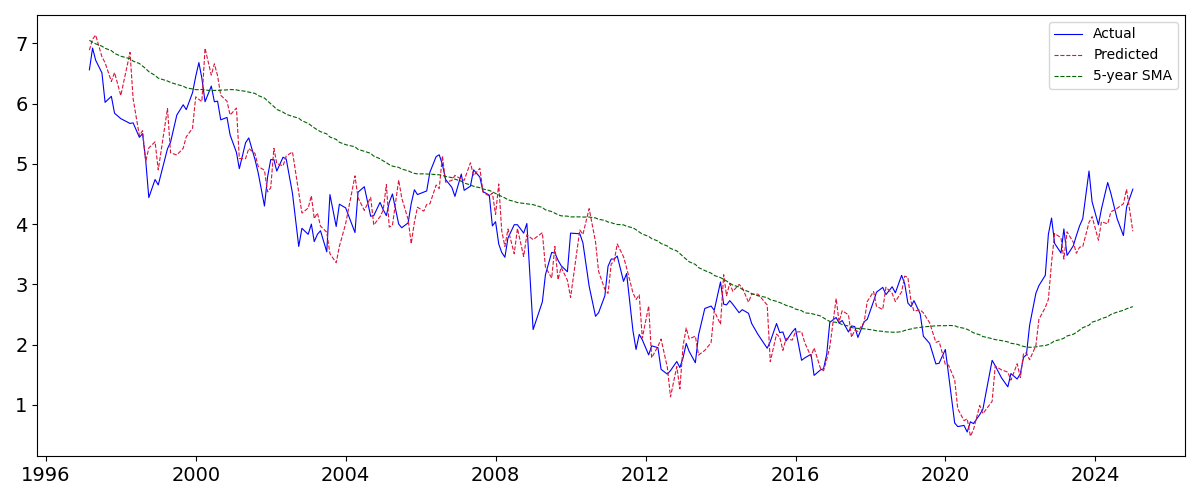}
    \caption{10-yr Treasury Rate (Rolling Window)}
    \label{fig:dgs10_rolling}
  \end{subfigure}
  \hfill
  \begin{subfigure}[b]{0.58\textwidth}
    \includegraphics[width=\textwidth]{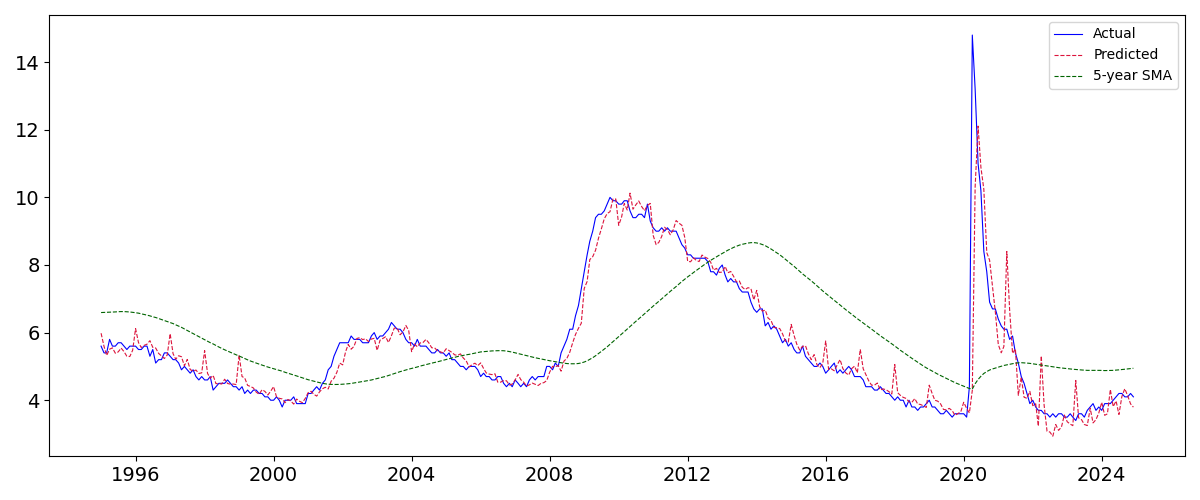}
    \caption{Unemployment (Rolling Window)}
    \label{fig:unrate_rolling}
  \end{subfigure}

\end{figure}

Using the text-embedding-3-large model, we first embed a series of prompts that omit a particular numeric value at a particular date for a macro indicator, for example, “In Q4 2020, the earliest estimate of the US GDP growth rate was.”\footnote{For more information about the embedding model, see: https://platform.openai.com/docs/models/text-embedding-3-large.} We then train a Ridge regression model with regularization strength $\lambda$ = 0.01 on those embeddings and the corresponding economic data to predict four economic indicators: GDP growth rate, inflation rate, 10-year treasury rate, and unemployment rate. Using a Ridge regression model allows us to control for multicollinearity among high-dimensional embedding features. The rolling window trains on the most recent five years of data (monthly for inflation, 10-year treasury rate, and unemployment; quarterly for GDP growth) and predicts the next time point. Figure \ref{fig:embeddings} shows, for each indicator, the comparison of actual values versus predicted values using the rolling windows and the 5-year simple moving average. \footnote{We show a similar pattern occurs when using an expanding window. Figure \ref{fig:embeddings_appendix} shows the comparison using the expanding windows and using a simple moving average with a five-year window. The expanding approach uses ten folds with a one-fold gap between training and testing data, i.e., for the nth fold, the model is trained on folds 1 through n and tested on fold n + 1. In addition, we also show that the results are not significantly changed using regularization strength $\lambda$ = 0.001 in Figure \ref{fig:embeddings_small} and $\lambda$ = 0.1 in Figure \ref{fig:embeddings_big}.} 

Table \ref{tab:embedding_corr_2} reports the correlations between the predicted and actual values, comparing the 5-year rolling-window model with a 5-year simple moving-average (SMA) benchmark.\footnote{We use the 5-year SMA as the benchmark because, under standard practice, Ridge regression on a rolling window collapses to the window mean (SMA) when the predictors carry no signal.} We find evidence of memorization in the embeddings of inflation and unemployment rates. In particular, the correlation between actual and rolling-window predicted values for inflation (0.892) is a large improvement over the SMA benchmark (0.333). Unemployment rate also shows high coherence (0.927 rolling, 0.385 SMA), particularly when leveraging recent history. GDP growth consistently registers near zero correlation (-0.114 rolling, –0.074 SMA). The embeddings provide almost no signal for GDP growth, but is slightly better than the SMA benchmark when training with the expanding window (with a correlation of 0.169). Taken together, these results suggest that embeddings do carry some associations, but not consistently across economic indicators.\footnote{We find similar results using several open-source embedding models, including (i) \emph{SFR-Embedding-Mistral} (\url{https://huggingface.co/Salesforce/SFR-Embedding-Mistral}), (ii) \emph{nomic-embed-text-v1.5} (\url{https://huggingface.co/nomic-ai/nomic-embed-text-v1.5}), and (iii) \emph{all-MiniLM-L6-v2} (\url{https://huggingface.co/sentence-transformers/all-MiniLM-L6-v2}) (see Table \ref{tab:embedding_corr_open} in the Appendix).}

\begin{table}[H]
\centering
\caption{Recall through Embeddings}
\label{tab:embedding_corr_2}
\caption*{\scriptsize 
  This table reports correlations between actual values and predicted values for GDP growth, inflation, 10-year treasury yield, and unemployment rate using Ridge regression trained on embeddings of textual prompts such as ``In Q4 2020, the earliest estimate of the US GDP growth rate was'', which we call the \textit{Date and Variable Embeddings}, and the corresponding values in the data, either rates or levels. The data covers January 1990 to December 2024. \textit{Rolling Window} retrains the Ridge regression each period on the most recent five years of data to predict the next value. \textit{SMA} is a simple moving average benchmark computed over the past five years (20 quarters for GDP growth, 60 months for the other series). We report t-statistics for the hypothesis “corr$_i$ = corr$_j$” for each pair of methods using Williams’s test. We include placebo tests using various embeddings: \textit{Date Embeddings} which just include the date of the value and leaves out the variable name, \textit{Shuffled Date and Variable Embeddings} which is a random reordering of the \textit{Date and Variable Embeddings}, and \textit{Random Numerical Vectors}.
}
\scriptsize
\begin{tabularx}{\linewidth}{l 
    *{4}{>{\centering\arraybackslash}X}
}
\toprule
    & \multicolumn{2}{c}{Correlation} 
    & \multicolumn{2}{c}{Difference vs SMA} \\
\cmidrule(lr){2-3}\cmidrule(lr){4-5}
Series  
    & Rolling Window Embeddings
    & SMA 
    & Roll--SMA 
    & Roll--SMA t--stat \\
\midrule
\multicolumn{5}{l}{\textit{Date and Variable Embeddings}} \\[0.5ex]
\midrule
Inflation         
    &  0.892   
    &  0.333      
    &  0.559
    &  $t=20.07$   \\[0.5ex]
10-Yr Treasury Yield              
    &  0.965      
    &  0.857      
    &  0.108
    &  $t=15.76$   \\[0.5ex]
Unemployment Rate              
    &  0.927      
    &  0.385      
    &  0.542
    &  $t=23.26$   \\[0.5ex]
GDP Growth         
    &  -0.114      
    &  -0.074      
    &  -0.040
    &  $t=-0.35$   \\
\midrule
\multicolumn{5}{l}{\textit{Date Embeddings}} \\[0.5ex]
\midrule
Inflation         
    &  0.882   
    &  0.333      
    &  0.549
    &  $t=18.45$     \\[0.5ex]
10-Yr Treasury Yield              
    &  0.966      
    &  0.857      
    &  0.109
    &  $t=15.85$     \\[0.5ex]
Unemployment Rate              
    &  0.915      
    &  0.385      
    &  0.530
    &  $t=21.77$     \\[0.5ex]
GDP Growth         
    &  -0.124      
    &  -0.074      
    &  -0.050
    &  $t=-0.44$     \\
\midrule
\multicolumn{5}{l}{\textit{Shuffled Date and Variable Embeddings}} \\[0.5ex]
\midrule
Inflation         
    &  0.118   
    &  0.333      
    &  -0.215
    &  $t=-4.17$   \\[0.5ex]
10-Yr Treasury Yield              
    &  0.806      
    &  0.857      
    &  -0.051
    &  $t=-4.86$   \\[0.5ex]
Unemployment Rate              
    &  0.301      
    &  0.385      
    &  -0.084
    &  $t=-2.06$   \\[0.5ex]
GDP Growth         
    &  0.001      
    &  -0.074      
    &  0.075
    &  $t=0.64$   \\
\midrule
\multicolumn{5}{l}{\textit{Random Numerical Vectors}} \\[0.5ex]
\midrule
Inflation         
    &  0.308  
    &  0.333      
    &  -0.025
    &  $t=-2.05$   \\[0.5ex]
10-Yr Treasury Yield              
    &  0.853     
    &  0.857      
    &  -0.004
    &  $t=-2.33$   \\[0.5ex]
Unemployment Rate              
    &  0.365      
    &  0.385      
    &  -0.020
    &  $t=-2.66$  \\[0.5ex]
GDP Growth         
    &  -0.135
    &  -0.074      
    &  -0.061
    &  $t=-1.28$  \\

\bottomrule
\end{tabularx}
\end{table}

Naturally, a concern that follows this result is whether the ridge regressions trained on embeddings are capturing time-based patterns rather than exhibiting memorization. Table \ref{tab:embedding_corr_2} also provides a set of placebo tests in which we modify the pertinent information contained in the embeddings. We conduct three placebo tests: (1) using embeddings that only contain date information, (2) using embeddings that are a random reordering of the date and variable embeddings, and (3) using a set of vectors containing randomly generated numbers. These tests show that there is still predictive power when the temporal information is kept in the embeddings, but disappears when we remove it. For the date embeddings, we still see a significant improvement in the rolling window results compared to the SMA benchmark for all of the variables except GDP growth. However, the rolling window results for the shuffled date and variable embeddings and the random numerical vectors are worse than the SMA. This suggests that part of the predictive power is driven by the embeddings' encoding of dates, even when the economic indicator is not included. Rather than directly weakening the existence of memorization, these placebo results are consistent with temporal information leakage within the embeddings, implying that the model contains time-indexed signals. In such cases, the forecasting performance may reflect temporal memorization rather than genuine economic reasoning. 

\begin{table}[H]
\centering
\caption{Cosine Similarities and Differences Across Embedding Types}\label{tab:cosine_sim}
\caption*{\scriptsize 
This table compares cosine similarities between value embeddings and several embedding constructions: \textit{Date and Variable Embeddings}, \textit{Date Embeddings}, \textit{Shuffled Date and Variable Embeddings}, and \textit{Random Numerical Vectors}. Value embeddings refer to the embeddings of the actual values of the economic variables. The table reports the mean cosine similarities between value embeddings and each alternative embedding type. T-statistics from paired mean difference tests with true mean equal to zero are shown in parentheses.
}
\scriptsize

\begin{tabularx}{\linewidth}{lYYYY}
\toprule
\multicolumn{5}{l}{\textit{ Mean Cosine Similarities with Value Embedding}} \\
\midrule
                    & Date and Variable Embeddings & Date Embeddings & Shuffled Date and Variable Embeddings & Random Numerical Vectors \\
\midrule
Inflation           & 0.223    & 0.219      & 0.220     & 0.000 \\
                    & (129.61) & (114.03)   & (123.96)  & (0.25)   \\
10-yr Interest Rate & 0.276    & 0.193      & 0.262     & 0.001 \\
                    & (192.02) & (110.40)   & (149.13)  & (1.36)   \\
Unemployment Rate   & 0.175    & 0.208      & 0.169     & 0.001 \\
                    & (98.21)  & (98.91)    & (104.15)  & (1.23)   \\
GDP Growth          & 0.257    & 0.267      & 0.252     & 0.001 \\
                    & (90.30)  & (69.24)    & (99.10)   & (0.69)   \\
\bottomrule
\end{tabularx}


\end{table}

This interpretation is reinforced by Table \ref{tab:cosine_sim}, which compares cosine similarities between embeddings constructed from the date and variable prompts (or placebo embeddings) and embeddings of the actual values of the economic indicators. Across the set of economic variables, the date and variable embeddings are generally closer to the value embeddings than to the shuffled or random placebo embeddings. 
These patterns suggest that the model has internalized associations between specific dates and the corresponding economic states associated with those dates. 

\begin{table}[H]
\centering
\caption{Recall Through Embeddings - vec2text Inversion}\label{tab:vec2text_width2}
\caption*{\scriptsize This table reports the results of using the vec2text library \citep{morris2023text, morris2023language} to invert embeddings of prompts we generate asking for the S\&P 500 month end closing values from January 2019 to December 2024. An example of a prompt is ``On 08/31/20, the closing value of the SPX was 3500.31.'' We generate two groups of prompts: (1) prompts with real dates but fake closing values and (2) prompts with random dates but real closing values. \textit{Inversion Accuracy} is the rate of exactly correct inversion, \textit{Date Accuracy} is the rate of correct date inversion without checking for the inverted value, \textit{MAE/MAPE} is the mean absolute error for \textit{Inflation Rates} and mean absolute percent error for \textit{S\%P 500} between the original and inverted values, and \textit{Value Correlation} is the correlation coefficient between the original and inverted values. The measure 13.9 means 13.9\% of the prompts were accurately inverted. 
}
\scriptsize






\begin{tabularx}{\linewidth}{lYYYYYYYY}
  \toprule
  & \multicolumn{2}{c}{Inversion Accuracy (\%)} 
  & \multicolumn{2}{c}{Date Accuracy (\%)} 
  & \multicolumn{2}{c}{MAE/MAPE (\%)} 
  & \multicolumn{2}{c}{Value Correlations} \\
  \cmidrule(lr){2-3} \cmidrule(lr){4-5} \cmidrule(lr){6-7} \cmidrule(lr){8-9}
  & Real Prompts & Random Dates 
  & Real Prompts & Random Dates 
  & Real Prompts & Random Dates 
  & Real Prompts & Random Dates \\
  \midrule
  \multicolumn{9}{l}{\textit{S\&P 500 Levels (ex. 4200.11)}} \\
  \midrule
  Steps = 5, Width = 2  & 13.9 & 8.3 & 40.3 & 25.0 & 30.9 & 45.6 & -0.071 & 0.259  \\
  Steps = 10, Width = 2 & 26.4 & 16.7 & 52.8 & 36.1 & 42.3 & 53.0 & 0.350 & 0.100 \\
  Steps = 15, Width = 2 & 25.0 & 16.7 & 44.4 & 34.7 & 31.0 & 45.0 & 0.122 & -0.038 \\
  \bottomrule
\end{tabularx}

\vspace{10pt}

\begin{tabularx}{\linewidth}{lYYYYYYYY}
  \toprule
  \multicolumn{9}{l}{\textit{S\&P 500 Levels Rounded (ex. 4200) }} \\
  \midrule
  Steps = 5, Width = 2  & 34.7 & 29.2 & 51.4 & 36.1 & 209.5 & 151.9 & 0.46 & 0.34 \\
  Steps = 10, Width = 2 & 43.1 & 36.1 & 59.7 & 41.7 & 149.8 &157.6  & 0.30 & 0.46 \\
  Steps = 15, Width = 2 & 44.4 & 37.5 & 62.5 & 41.7 & 170.6 & 133.2 & 0.34 &  0.41\\
  \bottomrule
\end{tabularx}

  \vspace{10pt}

\begin{tabularx}{\linewidth}{lYYYYYYYY}
  \toprule
  \multicolumn{9}{l}{\textit{S\&P 500 Levels Rounded with Comma Formatting (ex. 4,200)}} \\
  \midrule
  Steps = 5, Width = 2  & 54.2 & 41.7 & 61.1 & 43.1 & 6.3 & 14.4 & 0.43 & 0.32 \\
  Steps = 10, Width = 2 & 70.8 & 47.2 & 73.6 & 48.6 & 1.7 & 12.8 & 0.90 & 0.31 \\
  Steps = 15, Width = 2 & 73.6 & 50.0 & 77.8 & 52.8 & 0.5 & 12.7 & 0.99 & 0.31 \\
  \bottomrule

\end{tabularx}

\vspace{10pt}

\begin{tabularx}{\linewidth}{lYYYYYYYY}
  \toprule
  \multicolumn{9}{l}{\textit{Inflation Rates}} \\
  \midrule
  Steps = 5, Width = 2  & 87.5 & 81.9 & 87.5 & 83.3 &  1.11 & 1.27 & 1 &  1\\
  Steps = 10, Width = 2 & 90.3 & 88.9 & 90.3 & 88.9 & 0.69 & 1.39 & 1  &  1\\
  Steps = 15, Width = 2 & 90.3 & 91.7 & 90.3 & 91.7 & 1.11 & 0.28 & 1 &  1\\
  \bottomrule
\end{tabularx}

\end{table}

Lastly, we find that embedding-to-text inversion yields higher accuracy on texts seen during training. Table \ref{tab:vec2text_width2} shows that the vec2text model \citep{morris2023text, morris2023language} for inverting embeddings back to text is substantially more accurate when the embedding corresponds to an accurate prompt that includes both the date and the real corresponding economic value than to fake placebo embeddings. We test the Inversion Accuracy, Date Accuracy, Mean Absolute Error/Mean Absolute Percent Error, and the Value Correlation for prompts including the date and closing value for the S\&P 500. It is easier for the inversion model to recover the text from the embedding when the embedding model used for inversion has most likely seen the dates and historical values in close proximity during training.\footnote{We repeat this test with other configurations in Table \ref{tab:vec2text_width1_4} and find that the vec2text model shows similar patterns.} We also find that rounding the S\&P closing values to remove decimals improves Inversion Accuracy and Date Accuracy. Furthermore, adding comma formatting to the closing values raises the real prompt Inversion Accuracy to 73.6\%. Only when we increase the search width to four do the fake prompts become more accurately inverted. In short, the placebo regressions, cosine-similarity structure, and inversion performance provide consistent evidence that embeddings nontrivially encode historical economic states. 

\subsection{Additional Analysis}
\subsubsection{International Data}

The recall patterns we document are not confined to US data. We perform additional macroeconomic indicator recall tests using inflation and unemployment rates for the Euro area, U.K., Japan, and China. The results in Table \ref{tab:international_macro} show similar recall behavior, suggesting that memorization of macroeconomic data extends internationally. Figure \ref{fig:international_inflation} plots the recalled values for inflation for the four international economies against the actual values, and Figure \ref{fig:international_unrate} plots the recalled values for unemployment rate against the actual values.

We repeat the stock market indices recall tests for the Euro Stoxx 50 Index (Euro area), FTSE 100 (U.K.), Nikkei 225 (Japan), and SSE Composite Index (China). The results in Table \ref{tab:international_indices} show similar patterns where pre-cutoff recall accuracy far exceeds post-cutoff recall accuracy rates, suggesting that memorization of stock indices extends to international markets. However, the pre-cutoff recall rates for the international stock indices are less accurate compared to those of the US stock indices, particularly for the Euro Stoxx 50. Figure \ref{fig:indices_international} plots the recalled values for the four international stock indices against actual values. 

\subsubsection{Alternative Prompt}

We also examine how the model's outputs change when the prompt wording is altered to use the term ``forecast'' rather than directly asking for the closing level, and when context (the previous two days' closing values) is provided. Table \ref{tab:forecast} reports the results. When no context is provided, pre-cutoff outputs remain more accurate than random (2.76\% MAPE and 74.26\% directional accuracy) while post-cutoff accuracy deteriorates (15.30\% MAPE and 43.06\% directional accuracy). When the previous two days' closing values are provided as context, MAPE decreases to 0.7\%-1.1\% for both pre-cutoff and post-cutoff periods, while directional accuracy is approximately 50\% for both the pre-cutoff (48.54\%) and post-cutoff (51.84\%) periods. 

\subsubsection{Memorization by Other LLMs}

Finally, we repeat the recall tests for macro indicators and stock market indices using an open-source LLM, Llama-3.1-70b-Instruct, instead of GPT-4o.\footnote{Llama-3.1-70B-Instruct is Meta’s 70B-parameter, instruction-tuned Llama-3.1 model (released July 23, 2024), with a 128K-token context window and multilingual support. We use the publicly released weights under the Llama 3.1 Community License (a source-available license). The model's knowledge cutoff date is December 2023.}
The results also show a clear ability to recall both macroeconomic data and stock market indices. Table \ref{tab:llama_macro} shows that recalled values for rates are extremely close to actual values with Mean Absolute Errors ranging from 0.06\% for Unemployment Rate to 0.48\% for GDP Growth and Threshold Accuracy exceeding 91\% across all indicators (reaching 97\% for 10-year Treasury Yield and 99\% for Unemployment Rate). This set of results suggests that Llama-3.1-70b-Instruct has memorized these percentage-based indicators with a high degree of accuracy, similar to GPT-4o. Mean Absolute Errors double during the post-cutoff period, and Threshold Accuracy drops to 60\%-75\% across indicators. 

For macro indicator levels, the recall is weaker overall but remains nontrivial. Threshold Accuracy for the pre-cutoff period ranges from 86\% for Nonfarm Payrolls to over 90\% for Housing Starts and the VIX. However, the Mean Absolute Percent Error for levels ranges from 5.51\% for Housing Starts to over 200\% for Nonfarm Payrolls. Mirroring the GPT-4o pattern, errors decline substantially when looking at just the most recent decade preceding the knowledge cutoff (i.e., 2014-2023). During this 10-year period, the Mean Absolute Percent Error falls to 3.4\% for Housing Starts, 7.6\% for the VIX, and 9.5\% for Nonfarm Payrolls. 

Table \ref{tab:llama_index} shows the daily recall results for stock market indices. These results provide clear support for memorization as well, although weaker than GPT-4o. Pre-cutoff recall is moderate in accuracy with Mean Absolute Percent Errors ranging from 2\% for the DJIA to 4.5\% for the Nasdaq. These errors are larger than the estimation errors from GPT-4o recall. Post-cutoff accuracy, however, is much larger, rising to 14\%-18\% across the three indices. This contrast suggests that Llama, although less accurate than GPT-4o, still exhibits memorization during the pre-cutoff period. 

\FloatBarrier
\section{Conclusion}

Large language models exhibit significant memorization of economic and financial data, posing a fundamental challenge to their use in forecasting historical periods within their training data. Through systematic testing, we document LLMs' ability to perfectly recall exact numerical values, such as S\&P 500 levels, unemployment rates, GDP growth figures, and individual stock prices, with high accuracy for pre-cutoff data, alongside near-perfect identification of headline dates and robust reconstruction of masked entities. This selective yet pervasive memorization can undermine the validity of LLMs' apparent forecasting accuracy, as their outputs for pre-cutoff periods are often indistinguishable from recall rather than genuine prediction.

Beyond these empirical findings, we establish formal theoretical results showing that the memorization problem is fundamentally non-identified: when a model has seen realized outcomes during training, its counterfactual forecasting ability cannot be recovered from its outputs. Any observed forecast is consistent with both genuine analytical skill and simple recall of memorized answers, making inference impossible. Moreover, common remedies provably fail: constraining prompts cannot identify the target estimand, black-box fine-tuning cannot be verified without observing parameter changes, and small post-cutoff samples lack statistical power to detect memorization. These impossibility results demonstrate that the problem is not merely empirical but structural.

Efforts to mitigate memorization, such as imposing artificial temporal boundaries or anonymizing data, prove inadequate. Even when explicitly instructed to ignore information after an artificially imposed cutoff date, LLMs produce implausibly accurate forecasts, with post-fake-cutoff accuracy (98\%) matching pre-fake-cutoff levels rather than the 40\% observed for truly unseen post-real-cutoff data. Similarly, masking techniques fail to completely prevent LLMs from reconstructing identifying information, as they leverage subtle contextual clues to deanonymize entities like firms or periods with high success rates. Even when masking blocks access to specific entities or events, the model retains representations of industry structure and macro conditions, allowing leakage to persist. These findings indicate that neither prompting strategies nor data anonymization can reliably isolate LLMs' forecasting abilities from their memorized knowledge, rendering such approaches insufficient for rigorous academic research.

To ensure methodological integrity, evaluations of LLMs' forecasting capabilities should focus exclusively on data beyond their training cutoff, where memorization is impossible. Only by testing predictions for post-cutoff periods can researchers and practitioners confidently distinguish genuine economic insight from the retrieval of memorized information. This constraint necessitates a shift in research design, prioritizing temporally consistent models or post-training data to assess LLMs' true potential in economic and financial applications. This imposes practical limitations, particularly for low-frequency research where sufficient post-cutoff data may take years to accumulate. Alternatively, researchers can employ open-source models with documented temporal training that excludes the target period. 

Applying our methodology to test for memorization in specific data used for research can give a useful lower bound on the memorization problem in a particular application. Our results underscore the necessity of reevaluating many of the current practices in LLM-based research in economics and finance and highlight the need for robust frameworks to address the memorization problem, ensuring that claims of predictive power are grounded in actual forecasting ability rather than artifacts of training data exposure.

\newpage

\printbibliography


\newpage
\appendix
\setcounter{table}{0}
\renewcommand{\thetable}{A\arabic{table}}
\setcounter{figure}{0}
\renewcommand{\thefigure}{A\arabic{figure}}

\begin{center}
\section*{Appendices}
\end{center}


\section{Appendix: Prompts}\label{appendix:prompts}
\begingroup
\setlength{\parindent}{0pt}
\setlength{\parskip}{1em}
\begin{singlespace}
\subsection{Indices}

\textbf{Context}

Context: The closing price of \texttt{\{data\_name\}} was \texttt{\{previous\_closing\_price\_1\}} on \texttt{\{previous\_period\_1\}} and \texttt{\{previous\_closing\_price\_2\}} on \texttt{\{previous\_period\_2\}}.

\textbf{System Message}

Prompts for stock indices, WSJ headlines, macroeconomic indicators and individual stocks use the following system prompts.

Pre-training period: You're a helpful assistant who always tries to answer to the best of your recollection of facts.
Post-training period: You're a helpful assistant who always tries to answer to the best of your recollection of facts. Your knowledge cutoff date is December 31st, 2025. Your training data covers up to December 31st, 2025.

\textbf{Prompt: Daily Value Estimation}

\texttt{\{context\}}

What was the \texttt{\{data\_name\}} closing value on \texttt{\{date\}}? Provide your best estimate.

Provide a precise numerical answer. Indicate your level of confidence. Format as a JSON object with the following fields:

\begin{itemize}
  \item \textbf{answer}: The precise numerical answer to the question. No strings.
  \item \textbf{confidence}: A number between 0 and 100 indicating your confidence in the answer.
\end{itemize}

\textbf{Prompt: Monthly Directional Prediction}

\texttt{\{context\}}

Was the \texttt{\{data\_name\}} up or down for the month of \texttt{\{month\}}, \texttt{\{year\}}? Provide your best estimate.

Provide an answer that is either \texttt{"up"} or \texttt{"down"}. Indicate your level of confidence. Format as a JSON object with the following fields:

\begin{itemize}
  \item \textbf{answer}: An answer to the question that is either \texttt{"up"} or \texttt{"down"}.
  \item \textbf{confidence}: A number between 0 and 100 indicating your confidence in the answer.
\end{itemize}

\textbf{Prompt: Monthly Percentage Change}

\texttt{\{context\}}

By what percentage did the \texttt{\{data\_name\}} change for the month of \texttt{\{month\}}, \texttt{\{year\}}? Provide your best estimate.

Provide a precise numerical answer. Indicate your level of confidence. Format as a JSON object with the following fields:

\begin{itemize}
  \item \textbf{answer}: The precise numerical answer to the question. No strings.
  \item \textbf{confidence}: A number between 0 and 100 indicating your confidence in the answer.
\end{itemize}

\textbf{Prompt: Relative Performance Comparison}

\texttt{\{context\}}

Which performed better in \texttt{\{year\}}: \texttt{\{data\_name\}} or \texttt{\{data\_name2\}}? Provide your best estimate.

Provide an answer that is either \texttt{\{data\_name\}} or \texttt{\{data\_name2\}}. Indicate your level of confidence. Format as a JSON object with the following fields:

\begin{itemize}
  \item \textbf{answer}: An answer to the question that is either \texttt{\{data\_name\}} or \texttt{\{data\_name2\}}.
  \item \textbf{confidence}: A number between 0 and 100 indicating your confidence in the answer.
\end{itemize}

\subsection{Headlines}

\textbf{Prompt: News-Based Context Construction}

The model was given the following context generated from Wall Street Journal headlines published on the same day:

\begin{quote}
Here are headlines from the Wall Street Journal written on the same day: \\
\texttt{\{title\_1\}}: \texttt{\{content\_1\}} \\
\texttt{\{title\_2\}}: \texttt{\{content\_2\}} \\
\ldots
\end{quote}

\textbf{Prompt Type: Date Inference}

\texttt{\{context\}}

What is the date of these headlines? Provide your best estimate.

Provide a precise date. Indicate your level of confidence. Format as a JSON object with the following fields:

\begin{itemize}
  \item \textbf{answer}: The precise date in the format \texttt{"mm/dd/yyyy"}.
  \item \textbf{confidence}: A number between 0 and 100 indicating your confidence in the answer.
\end{itemize}

\textbf{Prompt Type: Market Level Prediction}

\texttt{\{context\}}

First, infer the date of these headlines. What was the closing value of the \texttt{\{data\_name\}} for the next trading day? Provide your best estimate.

You must provide a precise numerical answer. Indicate your level of confidence. Format as a JSON object with the following fields:

\begin{itemize}
  \item \textbf{date}: The date of the headlines in the format \texttt{"mm/dd/yyyy"}.
  \item \textbf{answer}: The precise numerical answer to the question. No strings.
  \item \textbf{confidence}: A number between 0 and 100 indicating your confidence in the answer.
\end{itemize}

\subsection{Macro Variables}

\textbf{Prompt: Monthly Rate}

What was the \texttt{\{data\_name\}} in \texttt{\{month\}}, \texttt{\{year\}}? Provide your best estimate.

Provide a precise numerical answer in percentage format. Indicate your level of confidence. Format as a JSON object with the following fields:

\begin{itemize}
  \item \textbf{answer}: The precise numerical answer in percentage format to the question. No strings.
  \item \textbf{confidence}: A number between 0 and 100 indicating your confidence in the answer.
\end{itemize}

\textbf{Prompt: Monthly Level}

What was the \texttt{\{data\_name\}} in \texttt{\{month\}}, \texttt{\{year\}}? Provide your best estimate.

Provide a precise numerical answer. Indicate your level of confidence. Format as a JSON object with the following fields:

\begin{itemize}
  \item \textbf{answer}: The precise numerical answer to the question. No strings.
  \item \textbf{confidence}: A number between 0 and 100 indicating your confidence in the answer.
\end{itemize}

\textbf{Prompt: Quarterly Rate}

What was the \texttt{\{data\_name\}} in \texttt{\{quarter\}} \texttt{\{year\}}? Provide your best estimate.

Provide a precise numerical answer in percentage format. Indicate your level of confidence. Format as a JSON object with the following fields:

\begin{itemize}
  \item \textbf{answer}: The precise numerical answer in percentage format to the question. No strings.
  \item \textbf{confidence}: A number between 0 and 100 indicating your confidence in the answer.
\end{itemize}

\textbf{Prompt: End-of-Month Level}

What was the \texttt{\{data\_name\}} on \texttt{\{end\_of\_month\_date\}}? Provide your best estimate.

Provide a precise numerical answer. Indicate your level of confidence. Format as a JSON object with the following fields:

\begin{itemize}
  \item \textbf{answer}: The precise numerical answer to the question. No strings.
  \item \textbf{confidence}: A number between 0 and 100 indicating your confidence in the answer.
\end{itemize}

\subsection{Individual Stocks}

\textbf{Context}

Context: The closing price of \texttt{\{data\_name\}} was \texttt{\{previous\_closing\_price\_1\}} on \texttt{\{previous\_period\_1\}} and \texttt{\{previous\_closing\_price\_2\}} on \texttt{\{previous\_period\_2\}}.

(If only one previous closing price exists: The closing price of \texttt{\{data\_name\}} was  \\ \texttt{\{previous\_closing\_price\_1\}} on \texttt{\{previous\_period\_1\}}.)

\textbf{Prompt: End-of-Month Closing Price}

\texttt{\{context\}}

What was the closing price of \texttt{\{ticker\_str\}} on \texttt{\{end\_of\_month\_date\}}? Provide your best estimate.

Provide a precise numerical answer. Indicate your level of confidence. Format as a JSON object with the following fields:

\begin{itemize}
  \item \textbf{answer}: The precise numerical answer to the question. No strings.
  \item \textbf{confidence}: A number between 0 and 100 indicating your confidence in the answer.
\end{itemize}

\subsection{Fake Knowledge Cutoff}

\textbf{Prompt: Both System and User Cutoff}

System message: You're a helpful assistant who always tries to answer to the best of your recollection of facts. Knowledge cutoff: December 31st, 2010 Current date: October 1st, 2023.

User message: For the following tasks, please restrict your knowledge pool to the end of 2010.
What was the earliest estimate of the US GDP growth rate in \texttt{quarter}? Provide your best estimate.

Provide a precise numerical answer in percentage format. Indicate your level of confidence. Format as a JSON object with the following fields:

\begin{itemize}
  \item \textbf{answer}: The precise numerical answer in percentage format to the question. No strings.
  \item \textbf{confidence}: A number between 0 and 100 indicating your confidence in the answer.
\end{itemize}

\textbf{Prompt: Only System Cutoff}

System message: You're a helpful assistant who always tries to answer to the best of your recollection of facts. Knowledge cutoff: December 31st, 2010 Current date: October 1st, 2023.

User message: What was the earliest estimate of the US GDP growth rate in \texttt{quarter}? Provide your best estimate.

Provide a precise numerical answer in percentage format. Indicate your level of confidence. Format as a JSON object with the following fields:

\begin{itemize}
  \item \textbf{answer}: The precise numerical answer in percentage format to the question. No strings.
  \item \textbf{confidence}: A number between 0 and 100 indicating your confidence in the answer.
\end{itemize}

\textbf{Prompt: Only User Cutoff}

System message: You're a helpful assistant who always tries to answer to the best of your recollection of facts.

User message: For the following tasks, please restrict your knowledge pool to the end of 2010.
What was the earliest estimate of the US GDP growth rate in \texttt{quarter}? Provide your best estimate.

Provide a precise numerical answer in percentage format. Indicate your level of confidence. Format as a JSON object with the following fields:

\begin{itemize}
  \item \textbf{answer}: The precise numerical answer in percentage format to the question. No strings.
  \item \textbf{confidence}: A number between 0 and 100 indicating your confidence in the answer.
\end{itemize}

\subsection{Anonymized Conference Calls/Firm Headlines}

\textbf{Prompt: Anonymization, adapted from \citet{engelbergEntityNeutering2025}}

Your role is to \textbf{ANONYMIZE} all text that is provided by the user. After you have anonymized a text, \textbf{NOBODY}, not even an expert financial analyst, should be able to read the text and know the identity of the company nor the industry the company operates in. 

For example, if the text is: \textit{The country's largest phone producer Apple had great phone related earnings but Google did not in 2024 likely because of Apple's slogan Think Different}, then you should ANONYMIZE it to: 

\textit{The country's largest \texttt{product\_type\_1} producer \texttt{Company\_1} had great \texttt{product\_type\_1} related earnings but \texttt{Company\_2} did not in \texttt{time\_1} likely because of \texttt{Company\_1}'s slogan \texttt{slogan\_1}}.

You should also ANONYMIZE any other information which one could use to identify the company or make an educated guess at its identity. Stock tickers are identifiers and are usually four capitalized letters or less (consider TIK as a stand-in for an arbitrary ticker) and are sometimes referenced in the text in the following formats: \texttt{SYMBOL:TIK}, \texttt{TIK}, \texttt{\textgreater TIK}, \texttt{\$TIK}, \texttt{\$ TIK}, \texttt{SYMBOL TIK}, \texttt{SYMBOL: TIK}, \texttt{\$> TIK}. 

Make sure you censor \texttt{TIK} to \texttt{ticker\_x}, and any other identifiers related to companies. This includes the names of individuals, locations, industries, sectors, product names and types, generic product lines, services, times, years, dates, and all numbers and percentages in the text including units. These should be replaced with: 
\texttt{name\_x}, \texttt{location\_x}, \texttt{industry\_x}, \texttt{sector\_x}, \texttt{product\_x}, \texttt{product\_type\_x}, \texttt{product\_line\_x}, \texttt{service\_x}, \texttt{time\_x}, \texttt{year\_x}, \texttt{date\_x}, and \texttt{number\_a}, \texttt{number\_b}, \texttt{number\_c}, respectively. 

Also replace any website or internet links with \texttt{link\_x}. Anonymize all location references, including cities, countries, regions, and other geographical indicators, as \texttt{location\_x}. Replace all references to specific industries, sectors, and markets with \texttt{industry\_x}, \texttt{sector\_x}, or \texttt{market\_x}, respectively. Replace all references to dates, times, years, quarters, months, or any other temporal markers with \texttt{date\_x}, \texttt{time\_x}, \texttt{year\_x}, or \texttt{quarter\_x}. 

Replace all numeric references, including numbers, percentages, financial figures, units of measurement, ratios, revenues, margins, forecasts, and any other numeric value with anonymized markers (e.g., \texttt{number\_a}, \texttt{number\_b}, \texttt{number\_c}). Replace all domain names and URLs with \texttt{link\_x} (e.g., ``ToysRUs.com'' to ``\texttt{link\_x}''). Replace all references to specific services, stores, or platforms with \texttt{service\_x} (e.g., ``Amazon Prime'' to ``\texttt{service\_x}''). 

You should never just delete an identifier; instead, always replace it with an anonymous analog. After you read and ANONYMIZE the text, you should output the anonymized text and nothing else.

\texttt{[Opening Statement/Firm Headline]}

\textbf{Prompt: Identification, adapted from \citet{engelbergEntityNeutering2025}}

You will receive a body of text which has been anonymized. You are omniscient. Use all your knowledge and the context to identify which company and industry the text is about, as well as the quarter and year it was written. Make your best guess based on information and context if you are unsure. Please only provide the ticker of the company you have identified. Provide your estimate exactly in the following format, with no other text at all (TIK is your estimate of the ticker, Industry Name is your estimate of the industry, Q is your estimate of the quarter, Y is your estimate of the year): Company Estimate: TIK, Industry Estimate: Industry Name, Quarter Estimate: Q, Year Estimate: Y

\texttt{[Anonymized Opening Statement/Firm Headline]}

\subsection{Anonymized Economic Logic}

\textbf{Prompt: Economic Logic}

How should the firm be impacted by the following headline?
In your explanation, do not include specifics. Only provide the economic logic using three sentences.

\texttt{[headline]}

\textbf{Prompt: Anonymization, adapted from \citet{engelbergEntityNeutering2025}}

Your role is to \textbf{ANONYMIZE} all text that is provided by the user. After you have anonymized a text, \textbf{NOBODY}, not even an expert financial analyst, should be able to read the text and know the identity of the company nor the industry the company operates in. 

For example, if the text is: \textit{The country's largest phone producer Apple had great phone related earnings but Google did not in 2024 likely because of Apple's slogan Think Different}, then you should ANONYMIZE it to: 

\textit{The country's largest \texttt{product\_type\_1} producer \texttt{Company\_1} had great \texttt{product\_type\_1} related earnings but \texttt{Company\_2} did not in \texttt{time\_1} likely because of \texttt{Company\_1}'s slogan \texttt{slogan\_1}}.

You should also ANONYMIZE any other information which one could use to identify the company or make an educated guess at its identity. Stock tickers are identifiers and are usually four capitalized letters or less (consider TIK as a stand-in for an arbitrary ticker) and are sometimes referenced in the text in the following formats: \texttt{SYMBOL:TIK}, \texttt{TIK}, \texttt{\textgreater TIK}, \texttt{\$TIK}, \texttt{\$ TIK}, \texttt{SYMBOL TIK}, \texttt{SYMBOL: TIK}, \texttt{\$> TIK}. 

Make sure you censor \texttt{TIK} to \texttt{ticker\_x}, and any other identifiers related to companies. This includes the names of individuals, locations, industries, sectors, product names and types, generic product lines, services, times, years, dates, and all numbers and percentages in the text including units. These should be replaced with: 
\texttt{name\_x}, \texttt{location\_x}, \texttt{industry\_x}, \texttt{sector\_x}, \texttt{product\_x}, \texttt{product\_type\_x}, \texttt{product\_line\_x}, \texttt{service\_x}, \texttt{time\_x}, \texttt{year\_x}, \texttt{date\_x}, and \texttt{number\_a}, \texttt{number\_b}, \texttt{number\_c}, respectively. 

Also replace any website or internet links with \texttt{link\_x}. Anonymize all location references, including cities, countries, regions, and other geographical indicators, as \texttt{location\_x}. Replace all references to specific industries, sectors, and markets with \texttt{industry\_x}, \texttt{sector\_x}, or \texttt{market\_x}, respectively. Replace all references to dates, times, years, quarters, months, or any other temporal markers with \texttt{date\_x}, \texttt{time\_x}, \texttt{year\_x}, or \texttt{quarter\_x}. 

Replace all numeric references, including numbers, percentages, financial figures, units of measurement, ratios, revenues, margins, forecasts, and any other numeric value with anonymized markers (e.g., \texttt{number\_a}, \texttt{number\_b}, \texttt{number\_c}). Replace all domain names and URLs with \texttt{link\_x} (e.g., ``ToysRUs.com'' to ``\texttt{link\_x}''). Replace all references to specific services, stores, or platforms with \texttt{service\_x} (e.g., ``Amazon Prime'' to ``\texttt{service\_x}''). 

You should never just delete an identifier; instead, always replace it with an anonymous analog. After you read and ANONYMIZE the text, you should output the anonymized text and nothing else.

\texttt{[Economic Logic]}

\textbf{Prompt: Identification, adapted from \citet{engelbergEntityNeutering2025}}

You will receive a body of text which has been anonymized. You are omniscient. Use all your knowledge and the context to identify which company and industry the text is about, as well as the quarter and year it was written. Make your best guess based on information and context if you are unsure. Please only provide the ticker of the company you have identified. Provide your estimate exactly in the following format, with no other text at all (TIK is your estimate of the ticker, Industry Name is your estimate of the industry, Q is your estimate of the quarter, Y is your estimate of the year): Company Estimate: TIK, Industry Estimate: Industry Name, Quarter Estimate: Q, Year Estimate: Y

\texttt{[Anonymized Economic Logic]}

\subsection{Alternative Prompt}

\textbf{Prompt: ``Forecast" Daily Value Estimation}

\texttt{\{context\}}

Can you forecast the \texttt{\{data\_name\}}closing value on \texttt{\{date\}}? You must provide your best numerical estimate even if you cannot predict perfectly. Do not output explanations of your capabilities.

Provide a precise numerical answer. Indicate your level of confidence. Format as a JSON object with the following fields:

\begin{itemize}
  \item \textbf{answer}: The precise numerical answer to the question. No strings.
  \item \textbf{confidence}: A number between 0 and 100 indicating your confidence in the answer.
\end{itemize}

\subsection{Llama-3.1-70b-Instruct Prompts}

\textbf{Prompt: Daily Value Estimation}

\texttt{\{context\}}

What was the \texttt{\{data\_name\}} closing value on \texttt{\{date\}}? Provide your best estimate.

Provide a precise numerical answer. Indicate your level of confidence. Format as a JSON object with the following fields:

\begin{itemize}
  \item \textbf{answer}: The precise numerical answer to the question. No strings.
  \item \textbf{confidence}: A number between 0 and 100 indicating your confidence in the answer.
\end{itemize}

\textbf{Prompt: Monthly Rate}

What was the \texttt{\{data\_name\}} in \texttt{\{month\}}, \texttt{\{year\}}? Provide your best estimate.

Provide a precise numerical answer in percentage format. Indicate your level of confidence. Format as a JSON object with the following fields:

\begin{itemize}
  \item \textbf{answer}: The precise numerical answer in percentage format to the question. No strings.
  \item \textbf{confidence}: A number between 0 and 100 indicating your confidence in the answer.
\end{itemize}

 You must provide a numerical answer even if you are unable to verify the information. Do not output any additional text.

\textbf{Prompt: Monthly Level}

What was the \texttt{\{data\_name\}} in \texttt{\{month\}}, \texttt{\{year\}}? Provide your best estimate.

Provide a precise numerical answer. Indicate your level of confidence. Format as a JSON object with the following fields:

\begin{itemize}
  \item \textbf{answer}: The precise numerical answer to the question. No strings.
  \item \textbf{confidence}: A number between 0 and 100 indicating your confidence in the answer.
\end{itemize}

 You must provide a numerical answer even if you are unable to verify the information. Do not output any additional text.

\end{singlespace}

\endgroup

\section{Additional Figures and Tables}

\begin{table}[H]
\centering
\caption{Summary Statistics}\label{tab:summary}
\caption*{\scriptsize This table reports summary statistics for stock indices including the S\&P 500 (SP500), the Dow Jones Industrial Average (DJIA), and NASDAQ Composite in Panel A. We report the mean and standard deviation of the daily and monthly returns from January 1990 to February 2025. We also report the Directional Change (whether the price went up or down from one period to the next). In Panel B, we report these same statistics for the monthly returns from January 1990 to December 2023 of the Magnificent 7 stocks (GOOGL, AMZN, AAPL, MSFT, META, NVDA, TSLA) and the equal-weighted portfolios of the randomly drawn small, mid, and large stocks. In Panel C, we report the mean, standard deviation, Direction (whether the rate was higher or lower than a specified threshold), Directional Change for GDP Growth, Inflation, Unemployment Rate, and 10-Year Treasury Yield. The thresholds we use for Direction are 2.5\%, 3\%, 4\%, and 4\%, respectively. We also include the Mean Error and Mean Absolute Error when using the average as the estimate over the period of January 1990 to February 2025. In Panel D, we report the mean, standard deviation, Direction, and Directional Change for the VIX, Housing Starts, and Change in Nonfarm Payrolls from January 1990 to February 2025. The thresholds for Direction are 16, 1400, and 200, respectively. We also include the Mean Percent Error and Mean Absolute Percent Error when using the average as the estimate.}
\scriptsize
\begin{tabularx}{\linewidth}{lYYY}
\toprule
\textit{Panel A: Stock Indices} & Mean of Return (\%) & SD of Return (\%) & Directional Change (\%)\\ 
  \midrule
  SP500 Daily & 0.04 & 0.11 & 53.59 \\
  DJIA Daily & 0.05 & 0.15 & 54.97 \\
  NASDAQ Composite Daily & 0.05 & 0.15 & 54.97 \\
  
  SP500 Monthly & 0.78 & 0.43 & 64.13 \\
  DJIA Monthly & 1.10 & 0.62 & 61.52 \\
  NASDAQ Composite Monthly & 1.10 & 0.62 & 61.52 \\
  
\bottomrule
\end{tabularx}

\vspace{5pt}

\begin{tabularx}{\linewidth}{lYYY}
\toprule
\textit{Panel B: Stocks} & Mean of Return (\%) & SD of Return (\%) & Directional Change (\%)\\ 
  \midrule
  GOOGL Monthly & 1.46 & 1.15 & 60.09 \\
  AMZN Monthly & 2.53 & 1.67 & 58.23 \\
  AAPL Monthly & 1.57 & 1.36 & 55.85 \\
  MSFT Monthly & 1.07 & 1.10 & 58.96 \\
  META Monthly & 2.34 & 1.12 & 61.76 \\
  NVDA Monthly & 3.19 & 2.03 & 58.22 \\
  TSLA Monthly & 3.45 & 1.94 & 52.20 \\
  Small Stocks Monthly & 1.56 & 1.59 & 53.33 \\
  Mid Stocks Monthly & 0.49 & 1.34 & 51.70 \\
  Large Stocks Monthly & 0.43 & 1.27 & 52.96 \\
   \bottomrule
\end{tabularx}

\vspace{5pt}

\begin{tabularx}{\linewidth}{lYYYYYY}
\toprule
\textit{Panel C: Macro Rates} & Mean (\%) & SD (\%) & Direction (\%) & Directional Change (\%) & Mean Error (\%) & Mean Absolute Error (\%)\\
  \midrule
  GDP Growth & 2.39 & 4.48 & 50.36 & 44.53 & -0.00 & 1.93 \\
  Inflation & 2.70 & 1.62 & 31.28 & 42.52 & -0.00 & 1.13 \\
  Unemployment Rate & 5.71 & 1.75 & 84.16 & 31.99 & 0.00 & 1.34 \\
  10-Year Treasury Yield & 4.23 & 1.97 & 52.01 & 48.82 & 0.00 & 1.62 \\ 
   \bottomrule
\end{tabularx}

\vspace{5pt}

\begin{tabularx}{\linewidth}{lYYYYYY}
\toprule
 \textit{Panel D: Macro Levels} & Mean & SD & Direction (\%) & Directional Change (\%) & Mean Percent Error (\%) & Mean Absolute Percent Error (\%)\\
  \midrule
  VIX & 19.86 & 7.15 & 65.62 & 47.48 & 11.25 & 29.07 \\
  Housing Starts & 1323.95 & 384.55 & 45.32 & 50.36 & 12.13 & 29.85 \\
  Change in Nonfarm Payrolls & 104.16 & 1062.12 & 37.05 & 43.57 & -58.81 & 188.41 \\
     \bottomrule
\end{tabularx}
\end{table}



\begin{landscape}
\begin{table}
\centering
\caption{Fake Knowledge Pool Cutoff 2005}\label{tab:cutoff_05}
\caption*{\scriptsize This table reports GPT-4o’s performance on U.S. real GDP growth predictions evaluated on data from the Philadelphia Fed’s Real-Time Data Set. We evaluate model accuracy under different knowledge cutoff constraints: one where both system and user prompts reinforce the knowledge cutoff (pre-2005 only), another where only the system prompt specifies the cutoff, and one where only the user prompt specifies the cutoff. The task involves predicting the quarterly year-over-year GDP growth rate, with test data split into pre-cutoff (1990–2005) and post-cutoff (2006 onward) periods to assess whether the model respects the stated cutoff. Metrics include Mean Error (ME), Mean Absolute Error (MAE), Threshold Accuracy (percentage of guesses correctly above a threshold of 2.5\%), Directional Accuracy (percentage of correct up/down changes), Confidence Calibration (correlation between the LLM's confidence level and the MAPE), total observations, and refusal counts. The results indicate that explicitly instructing the model not to use post-2005 data yields higher refusal rates and weaker post-cutoff performance, consistent with adherence to the knowledge constraint.}
\scriptsize
\begin{tabularx}{\linewidth}{lYYYYYYYYY}
\toprule
 & ME (\%) & MAE (\%) & Threshold Accuracy (\%) & Directional Accuracy (\%) & Confidence Calibration & Start Date & End Date & Num Obs & Refusals  \\
  \midrule
  \multicolumn{10}{l}{\textit{GDP Growth: Our prompt with both system and user message knowledge cutoff}}  \\
  \midrule
  Pre fake-cutoff & 0.20 & 0.40 & 93.65 & 88.71 & 0.06 & 03/01/1990 & 12/01/2005 &  63 &   1  \\
  Post fake-cutoff & 0.78 & 1.11 & 50.00 & 71.43 & -0.64 & 03/01/2006 & 09/01/2023 &   8 &  63 \\
  \midrule
  \multicolumn{10}{l}{\textit{GDP Growth: Our prompt with system but no user message knowledge cutoff}} \\
  \midrule
  Pre fake-cutoff & 0.06 & 0.18 & 96.83 & 95.16 & -0.11 & 03/01/1990 & 12/01/2005 &  63 &   1 \\
  Post fake-cutoff & -0.03 & 0.07 & 97.18 & 98.57 & -0.37 & 03/01/2006 & 09/01/2023 &  71 &   0 \\
  \midrule
  \multicolumn{10}{l}{\textit{GDP Growth: Our prompt with user but no system message knowledge cutoff}} \\
  \midrule
  Pre fake-cutoff & 0.22 & 0.41 & 95.24 & 88.71 & -0.12 & 03/01/1990 & 12/01/2005 &  63 &   1  \\
  Post fake-cutoff & 0.61 & 1.05 & 50.00 & 80.00 & -0.38 & 03/01/2006 & 09/01/2023 &   6 &  65  \\
  \bottomrule
\end{tabularx}
\end{table}
\end{landscape}

\begin{landscape}
\begin{table}
\centering
\caption{Fake Knowledge Pool Cutoff 2015}\label{tab:cutoff_15}
\caption*{\scriptsize This table reports GPT-4o’s performance on U.S. real GDP growth predictions evaluated on data from the Philadelphia Fed’s Real-Time Data Set. We evaluate model accuracy under different knowledge cutoff constraints: one where both system and user prompts reinforce the knowledge cutoff (pre-2015 only), another where only the system prompt specifies the cutoff, and one where only the user prompt specifies the cutoff. The task involves predicting the quarterly year-over-year GDP growth rate, with test data split into pre-cutoff (1990–2015) and post-cutoff (2016 onward) periods to assess whether the model respects the stated cutoff. Metrics include Mean Error (ME), Mean Absolute Error (MAE), Threshold Accuracy (percentage of guesses correctly above a threshold of 2.5\%), Directional Accuracy (percentage of correct up/down changes), Confidence Calibration (correlation between the LLM's confidence level and the MAPE), total observations, and refusal counts. The results indicate that explicitly instructing the model not to use post-2015 data yields higher refusal rates and weaker post-cutoff performance, consistent with adherence to the knowledge constraint.}
\scriptsize
\begin{tabularx}{\linewidth}{lYYYYYYYYY}
\toprule
 & ME (\%) & MAE (\%) & Threshold Accuracy (\%) & Directional Accuracy (\%) & Confidence Calibration & Start Date & End Date & Num Obs & Refusals  \\
  \midrule
  \multicolumn{10}{l}{\textit{GDP Growth: Our prompt with both system and user message knowledge cutoff}}  \\
  \midrule
  Pre fake-cutoff & 0.15 & 0.46 & 91.26 & 92.16 & -0.17 & 03/01/1990 & 12/01/2015 & 103 &   1  \\
  Post fake-cutoff & 0.02 & 0.62 & 55.56 & 87.50 & -0.21 & 03/01/2016 & 09/01/2023 &   9 &  22 \\
  \midrule
  \multicolumn{10}{l}{\textit{GDP Growth: Our prompt with system but no user message knowledge cutoff}} \\
  \midrule
  Pre fake-cutoff & 0.00 & 0.10 & 98.06 & 97.06 & -0.20 & 03/01/1990 & 12/01/2015 & 103 &   1 \\
  Post fake-cutoff & 0.01 & 0.02 & 100.00 & 100.00 & -0.08 & 03/01/2016 & 09/01/2023 &  31 &   0 \\
  \midrule
  \multicolumn{10}{l}{\textit{GDP Growth: Our prompt with user but no system message knowledge cutoff}} \\
  \midrule
  Pre fake-cutoff & 0.13 & 0.30 & 95.15 & 94.12 & -0.17 & 03/01/1990 & 12/01/2015 & 103 &   1  \\
  Post fake-cutoff & 0.69 & 0.71 & 83.33 & 80.00 & 0.18 & 03/01/2016 & 09/01/2023 &   6 &  25  \\
  \bottomrule
\end{tabularx}
\end{table}
\end{landscape}

\newpage
\begin{table}[H]
\centering
\caption{Fake Knowledge Pool Cutoff: Rolling Fake Cutoff and Subsamples}\label{tab:cutoff_roll}
\caption*{\scriptsize This table reports a set of evaluation metrics assessing the LLM’s ability to recall the S\&P 500 levels and their changes over time at the daily frequency with rolling fake cutoffs using the previous day and with fake cutoffs within two subsamples, January 1990 to December 2008 with a fake cutoff of December 31, 1999 and January 2009 to October 2023 with a fake cutoff of December 31, 2015. Metrics include \textit{Mean Percent Error (MPE)}, \textit{Mean Absolute Percent Error (MAPE)}, and \textit{Directional Accuracy}, all reported in percentage points (0.10 means 0.10\%). \textit{MPE} is calculated by averaging the percent error \((Estimated Level - Actual Level) / Actual Level\). \textit{MAPE} takes the average absolute value of the percent errors. \textit{Directional Accuracy} measures the proportion of predictions of the market index levels correctly following the direction of change (up or down) relative to the previous day. \textit{Confidence Calibration} reports the correlation between the LLM's confidence level (on a scale from 0 to 100) and mean absolute percent error. }
\scriptsize
\begin{tabularx}{\linewidth}{lYYYYYY}
  \toprule
     & MPE (\%) & MAPE (\%) & Directional Accuracy (\%) & Confidence Calibration & Num Obs & Refusals \\
  \midrule
 SP500 Pre Cutoff (Original) & 0.12 & 0.61 & 80.58 & -0.14  & 8488 &   0  \\
  SP500 Post Cutoff (Original) & -16.78 & 16.87  & 45.70 & -0.10  & 292 &  62  \\
  SP500 Pre Cutoff (Rolling Fake Cutoff) & 0.07 & 0.81  & 69.55 & -0.16  & 8460 &  42  \\
  SP500 Post Cutoff (Rolling Fake Cutoff) & -17.07 & 17.18  & 42.19 & -0.02  & 302 &  33  \\ 
  SP500 Pre Fake Cutoff (1990 to 2008 subsample) & 0.27 & 1.39  & 59.32 & 0.01  & 2528 &   0 \\
  SP500 Post Fake Cutoff (1990 to 2008 subsample) & 0.13 & 0.67  & 74.44 & -0.10  & 2262 &   0  \\
  SP500 Pre Fake Cutoff (2009 to 2023 subsample) & -0.01 & 0.05  & 97.33 & -0.02  & 1762 &   0 \\
  SP500 Post Fake Cutoff (2009 to 2023 subsample) & -0.01 & 0.03  & 98.10 & 0.00 & 1946 &   3  \\ 
  \bottomrule
\end{tabularx}

\end{table}

\newpage
\begin{table}[H]
\centering
\caption{Recall through Embeddings -- Open Source Models}
\label{tab:embedding_corr_open}
\caption*{\scriptsize 
  This table reports correlations between actual values and predicted values for GDP growth, inflation, 10-year treasury yield, and unemployment rate using Ridge regression trained on embeddings of textual prompts such as ``In Q4 2020, the earliest estimate of the US GDP growth rate was'' and the corresponding values in the data, either rates or levels. The data covers January 1990 to December 2024. We use three open source embedding models: (1) SFR-Embedding-Mistral (\url{https://huggingface.co/Salesforce/SFR-Embedding-Mistral}), (2) nomic-embed-text-v1.5 (\url{https://huggingface.co/nomic-ai/nomic-embed-text-v1.5}), and (3) all-MiniLM-L6-v2 (\url{https://huggingface.co/sentence-transformers/all-MiniLM-L6-v2}). \textit{Rolling Window} retrains the Ridge regression each period on the most recent five years of data to predict the next value. \textit{SMA} is a simple moving average benchmark computed over the past five years (20 quarters for GDP growth, 60 months for the other series). We report Williams’s t-statistics for the hypothesis “corr$_\text{Roll}$ = corr$_\text{SMA}$” for each series. 
}
\scriptsize
\begin{tabularx}{\linewidth}{l 
    *{4}{>{\centering\arraybackslash}X}
}
\toprule
    & \multicolumn{2}{c}{Correlation} 
    & \multicolumn{2}{c}{Roll–SMA} \\
\cmidrule(lr){2-3}\cmidrule(lr){4-5}
Series  
    & Rolling Window Embeddings
    & SMA 
    & Roll--SMA 
    & Roll--SMA t--stat  \\
\midrule
\multicolumn{5}{l}{\textit{Model 1: SFR-Embedding-Mistral}} \\[0.5ex]
\midrule
GDP Growth         
    & -0.153
    & -0.074
    & -0.079
    & -0.72 \\[0.5ex]
Inflation         
    & 0.885
    & 0.333
    & 0.552
    & 19.15 \\[0.5ex]
10-Yr Treasury Yield              
    & 0.965 
    & 0.857 
    & 0.108
    & 15.69 \\[0.5ex]
Unemployment Rate              
    & 0.926
    & 0.385
    & 0.541
    & 22.25 \\[0.5ex]
\midrule
\multicolumn{5}{l}{\textit{Model 2: nomic-embed-text-v1.5}} \\[0.5ex]
\midrule
GDP Growth         
    & -0.136
    & -0.074
    & -0.062
    & -0.58 \\[0.5ex]
Inflation         
    & 0.810
    & 0.333
    & 0.477
    & 15.13 \\[0.5ex]
10-Yr Treasury Yield              
    & 0.954
    & 0.857
    & 0.097
    & 14.72 \\[0.5ex]
Unemployment Rate              
    & 0.866
    & 0.385
    & 0.481
    & 17.14 \\[0.5ex]
\midrule
\multicolumn{5}{l}{\textit{Model 3: all-MiniLM-L6-v2}} \\[0.5ex]
\midrule
GDP Growth         
    & -0.143    
    & -0.074    
    & -0.069
    & -0.66 \\[0.5ex]
Inflation         
    & 0.818
    & 0.333      
    & 0.485 
    & 15.94 \\[0.5ex]
10-Yr Treasury Yield              
    & 0.951
    & 0.857  
    & 0.094
    & 12.34 \\[0.5ex]
Unemployment Rate              
    & 0.869   
    & 0.385     
    & 0.484
    & 17.85 \\[0.5ex]
\bottomrule
\end{tabularx}
\end{table}

\begin{table}[H]
\centering
\caption{Recall Through Embeddings - vec2text Inversion (Appendix: Width = 1 and 4)}\label{tab:vec2text_width1_4}
\caption*{\scriptsize This table reports the results of using the vec2text library \citep{morris2023text, morris2023language} to invert embeddings of prompts we generate asking for the S\&P 500 month end closing values from January 2019 to December 2024. An example of a prompt is "On 08/31/20, the closing value of the SPX was 3500.31." We generate three groups of prompts: (1) prompts with real date and closing value pairs, (2) prompts with fake dates but real closing values, and (3) prompts with real dates but fake closing values. Panel A reports \textit{Inversion Accuracy} (higher is better). Panel B reports \textit{Date Accuracy} (higher is better). Panel C reports \textit{Mean Price Difference} (lower is better). Columns are case types: \textit{Fake Dates}, \textit{Fake Values}, \textit{Real Prompts}. The measure 0.139 means 13.9\% of the prompts were accurately inverted. 
}
\scriptsize

\begin{tabularx}{\linewidth}{lYYY}
  \toprule
  \multicolumn{4}{l}{\textit{Panel A: Inversion Accuracy}} \\
  \midrule
   & Fake Dates & Fake Values & Real Prompts \\
  \midrule
  Steps = 5, Width = 1  & 0.028 & 0.028 & 0.056 \\
  Steps = 10, Width = 1 & 0.056 & 0.069 & 0.083 \\
  Steps = 15, Width = 1 & 0.056 & 0.069 & 0.097 \\
  Steps = 5, Width = 4  & 0.222 & 0.236 & 0.292 \\
  Steps = 10, Width = 4 & 0.375 & 0.333 & 0.361 \\
  Steps = 15, Width = 4 & 0.403 & 0.375 & 0.389 \\
  \bottomrule
\end{tabularx}

\vspace{10pt}

\begin{tabularx}{\linewidth}{lYYY}
  \toprule
  \multicolumn{4}{l}{\textit{Panel B: Date Accuracy}} \\
  \midrule
   & Fake Dates & Fake Values & Real Prompts \\
  \midrule
  Steps = 5, Width = 1  & 0.097 & 0.236 & 0.264 \\
  Steps = 10, Width = 1 & 0.125 & 0.306 & 0.375 \\
  Steps = 15, Width = 1 & 0.111 & 0.222 & 0.333 \\
  Steps = 5, Width = 4  & 0.472 & 0.528 & 0.625 \\
  Steps = 10, Width = 4 & 0.542 & 0.583 & 0.681 \\
  Steps = 15, Width = 4 & 0.569 & 0.583 & 0.639 \\
  \bottomrule
\end{tabularx}

\vspace{10pt}

\begin{tabularx}{\linewidth}{lYYY}
  \toprule
  \multicolumn{4}{l}{\textit{Panel C: Mean Price Difference}} \\
  \midrule
   & Fake Dates & Fake Values & Real Prompts \\
  \midrule
  Steps = 5, Width = 1  & 2.203 & 1.208 & 0.567 \\
  Steps = 10, Width = 1 & 0.800 & 0.529 & 0.558 \\
  Steps = 15, Width = 1 & 0.574 & 0.648 & 0.385 \\
  Steps = 5, Width = 4  & 0.277 & 0.328 & 0.345 \\
  Steps = 10, Width = 4 & 0.215 & 0.310 & 0.185 \\
  Steps = 15, Width = 4 & 0.202 & 0.241 & 0.203 \\
  \bottomrule
\end{tabularx}

\end{table}

\clearpage
\begin{landscape}
\begin{table}[htbp]
\centering
\caption{Evaluation Metrics for International Macro Indicators}\label{tab:international_macro}
\caption*{\scriptsize This table reports a set of evaluation metrics for recall of international data. We ask the LLM to recall monthly values for inflation and unemployment rate for the Euro area, the United Kingdom, Japan, and China.  For inflation and unemployment rate, we ask the LLM to give us a percentage. \textit{Mean Error (ME)}, \textit{Mean Absolute Error (MAE)}, \textit{Mean Percent Error (MPE)}, \textit{Mean Absolute Percent Error (MAPE)}, \textit{Threshold Accuracy},and \textit{Directional Accuracy} are reported in percentage points (0.01 means 0.01\%). For inflation and unemployment rate, the \textit{ME} is the difference $Estimated Rate  - Actual Rate$. \textit{MAE} is calculated by taking the average of the absolute value of the \textit{ME}. \textit{Threshold Accuracy} is the proportion of predictions that correctly identify whether the rate or level is above a threshold value (3\% for Inflation, 4\% for Unemployment Rate). \textit{Directional Accuracy} is the proportion of predictions that correctly identify the direction of change (up or down) relative to the previous month. \textit{Confidence Calibration} is the correlation between the LLM's confidence level (on a scale from 0 to 100) and the MAPE.  \textit{Num Obs} is the number of observations used in the evaluation. \textit{Refusals} are the number of instances in which the model withheld a prediction by either answering "null" or 0. Refusal count also includes instances of missing data.}
\scriptsize
\begin{tabularx}{\linewidth}{lYYYYYYYYY}
  \toprule
\textit{Panel A: Macro Indicators} & ME (\%) & MAE (\%) & Threshold Accuracy (\%) & Directional Accuracy (\%) & Confidence Calibration & Start Month & End Month & Num Obs & Refusals \\
  \midrule
  \multicolumn{10}{l}{\textit{Pre-cutoff}} \\
  \midrule
  Euro Area Headline HCIP Inflation   & 0.02  & 0.06 & 96.57 & 95.00 & -0.25 & 01/1997 & 09/2023 & 321 &   2  \\
  UK CPIH Inflation                   & -0.05 & 0.36 & 85.15 & 86.10 & -0.48 & 01/1990 & 09/2023 & 404 &  13  \\
  Japan Headline CPI Inflation        & 0.02  & 0.24 & 98.77 & 86.78 & -0.41 & 01/1990 & 09/2023 & 405 &  22  \\
  China Headline CPI Inflation        & 0.06  & 0.58 & 97.28 & 95.53 & -0.35 & 01/1990 & 09/2023 & 405 &  11  \\
    \midrule
  \multicolumn{10}{l}{\textit{Post-cutoff}} \\
  \midrule
  Euro Area Headline HCIP Inflation & 2.03 & 2.03 & 35.29  & 37.50 & -0.08 & 10/2023 & 02/2025 &  17 &   0  \\
  UK CPIH Inflation                 & 1.89 & 2.03 & 100.00 & 56.25 & 0.69  & 10/2023 & 02/2025 &  17 &   0  \\
  Japan Headline CPI Inflation      & 0.26 & 0.62 & 64.71  & 62.50 &  -    & 10/2023 & 02/2025 &  17 &   0  \\
  China Headline CPI Inflation      & 0.89 & 1.02 & 100.00 & 56.25 & -0.29 & 10/2023 & 02/2025 &  17 &   0  \\
  \midrule
  \multicolumn{10}{l}{\textit{Pre-cutoff}} \\
  \midrule
    Euro Area Unemployment Rate         & -0.23 & 0.24 & 100.00 & 85.56 & -0.58 & 01/2000 & 09/2023 & 285 &   0  \\
  UK Unemployment Rate                & -0.03 & 0.15 & 96.54  & 92.57 & -0.24 & 01/1990 & 09/2023 & 405 &   0  \\
  Japan Unemployment Rate             & 0.01  & 0.06 & 98.75  & 93.47 & -0.07 & 01/1990 & 09/2023 & 399 &   0  \\
  China Unemployment Rate             & 0.04  & 0.07 & 89.95  & 84.62 & -0.21 & 01/1990 & 09/2023 & 378 &  27  \\ 
      \midrule
  \multicolumn{10}{l}{\textit{Post-cutoff}} \\
  \midrule
  Euro Area Unemployment Rate       & 0.06 & 0.11 & 100.00 & 75.00 & -0.16 & 10/2023 & 02/2025 &  17 &   0  \\
  UK Unemployment Rate              & -0.12 & 0.21 & 58.82 & 75.00 & -0.48 & 10/2023 & 02/2025 &  17 &   0  \\
  Japan Unemployment Rate           & 0.09 & 0.09 & 100.00 & 56.25 & 0.64  & 10/2023 & 02/2025 &  17 &   0 \\ 
  China Unemployment Rate           & 0.05 & 0.09 & 100.00 & 62.50 &  -    & 10/2023 & 02/2025 &  17 &   0  \\
  \bottomrule
\end{tabularx}
\end{table}
\end{landscape}

\clearpage
\begin{landscape}
\begin{table}[H]
\centering
\caption{Evaluation Metrics for International Stock Indices}\label{tab:international_indices}
\caption*{\scriptsize This table reports a set of evaluation metrics for recall of international data. We also ask the LLM to recall closing values of the Euro Stoxx 50 Index, the Nikkei 225 Index, the FTSE 100 Index, and the SSE Composite Index. For the stock indices, the \textit{MPE} is calculated by taking the average of the percent error $(Estimated Level  - Actual Level)/Actual Level$. \textit{MAPE} is calculated by taking the average of the absolute value of the percent error. \textit{Directional Accuracy} is the proportion of predictions that correctly identify the direction of change (up or down) relative to the previous month. \textit{Confidence Calibration} is the correlation between the LLM's confidence level (on a scale from 0 to 100) and the MAPE.  \textit{Num Obs} is the number of observations used in the evaluation. \textit{Refusals} are the number of instances in which the model withheld a prediction by either answering "null" or 0. Refusal count also includes instances of missing data.}
\scriptsize
\begin{tabularx}{\linewidth}{lYYYYYYYY}
  \toprule
     & MPE (\%) & MAPE (\%) & Directional Accuracy (\%) & Confidence Calibration & Num Obs & Start Month & End Month & Refusals \\
  \midrule
  \multicolumn{9}{l}{\textit{Pre-cutoff}} \\
  \midrule
  Euro STOXX 50 & -6.22  & 10.53   & 84.17 & -0.92 & 01/1990 & 09/2023 & 361 &   44  \\
  Nikkei 225    & -0.11  & 1.55    & 97.52 & -0.22 & 01/1990 & 09/2023 & 404 &   1  \\
  FTSE 100      & 0.09   & 1.22    & 97.77 & -0.37 & 01/1990 & 09/2023 & 405 &   0  \\
  SSE Composite & -3.27  & 6.68    & 81.42 & -0.19 & 12/1990 & 09/2023 & 394 &   0  \\ 
  \midrule
  \multicolumn{9}{l}{\textit{Post-cutoff}} \\
  \midrule
  Euro STOXX 50 & -27.21 & 27.54  & 46.67 & -0.79 & 10/2023 & 02/2025 &  16 &   1  \\
  Nikkei 225    & -32.77 & 32.77  & 43.75 & -0.86 & 10/2023 & 02/2025 &  17 &   0  \\
  FTSE 100      & -23.27 & 23.43  & 50.00 & -0.73 & 10/2023 & 02/2025 &  17 &   0  \\
  SSE Composite & -9.81  & 18.23  & 40.00 & -0.61 & 10/2023 & 02/2025 &  16 &   1  \\
  \bottomrule
\end{tabularx}
\end{table}
\end{landscape}

\begin{table}[htbp]
\centering
\caption{Evaluation Metrics for S\&P 500 Using Forecast Prompt}\label{tab:forecast}
\caption*{\scriptsize This table reports a set of evaluation metrics assessing the LLM’s ability to recall daily market index levels using an alternative prompt using the term ``forecast" instead of directly asking for the LLM to provide the S\&P 500 closing value. \textit{Mean Percent Error (MPE)}, \textit{Mean Absolute Percent Error (MAPE)}, and \textit{Directional Accuracy } are reported in percentage points (0.18 means 0.18\%). \textit{MPE} is calculated by taking the average of the percent error $(Predicted Price - Actual Price) / Actual Price$. \textit{MAPE} is calculated by taking the average of the absolute value of the percent error. \textit{Directional Accuracy} is the proportion of predictions that went in the correct direction (up or down) with respect to the previous month. \textit{Confidence Calibration} is the correlation between the LLM's confidence level (on a scale of 0 to 100) and the MAPE. \textit{Num Obs} is the number of observations used in the evaluation. \textit{Refusals} are the number of instances in which the model withheld a prediction by either answering "null" or 0. Results are provided for a prompt that contains an empty context in panel A and a prompt that provides the previous two day's closing prices as context in panel B. The pretraining period covers 01/02/1990 to 09/29/2023 and post-cutoff period covers 10/02/2023 to 02/28/2025.}
\scriptsize
\begin{tabularx}{\linewidth}{lYYYYYY}
  \toprule
     & MPE (\%) & MAPE (\%) & Directional Accuracy (\%) & Confidence Calibration & Num Obs  & Refusals \\
   \midrule
  \multicolumn{5}{l}{\textit{Panel A: No Context}} \\
  \midrule
  SP500 & -1.92 & 2.76  & 74.26 & -0.32& 8502   &   0 \\
  SP500 Post Cutoff & -15.10 & 15.30  & 43.06 & 0.10& 354   &   0 \\
  \midrule
  \multicolumn{5}{l}{\textit{Panel B: With Context}} \\
  \midrule
  SP500  & 0.03 & 1.10  & 48.54 & -0.06 & 8502 &   0  \\
  SP500 Post Cutoff & 0.09 & 0.74  & 51.84 & -0.10& 354 &   0  \\
   \bottomrule
\end{tabularx}
\end{table}

\clearpage
\begin{landscape}
\begin{table}[htbp]
\centering
\caption{Evaluation Metrics for Macro Indicators - Llama 3.1-70b-Instruct}\label{tab:llama_macro}
\caption*{\scriptsize This table reports a set of evaluation metrics for various macroeconomic indicators grouped into three panels: Rates, Levels, and Levels, Recent Period: Past 10 years. We ask the LLM to recall monthly values (quarterly for GDP, specific end of month date for 10-Year Treasury Yield and VIX) for each indicator. The indicators in the \textit{Rates} panel include GDP Growth, Inflation, Unemployment Rate, and the 10-Year Treasury Yield. For these indicators, we ask the LLM to give us a percentage. The \textit{Levels} panel includes Housing Starts, VIX, and Nonfarm Payrolls, evaluated over the full sample period. The \textit{Levels, Recent Pre-cutoff Period: Past 10 years} panel evaluates these same indicators over a more recent, shorter period. \textit{Mean Error (ME)}, \textit{Mean Absolute Error (MAE)}, \textit{Mean Percent Error (MPE)}, \textit{Mean Absolute Percent Error (MAPE)}, \textit{Threshold Accuracy},and \textit{Directional Accuracy} are reported in percentage points (0.01 means 0.01\%). For \textit{Rates}, the \textit{ME} is the difference $Estimated Rate  - Actual Rate$. \textit{MAE} is calculated by taking the average of the absolute value of the \textit{ME}. \textit{Threshold Accuracy} is the proportion of predictions that correctly identify whether the rate or level is above a threshold value (2.5\% for GDP Growth, 3\% for Inflation, 4\% for Unemployment Rate, 4\% for the 10-Year Treasury Yield, 16 for VIX, 1400 for Housing Starts, and 200 for Nonfarm Payrolls). For \textit{Levels}, the \textit{MPE} is calculated by taking the average of the percent error $(Estimated Level  - Actual Level)/Actual Level$. \textit{MAPE} is calculated by taking the average of the absolute value of the percent error. \textit{Directional Accuracy} is the proportion of predictions that correctly identify the direction of change (up or down) relative to the previous month. \textit{Confidence Calibration} is the correlation between the LLM's confidence level (on a scale from 0 to 100) and the MAPE.  \textit{Num Obs} is the number of observations used in the evaluation, \textit{Start Date} and \textit{End Date} indicate the period over which the metrics were computed. \textit{Refusals} are the number of instances in which the model withheld a prediction by either answering "null" or 0. Refusal count also includes instances of missing data.}
\scriptsize

\begin{tabularx}{\linewidth}{lYYYYYYY}
  \toprule
\textit{Panel A: Rates} & ME (\%) & MAE (\%) & Threshold Accuracy (\%) & Directional Accuracy (\%) & Confidence Calibration & Num Obs & Refusals \\
  \midrule
  \multicolumn{8}{l}{\textit{Pre-cutoff, 01/1990 to 11/2023}} \\
  \midrule
  GDP Growth & 0.07 & 0.48 & 92.59 & 91.04 & -0.11 & 135 & 1 \\
  Inflation & -0.15 & 0.39 & 91.87 & 66.67 & -0.24 & 406 & 1 \\ 
  Unemployment Rate & -0.03 & 0.06 & 99.26 & 76.49 & -0.17 & 405 & 2 \\
  10-Yr Treasury Yield & 0.02 & 0.13 & 97.24 & 81.61 & -0.34 & 398 & 9 \\ 
  \midrule
  \multicolumn{8}{l}{\textit{Post-cutoff, 12/2023 to 02/2025}} \\
  \midrule
  GDP Growth & -0.13  & 1.04 & 60.00 & 100.00  & -0.02 & 5  & 0 \\
  Inflation & -0.72 & 0.88 & 66.67 & 71.43 & -0.38 & 15 & 0 \\
  Unemployment Rate & -0.14 & 0.14 & 75.00  & 27.27 & -0.33 & 12 & 3 \\
  10-Yr Treasury Yield & 0.21 & 0.38 &  64.29 & 69.23 & 0.33 & 14 & 1 \\ 
  \bottomrule
\end{tabularx}

\vspace{5pt}

\begin{tabularx}{\linewidth}{lYYYYYYY}
  \toprule
  \textit{Panel B: Levels} & MPE (\%) & MAPE (\%) & Threshold Accuracy (\%) & Directional Accuracy (\%) & Confidence Calibration & Num Obs & Refusals \\
  \midrule
  \multicolumn{8}{l}{\textit{Pre-cutoff, 01/1990 to 11/2023}} \\
  \midrule
  VIX & 0.99 & 12.12 & 89.45 & 75.82 & -0.23 & 398 & 9 \\
  Housing Starts & -2.29 & 5.51 & 90.27 & 74.25 & -0.36 & 401 & 6 \\
  Nonfarm Payrolls & 42.85 & 208.61 & 86.67 & 84.90 & -0.11 & 405 & 2 \\
  \midrule
  \multicolumn{8}{l}{\textit{Recent Pre-cutoff Period, 12/2014 to 11/2023}} \\
  \midrule
  VIX & 1.68 & 7.58 & 91.51 & 87.62 & -0.11 & 106 & 2 \\
  Housing Starts & -1.54 & 3.37 & 91.51 & 81.90 & -0.11 & 106 & 2 \\ 
  Nonfarm Payrolls & -6.09 & 9.50 & 98.13 & 98.11 & 0.03 & 107 & 1 \\
  \midrule
  \multicolumn{8}{l}{\textit{Post-cutoff, 12/2023 to 2/2025}} \\
  \midrule
  VIX & 15.31 & 20.70 & 37.50 & 57.14 & 0.14 & 8 &   7 \\ 
  Housing Starts & -0.78 & 3.84 & 86.67 & 92.86 & 0.43 & 15 & 0 \\
  Nonfarm Payrolls & 207.67 & 222.88 & 42.86 & 76.92  & -0.16 &  15 & 0 \\
  \bottomrule
\end{tabularx}

\end{table}
\end{landscape}

\begin{table}[htbp]
\centering
\caption{Evaluation Metrics for Market Indices - Llama 3.1-70b-Instruct}\label{tab:llama_index}
\caption*{\scriptsize This table reports a set of evaluation metrics assessing the LLM’s ability to recall market index levels and their changes over time. These tests are done at the daily or monthly frequency. We ask the LLM to recall the closing value of the index each trading day. Panel A provides metrics for predictions of \textit{Daily Levels} and \textit{Daily Levels with context} (where the previous two days' index levels are provided). We ask the LLM to provide monthly returns for these indices as well. Metrics include \textit{Mean Percent Error (MPE)}, \textit{Mean Absolute Percent Error (MAPE)}, and \textit{Directional Accuracy}, all reported in percentage points (0.10 means 0.10\%). \textit{MPE} is calculated by averaging the percent error \((Estimated Level - Actual Level) / Actual Level\). \textit{MAPE} takes the average absolute value of the percent errors. \textit{Directional Accuracy} measures the proportion of predictions of the market index levels correctly following the direction of change (up or down) relative to the previous day. \textit{Confidence Calibration} reports the correlation between the LLM's confidence level (on a scale from 0 to 100) and mean absolute percent error. Panel B presents accuracy metrics related to predicting \textit{Directional Changes} and \textit{Relative Performance} between indices. \textit{Directional Changes} asks the LLM directly for an up or down answer for each month. \textit{Relative Performance} asks the LLM to answer which index of the index pair performed better during the month. \textit{Accuracy} reports the proportion of predictions correctly identifying either the direction of change or relative performance in percentage points. \textit{Confidence Calibration} in this panel reflects the correlation between the LLM’s confidence and the MAPE. Results are separately provided for the S\&P 500 (SP500), Dow Jones Industrial Average (DJIA), and Nasdaq Composite indices.}
\scriptsize
\begin{tabularx}{\linewidth}{lYYYYYY}
  \toprule
  \multicolumn{7}{l}{\textit{Panel A: Numerical Tests}} \\
  \midrule
     & MPE (\%) & MAPE (\%) & Directional Accuracy (\%) & Confidence Calibration & Num Obs & Refusals \\
  \midrule
  \multicolumn{7}{l}{\textit{Daily Levels: Pre-cutoff, 01/02/1990 to 11/30/2023}} \\
  \midrule
   SP500            & -0.19 & 2.03  & 60.57 & -0.15 & 8545  &   0  \\ 
   DJIA             & 0.03 & 1.97  & 61.01 & -0.41 & 8487  &  58  \\
   Nasdaq Composite & -0.28 & 4.48  & 58.05 & -0.26 & 8545  &   0  \\
   
   \midrule
   \multicolumn{7}{l}{\textit{Daily Levels: Post-cutoff, 12/01/2023 to 02/28/2025}} \\
   \midrule
   SP500            &  -14.65 & 14.74   & 43.96 & - & 92  &  234  \\ 
   DJIA             &  10.76 & 18.08 & 58.82 & - & 69  &  266  \\
   Nasdaq Composite & -14.10 & 14.51  & 45.05 & - & 92 &  219 \\
   \midrule
   \multicolumn{7}{l}{\textit{Daily Levels: Pre-cutoff with context, 01/02/1990 to 11/30/2023}}\\
   \midrule
   SP500             & 0.12 & 0.96  & 52.31 & -0.07  & 8544 &   1  \\
   DJIA              & 0.09 & 0.93  & 51.28 & -0.08  & 8544 &   1  \\
   Nasdaq Composite  & 0.15 & 1.27  & 53.76 & -0.06  & 8545 &   0  \\
   \midrule
   \multicolumn{7}{l}{\textit{Daily Levels: Post-cutoff with context, 12/01/2023 to 02/28/2025 }} \\
   \midrule
   SP500             & 0.05 & 0.81  & 50.32 & -0.11 & 309 &   2  \\
   DJIA              & 0.06 & 0.69  & 52.43 & -0.13 & 310 &   1  \\
   Nasdaq Composite  & 0.13 & 1.15  & 52.43 & -0.09 & 310 &   1  \\
  \bottomrule
\end{tabularx}


\end{table}

\begin{figure}[htbp]
    \centering
    \caption{Recall of exact numerical values of international inflation rates.} \label{fig:international_inflation}
    \caption*{\scriptsize This figure shows the LLM's estimated values of inflation in the Euro area, Japan, the United Kingdom, and China compared to the actual values. Panels A, C, E, and G graph the actual values against the estimated values. Panels B, D, F, and H show the estimation error. Estimation error is calculated as \textit{(Estimated - Actual)/Actual} and is shown in percentages (5 means 5\%). The post-cutoff period (10/2023 onward) is shaded gray.}
    \includegraphics[width=0.9\linewidth]{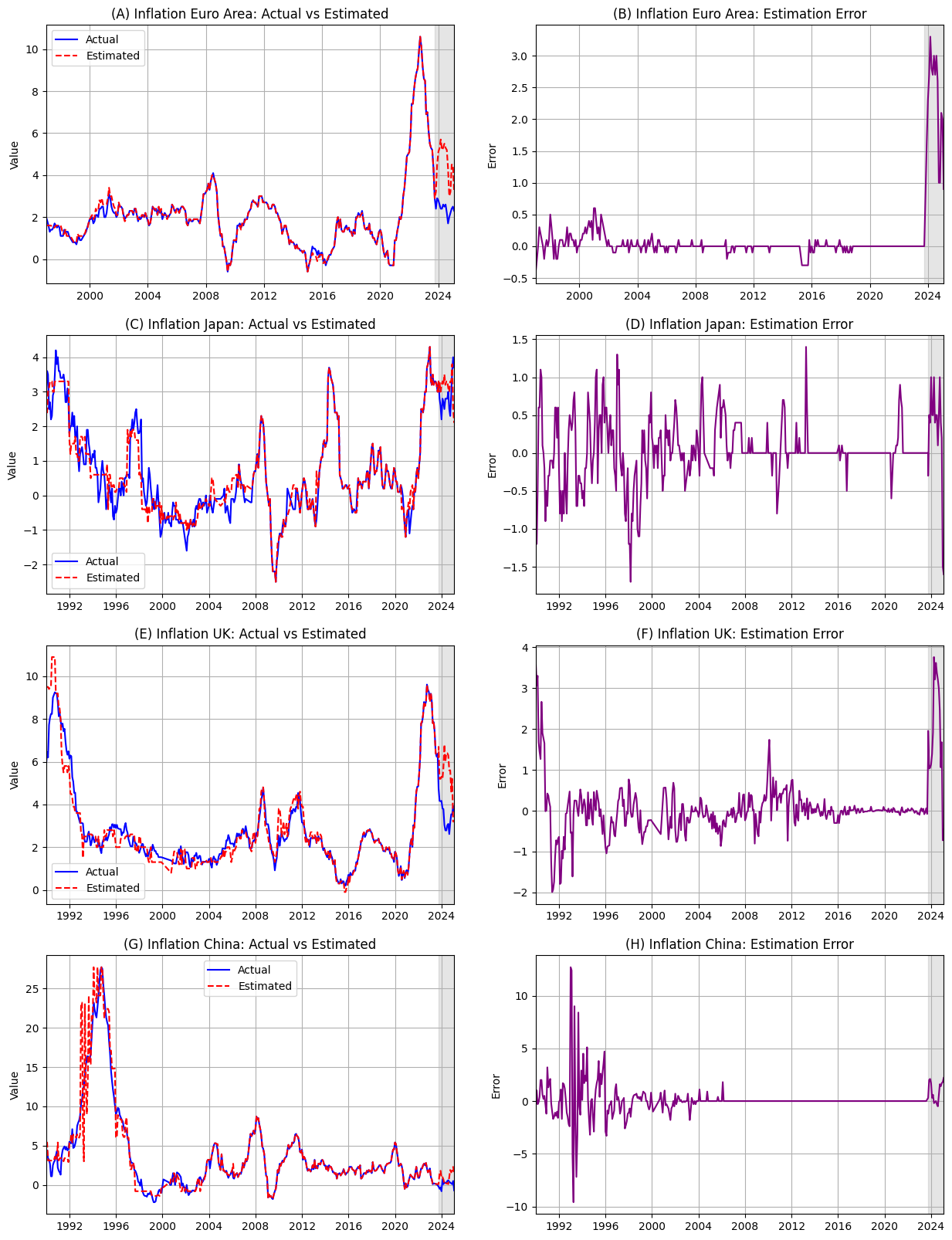}
\end{figure}

\begin{figure}[htbp]
    \centering
    \caption{Recall of exact numerical values of international unemployment rates.} \label{fig:international_unrate}
    \caption*{\scriptsize This figure shows the LLM's estimated values of unemployment rate in the Euro area, Japan, the United Kingdom, and China compared to the actual values. Panels A, C, E, and G graph the actual values against the estimated values. Panels B, D, F, and H show the estimation error. Estimation error is calculated as \textit{(Estimated - Actual)/Actual} and is shown in percentages (5 means 5\%). The post-cutoff period (10/2023 onward) is shaded gray.}
    \includegraphics[width=0.9\linewidth]{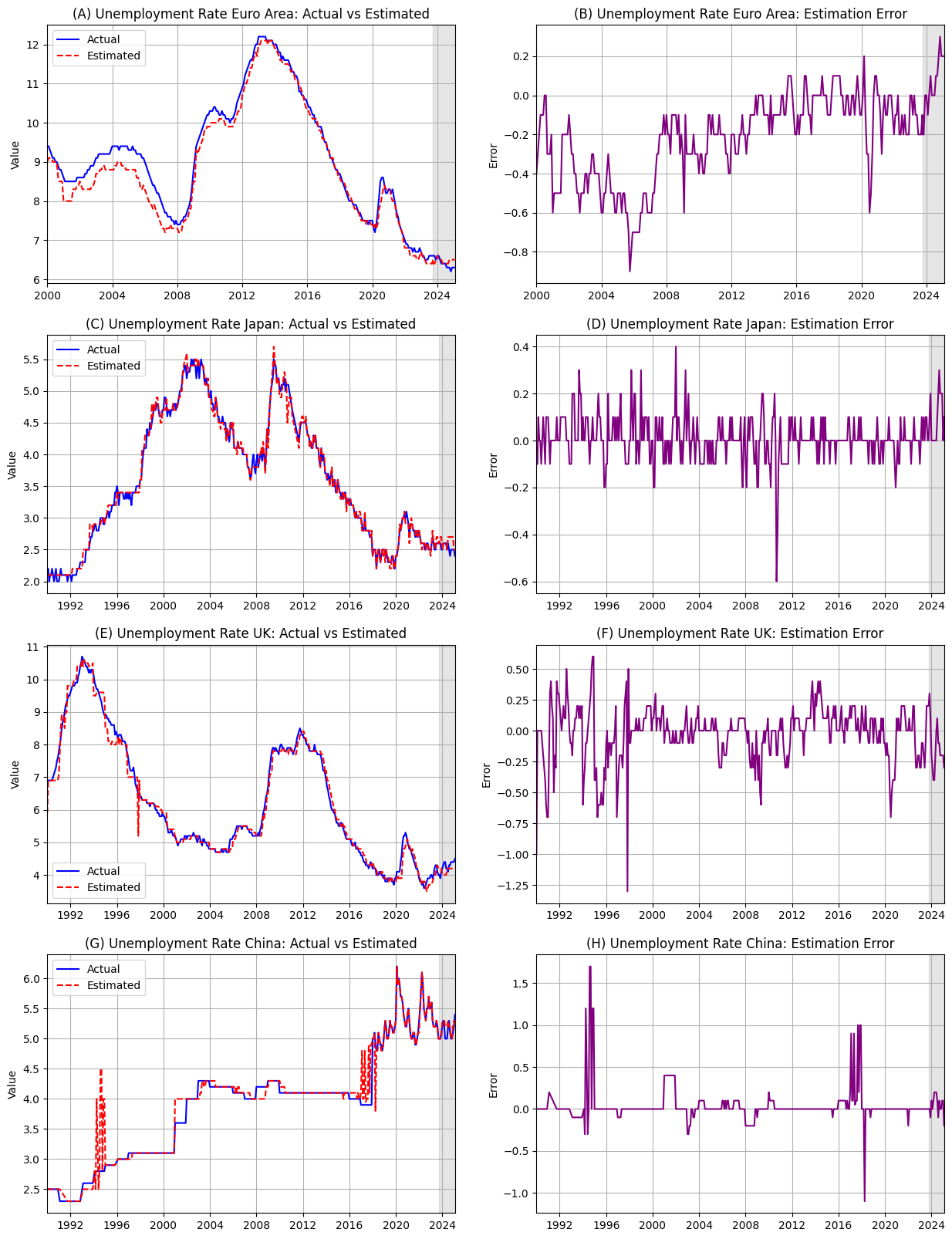}
\end{figure}

\begin{figure}[htbp]
    \centering
    \caption{Recall of exact numerical levels of international market indices.} \label{fig:indices_international}
    \caption*{\scriptsize This figure shows the LLM's estimated values of the international stock market indices compared to the actual values. Panels A, C, E, and G graph the actual values against the estimated values. Panels B, D, F and H show the estimation error for the Euro Stoxx 50 Index, the Nikkei 225 Index, the FTSE 100 Index, and the SSE Composite Index. Estimation error is calculated as \textit{(Estimated - Actual)/Actual} and is shown in percentage points (5 means 5\%). For the Nasdaq Composite panels, 10 outliers were removed for the ease of plotting. These values are still included in the evaluation metrics table. The post-cutoff period (10/2023 onward) is shaded gray.}
    \includegraphics[width=0.9\linewidth]{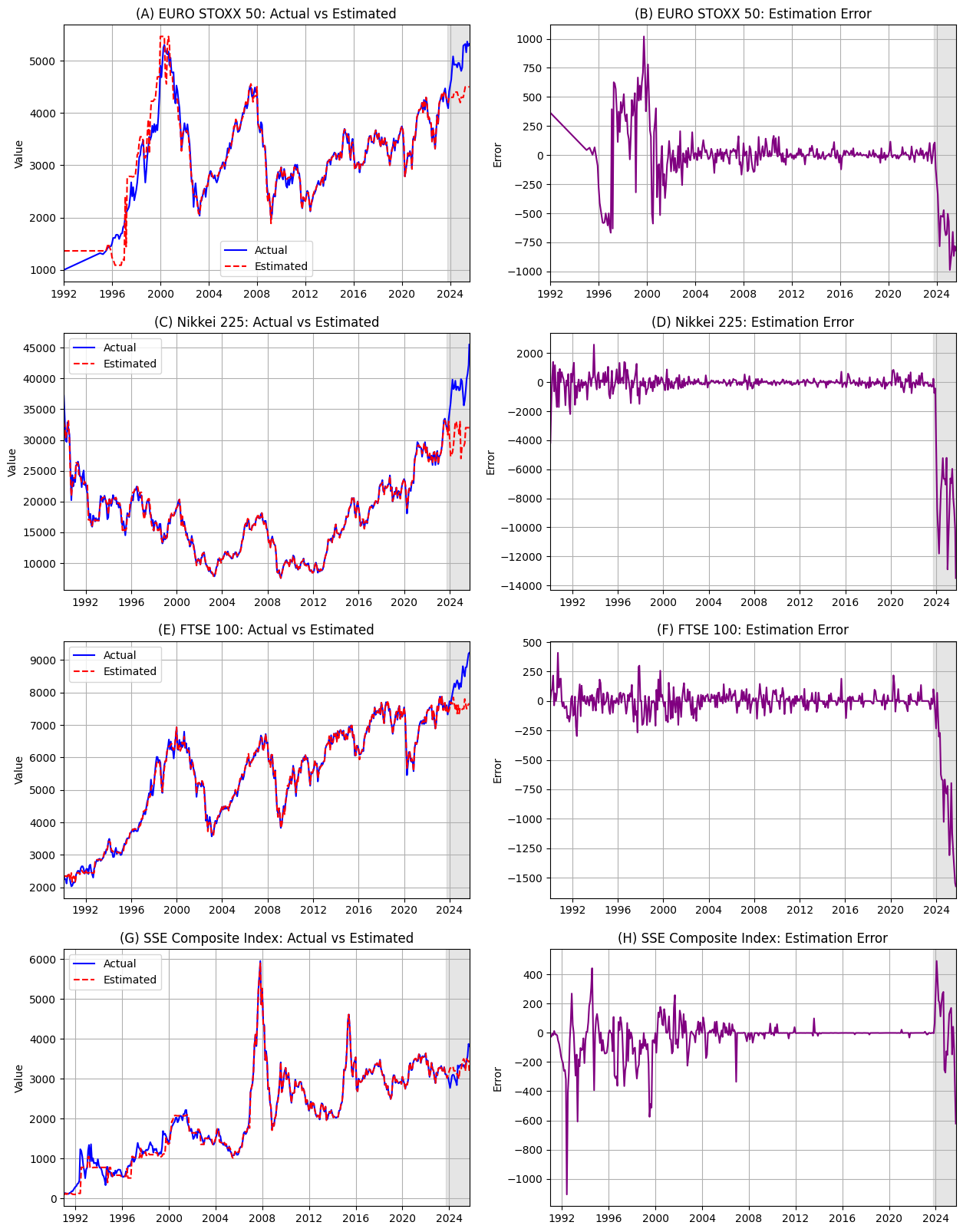}
\end{figure}

\begin{figure}[htbp]
    \centering
    \caption{Recall through Embeddings - Expanding Window and SMA} \label{fig:embeddings_appendix}
    \caption*{\scriptsize This figure shows the comparison of actual values and predicted values for GDP growth, inflation, 10-year treasury rate, and unemployment rate using Ridge regression with regularization parameter of 0.01 trained on embeddings of textual prompts such as ``In Q4 2020, the earliest estimate of the US GDP growth rate was" and the corresponding economic data, either rates or levels. This figure shows the results of using an expanding window of training data and a simple moving average with a 5 year window. The solid blue lines show the actual values and dashed red lines show the predicted values. }
  \begin{subfigure}[b]{0.48\textwidth}
    \includegraphics[width=\textwidth]{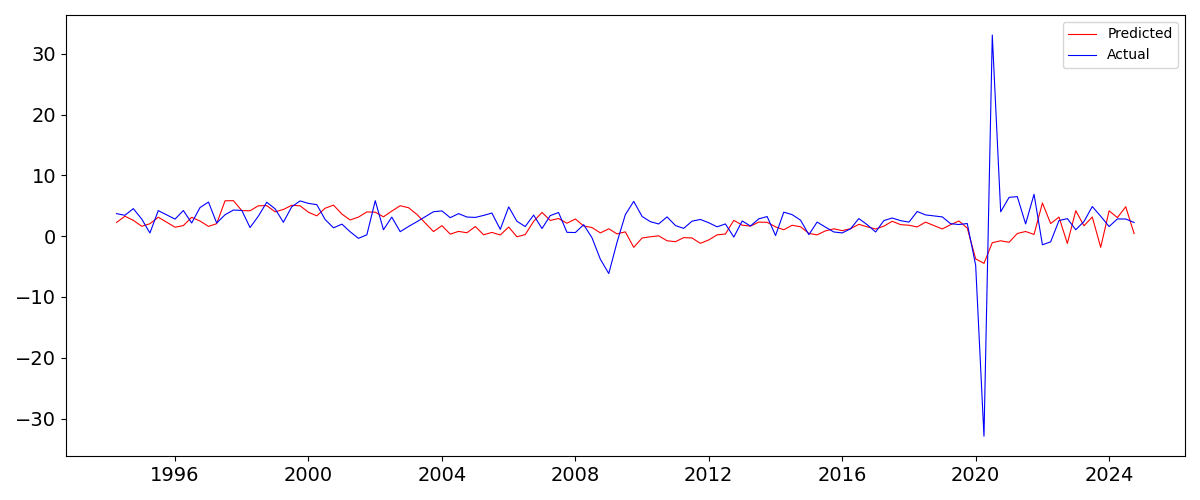}
    \caption{GDP Growth (Expanding Window)}
    \label{fig:gdp_expanding}
  \end{subfigure}
  \hfill
  \begin{subfigure}[b]{0.48\textwidth}
    \includegraphics[width=\textwidth]{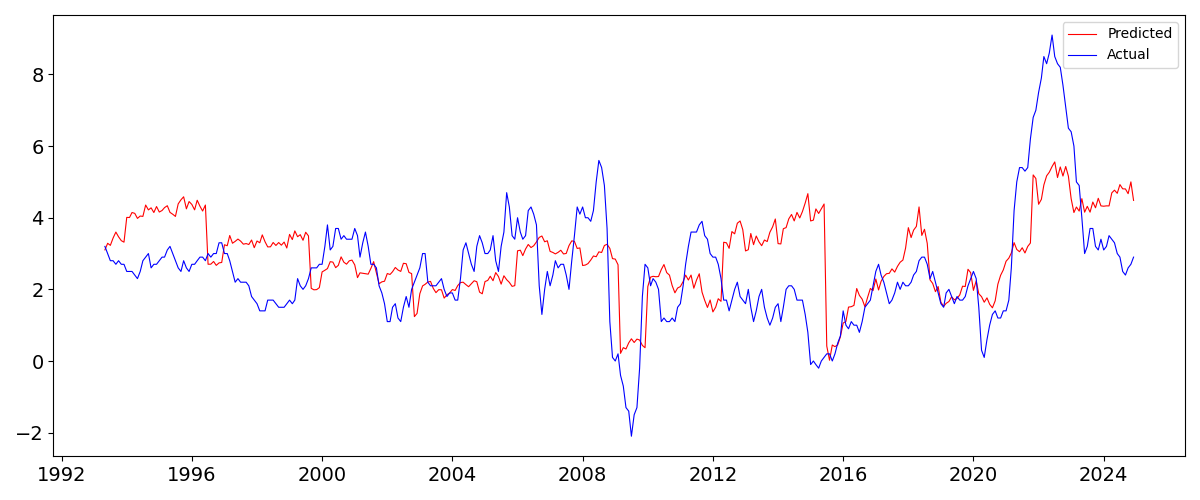}
    \caption{Inflation (Expanding Window)}
    \label{fig:inflation_expanding}
  \end{subfigure}

  \vspace{1em}

  \begin{subfigure}[b]{0.48\textwidth}
    \includegraphics[width=\textwidth]{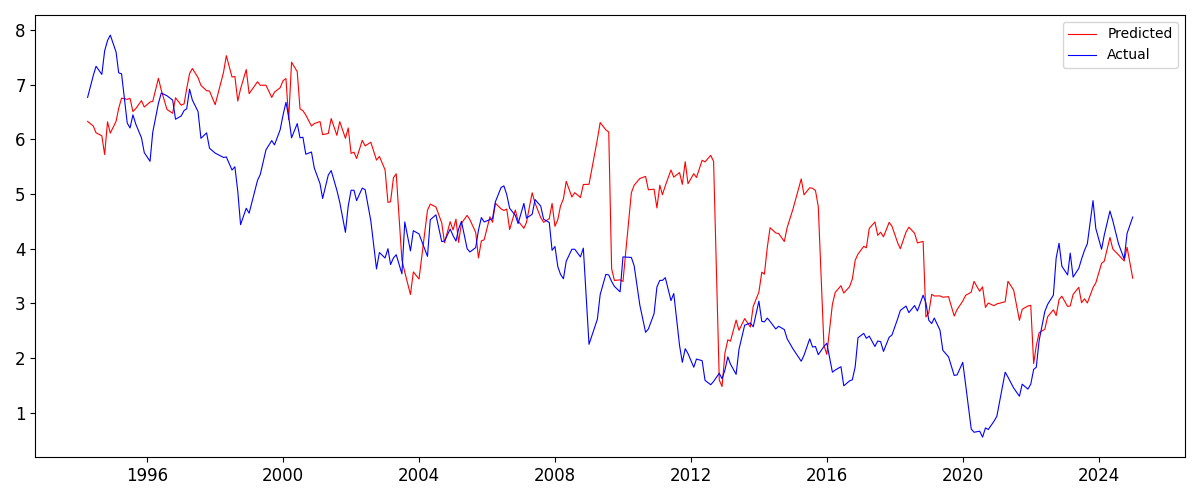}
    \caption{10-yr Treasury Rate (Expanding Window)}
    \label{fig:dgs10_expanding}
  \end{subfigure}
  \hfill
  \begin{subfigure}[b]{0.48\textwidth}
    \includegraphics[width=\textwidth]{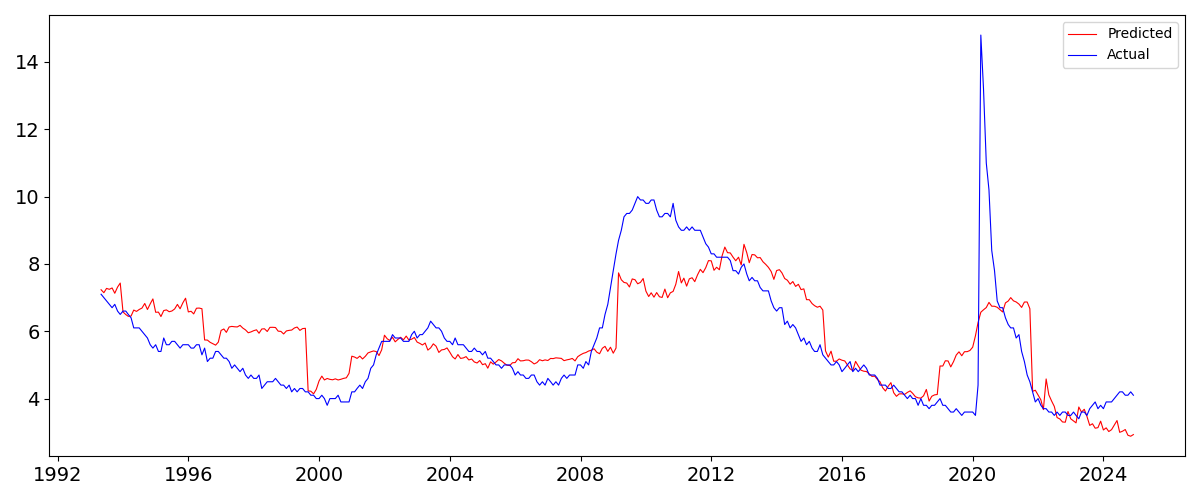}
    \caption{Unemployment (Expanding Window)}
    \label{fig:unrate_expanding}
  \end{subfigure}

  \vspace{1.5em}

  \begin{subfigure}[b]{0.48\textwidth}
    \includegraphics[width=\textwidth]{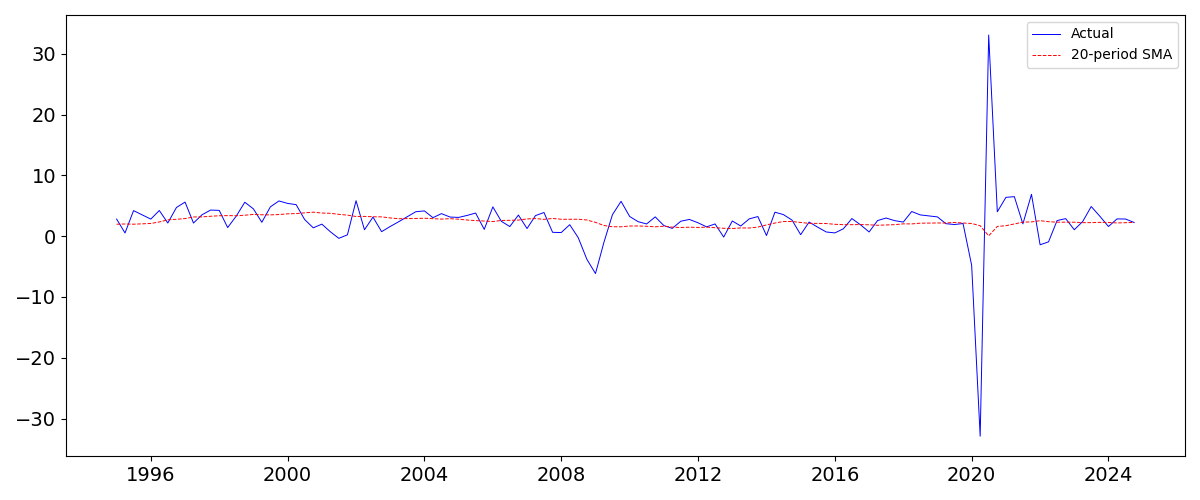}
    \caption{GDP Growth (SMA)}
    \label{fig:gdp_sma}
  \end{subfigure}
  \hfill
  \begin{subfigure}[b]{0.48\textwidth}
    \includegraphics[width=\textwidth]{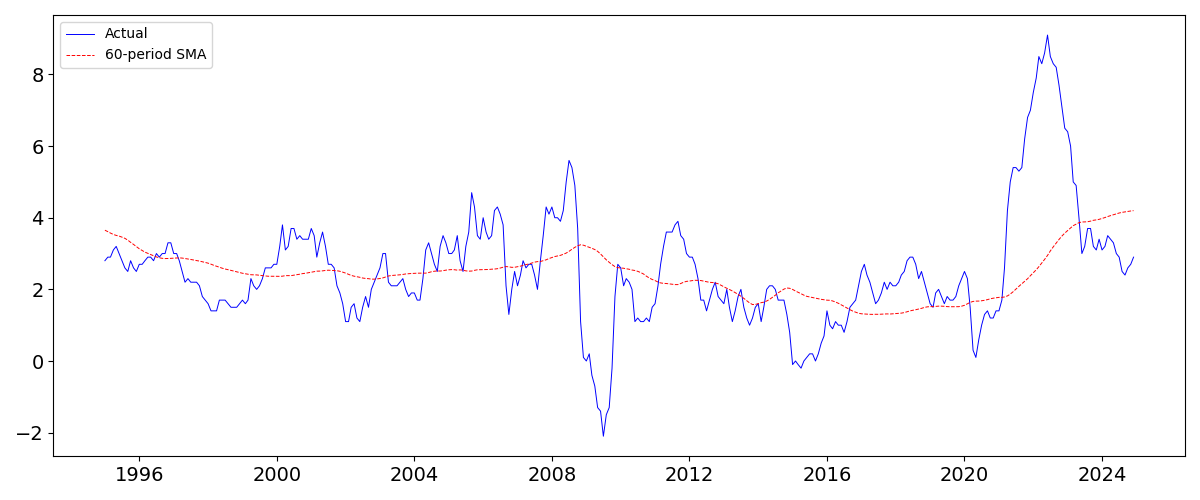}
    \caption{Inflation (SMA)}
    \label{fig:inflation_sma}
  \end{subfigure}

  \vspace{1em}

  \begin{subfigure}[b]{0.48\textwidth}
    \includegraphics[width=\textwidth]{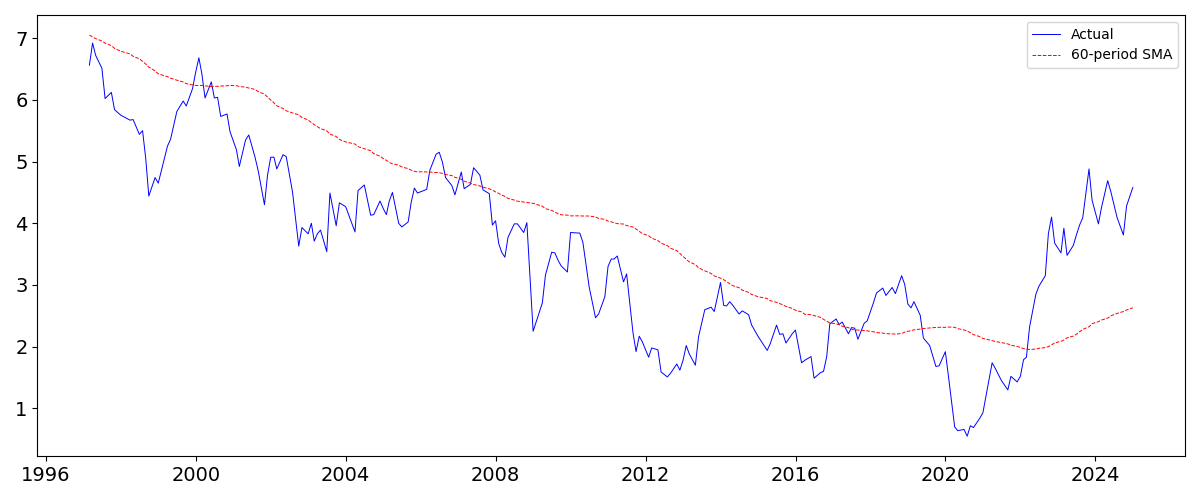}
    \caption{10-yr Treasury Rate (SMA)}
    \label{fig:dgs10_sma}
  \end{subfigure}
  \hfill
  \begin{subfigure}[b]{0.48\textwidth}
    \includegraphics[width=\textwidth]{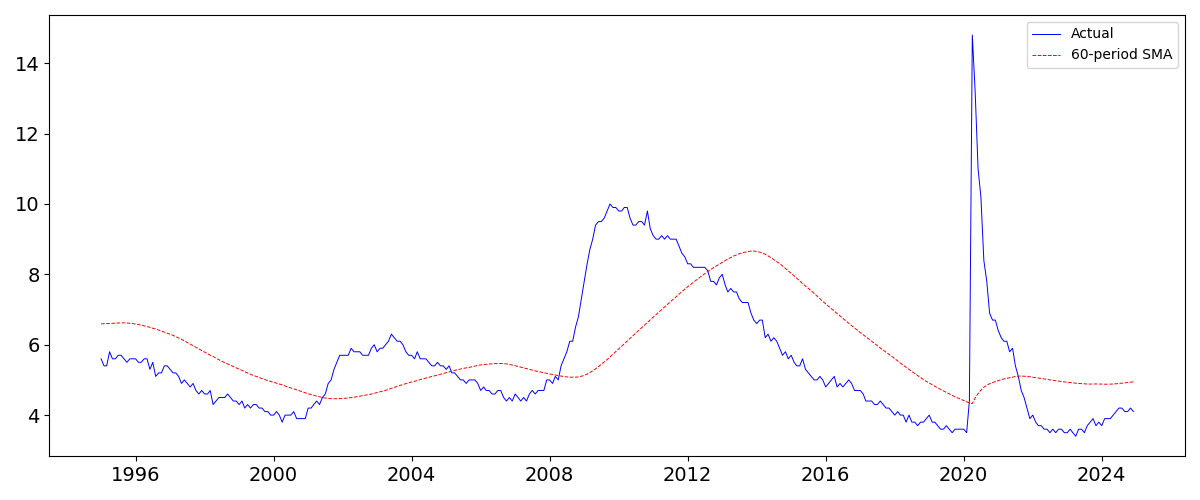}
    \caption{Unemployment (SMA)}
    \label{fig:unrate_sma}
  \end{subfigure}

\end{figure}

\begin{figure}[htbp]
    \centering
    \caption{Recall through Embeddings ($\lambda$ = 0.001)} \label{fig:embeddings_small}
    \caption*{\scriptsize This figure shows the comparison of actual values and predicted values for GDP growth, inflation, 10-year treasury rate, and unemployment rate using Ridge regression with regularization parameter of 0.001 trained on embeddings of textual prompts such as ``In Q4 2020, the earliest estimate of the US GDP growth rate was" and the corresponding economic data, either rates or levels. We use a 5 year rolling window to train the Ridge regression each period on the most recent five years of data to predict the next value. The solid blue lines show the actual values and dashed red lines show the predicted values.}

  \begin{subfigure}[b]{0.58\textwidth}
    \includegraphics[width=\textwidth]{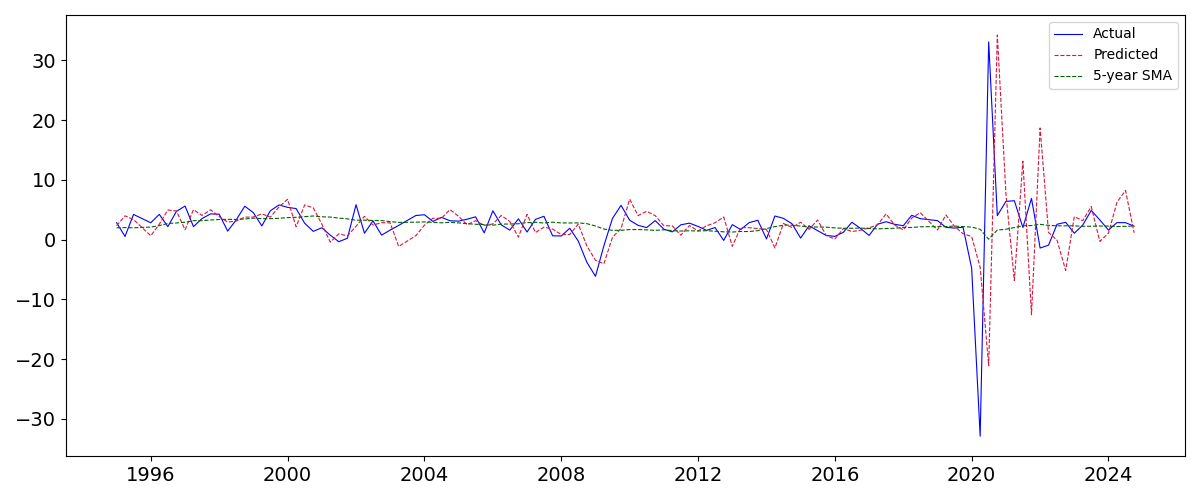}
    \caption{GDP Growth (Rolling Window)}
    \label{fig:gdp_rolling_small}
  \end{subfigure}
  \hfill
  \begin{subfigure}[b]{0.58\textwidth}
    \includegraphics[width=\textwidth]{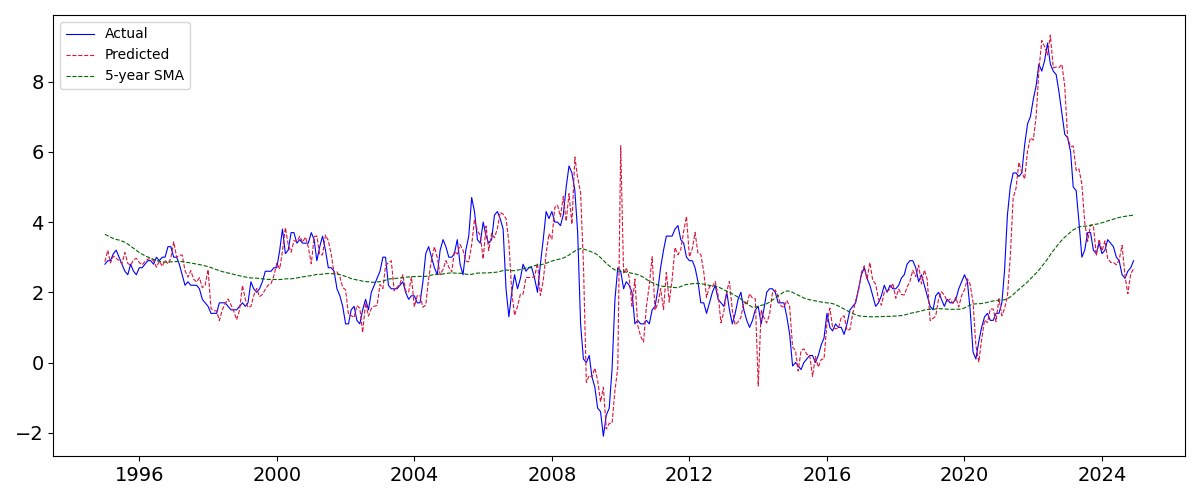}
    \caption{Inflation (Rolling Window)}
    \label{fig:inflation_rolling_small}
  \end{subfigure}

  \vspace{1em}

  \begin{subfigure}[b]{0.58\textwidth}
    \includegraphics[width=\textwidth]{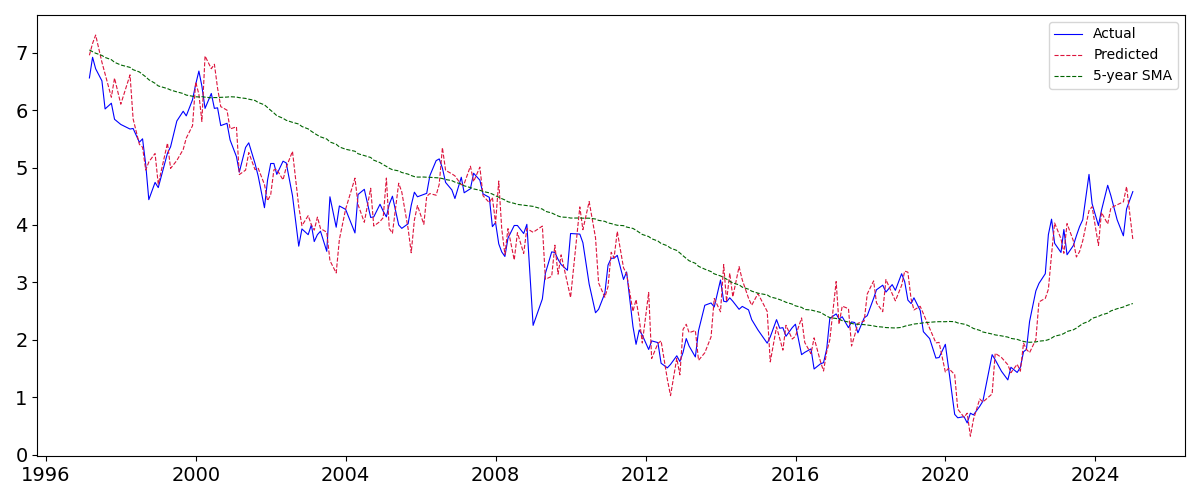}
    \caption{10-yr Treasury Rate (Rolling Window)}
    \label{fig:dgs10_rolling_small}
  \end{subfigure}
  \hfill
  \begin{subfigure}[b]{0.58\textwidth}
    \includegraphics[width=\textwidth]{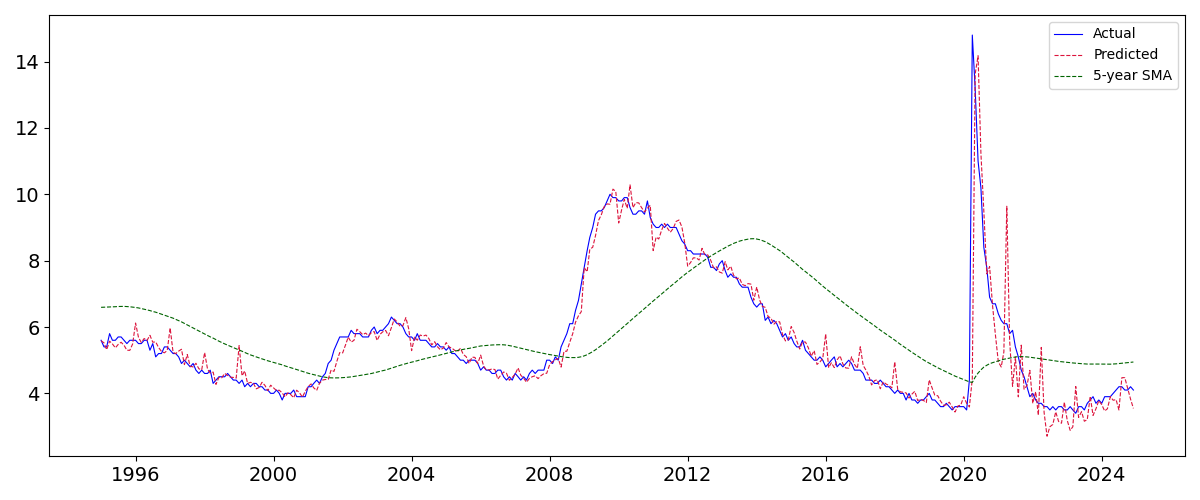}
    \caption{Unemployment (Rolling Window)}
    \label{fig:unrate_rolling_small}
  \end{subfigure}

\end{figure}

\begin{figure}[htbp]
    \centering
    \caption{Recall through Embeddings ($\lambda$ = 0.1)} \label{fig:embeddings_big}
    \caption*{\scriptsize This figure shows the comparison of actual values and predicted values for GDP growth, inflation, 10-year treasury rate, and unemployment rate using Ridge regression with regularization parameter of 0.1 trained on embeddings of textual prompts such as ``In Q4 2020, the earliest estimate of the US GDP growth rate was" and the corresponding economic data, either rates or levels. We use a 5 year rolling window to train the Ridge regression each period on the most recent five years of data to predict the next value. The solid blue lines show the actual values and dashed red lines show the predicted values. }

  \begin{subfigure}[b]{0.58\textwidth}
    \includegraphics[width=\textwidth]{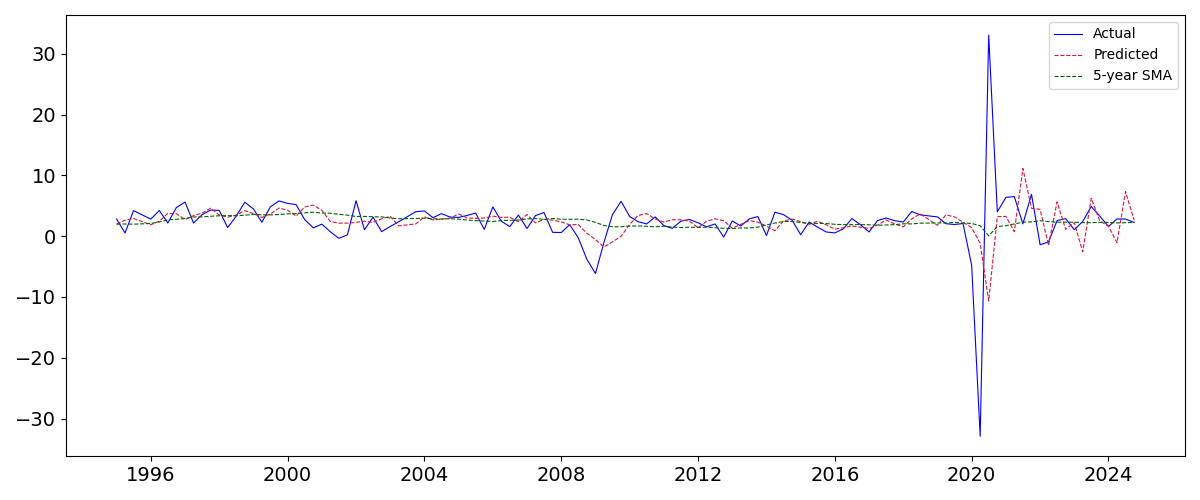}
    \caption{GDP Growth (Rolling Window)}
    \label{fig:gdp_rolling_big}
  \end{subfigure}
  \hfill
  \begin{subfigure}[b]{0.58\textwidth}
    \includegraphics[width=\textwidth]{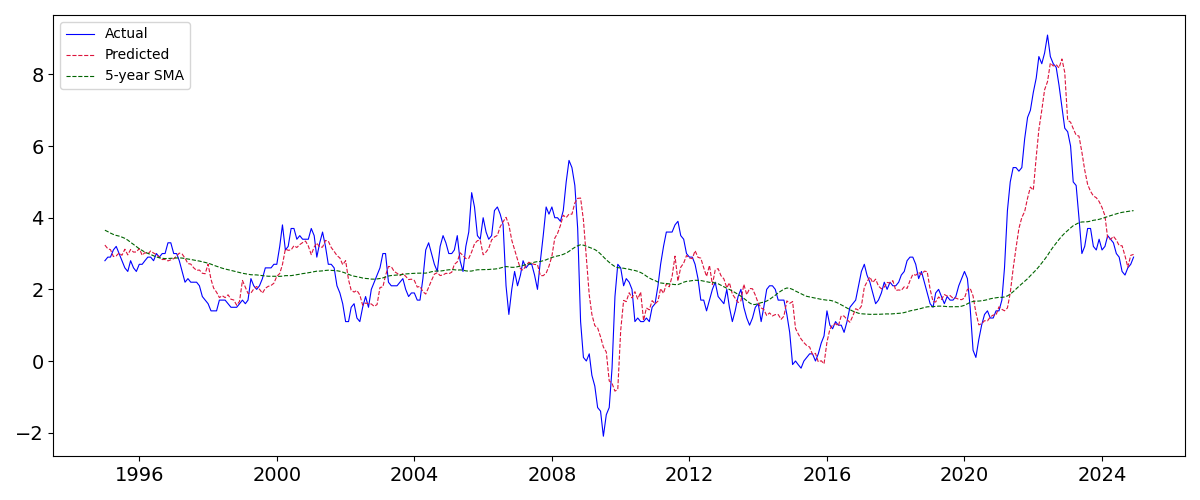}
    \caption{Inflation (Rolling Window)}
    \label{fig:inflation_rolling_big}
  \end{subfigure}

  \vspace{1em}

  \begin{subfigure}[b]{0.58\textwidth}
    \includegraphics[width=\textwidth]{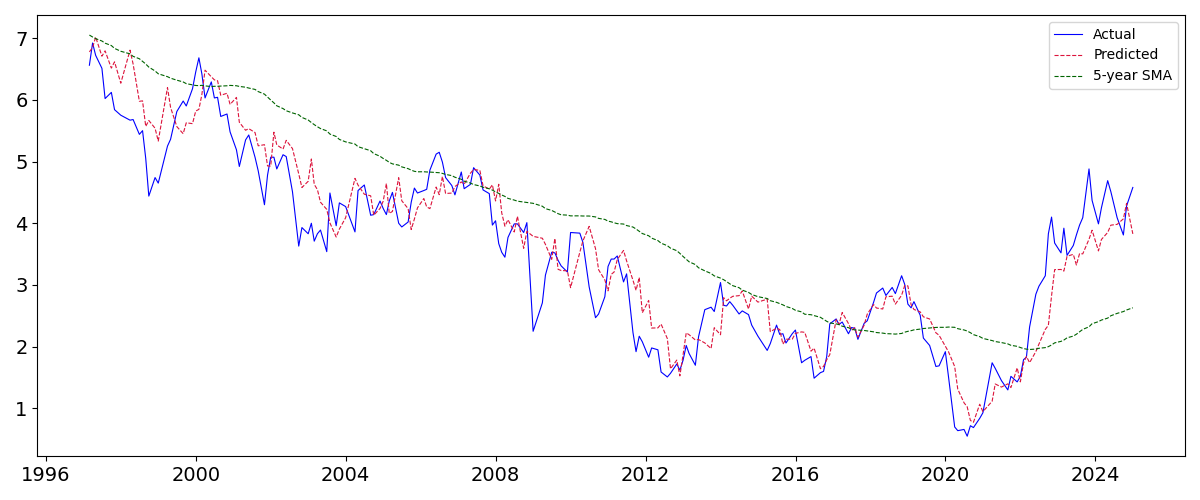}
    \caption{10-yr Treasury Rate (Rolling Window)}
    \label{fig:dgs10_rolling_big}
  \end{subfigure}
  \hfill
  \begin{subfigure}[b]{0.58\textwidth}
    \includegraphics[width=\textwidth]{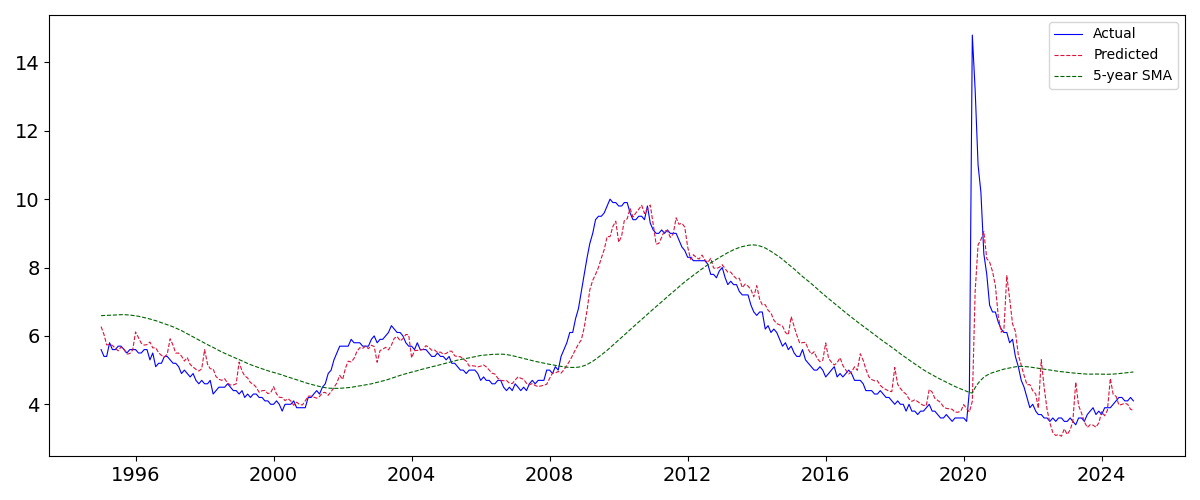}
    \caption{Unemployment (Rolling Window)}
    \label{fig:unrate_rolling_big}
  \end{subfigure}

\end{figure}

\FloatBarrier
\section{Proof of the Memorization Problem}
\label{sec:formalization-proof}

\subsection{Formal Definitions and Assumptions}

Let $\Phi$ be the space of all possible model parameters for the fixed architecture and training pipeline.
\begin{itemize}
    \item \textbf{Generic Parameters ($\phi$):} We use $\phi \in \Phi$ as a generic placeholder for a set of model parameters.
    \item \textbf{Factual Parameters ($\theta$):} $\theta = \theta(I_{\text{all}}) \in \Phi$ are the \textbf{factual} parameters of the fully trained model, which has seen all information $I_{\text{all}}$.
    \item \textbf{Counterfactual Parameters ($\theta_t$):} $\theta_t = \theta(I_t) \in \Phi$ are the \textbf{ideal} but unobservable counterfactual parameters that \textit{would have been obtained} by training on the restricted information set $I_t$.
\end{itemize}

Let $\mathcal{Y}$ be the finite, non-empty set of possible answers.
\begin{itemize}
    \item \textbf{Task:} $Q_t$ is the task posed ``as of time $t$''.
    \item \textbf{Additional instructions:} $P$ represents additional instructions that modify the task (e.g., constraining prompts). We use $P=\varnothing$ to denote no additional instructions beyond the task itself.
    \item \textbf{Scoring Function:} $S_\phi(y; Q, P)$ is the model's internal score for answer $y \in \mathcal{Y}$ given parameters $\phi$, task $Q$, and additional instructions $P$.
    \item \textbf{Decision Rule ($\delta$):} We define the model's deterministic decision rule as:
        \[
        \delta_\phi(Q,P) \coloneqq \argmax_{y\in\mathcal{Y}} S_\phi(y;Q,P)
        \]
        We assume a fixed, deterministic tie-breaking rule (e.g., lexicographical) to ensure $\delta$ is a well-defined function. The choice of tie-breaking rule does not affect the non-identification result.
\end{itemize}

With these definitions, we can state the core quantities:
\begin{itemize}
    \item \textbf{Ideal Estimand (Target):} $Y_t^\star \coloneqq \delta_{\theta_t}(Q_t, \varnothing)$
    \item \textbf{Observed Constrained Decision:} $Y_{\mathrm{constrained}}(P) \coloneqq \delta_{\theta}(Q_t, P)$
\end{itemize}

Finally, we state a mild assumption required for the proof's construction.

\begin{assumption}[Decision reachability]\label{ass:reach}
For any task $Q$ and the unconstrained case $P=\varnothing$, for every label $y\in\mathcal{Y}$ there exists a parameter $\phi_y\in\Phi$ such that $\delta_{\phi_y}(Q,\varnothing)=y$.\footnote{This assumption is extremely weak. For any specific task, the proof only requires this to hold for the specific labels involved in the counterfactual comparison. For neural networks with continuous, high-dimensional parameter spaces and finite answer sets, this is essentially uncontroversial.}
\end{assumption}

\subsection{The Prompt-as-Operator Framework}

The identification problem arises because the additional instructions $P$ do not alter the model's factual parameters $\theta$. Instead, they act as an unknown internal operator that shifts which parts of the parameter space govern the output. For each set of additional instructions $P$ and task $Q$, we can define an effective parameter mapping $T_{P,Q}:\Phi\to\Phi$ such that
\[
\delta_\theta(Q,P) \;=\; \delta_{\,T_{P,Q}(\theta)}(Q,\varnothing).
\]
This is definitional: by Assumption~\ref{ass:reach}, for any observed output $y=\delta_\theta(Q,P)$ there exists some $\phi_y$ such that $\delta_{\phi_y}(Q,\varnothing)=y$, so we simply define $T_{P,Q}(\theta) \coloneqq \phi_y$. This captures the observed behavior: prompting the model with $P$ produces the same output as some \textit{effective} parameter set $T_{P,Q}(\theta)$ without prompting. The prompt does not change the model's weights, but it modulates how those weights are used to generate outputs. Crucially, we impose no \textit{a priori} structure linking this unknown operator $T_{P,Q}$ to the target counterfactual $\theta_t$, which is what leads to non-identification.

\paragraph{Why not assume the prompt works?}
One might object: ``What if the constraining prompt actually achieves its intended purpose?'' This would require assuming $T_{P_{\mathrm{constraint}},Q_t}(\theta) = \theta_t$, that the prompt perfectly transforms the factual parameters into the counterfactual ones. However, this assumption is empirically unverifiable without white-box access (direct observation of the model's internal computations and parameter representations) to the model's internal mechanisms. The model's opaque decision-making process makes it impossible to confirm whether the prompt genuinely restricts the model to pre-$t$ information or whether post-$t$ knowledge remains embedded and influences outputs in undetectable ways (e.g., through the model drawing on memorized historical beliefs to simulate period-appropriate responses). Our agnostic framework reflects this epistemic limitation: absent verifiable evidence, we cannot identify which of many observationally equivalent processes generated the output.

\subsection{Non-Identification of the Ideal Estimand}

This framework leads to a general non-identification result.

\setcounter{proposition}{\numexpr\getrefnumber{prop:nonid}-1\relax}
\begin{proposition}[Non-identification]\label{prop:nonid-appendix}
Fix $Q_t$ and a constraining prompt $P$. Suppose we observe the constrained decision $Y_{\mathrm{constrained}}(P) = y \in \mathcal{Y}$. Then for any two distinct labels $y^\star, y^\dagger \in \mathcal{Y}$, there exist two data-generating worlds $W_\star$ and $W_\dagger$ such that:
\begin{enumerate}
    \item They produce the \textbf{same} observable under the \textbf{same} factual parameters $\theta$:
        \[
        \delta_{\theta}(Q_t,P)=y
        \]
    \item They disagree on the target estimand:
        \[
        \delta_{\theta_t^{(\star)}}(Q_t,\varnothing)=y^\star
        \quad\text{and}\quad
        \delta_{\theta_t^{(\dagger)}}(Q_t,\varnothing)=y^\dagger
        \]
\end{enumerate}
\end{proposition}

\begin{proof}
We construct two worlds $W_\star$ and $W_\dagger$ that produce identical observables but differ in the target estimand.

\textbf{Shared observable.} By Assumption~\ref{ass:reach}, pick $\tilde\theta \coloneqq \phi_{y}$ such that $\delta_{\tilde\theta}(Q_t,\varnothing)=y$. Both worlds share the same factual parameters $\theta$ and the same prompt operator $T_{P,Q_t}$ defined by $T_{P,Q_t}(\theta) = \tilde\theta$. This ensures that in both worlds, the observable matches:
\[
\delta_{\theta}(Q_t,P) \;=\; \delta_{T_{P,Q_t}(\theta)}(Q_t,\varnothing) \;=\; \delta_{\tilde\theta}(Q_t,\varnothing) \;=\; y.
\]

\textbf{Distinct counterfactuals.} The worlds differ only in their unobservable counterfactual parameters $\theta_t$. By Assumption~\ref{ass:reach}, we set:
\begin{itemize}
    \item In $W_\star$: $\theta_t^{(\star)} = \phi_{y^\star}$, so $Y_t^\star = \delta_{\theta_t^{(\star)}}(Q_t,\varnothing) = y^\star$
    \item In $W_\dagger$: $\theta_t^{(\dagger)} = \phi_{y^\dagger}$, so $Y_t^\star = \delta_{\theta_t^{(\dagger)}}(Q_t,\varnothing) = y^\dagger$
\end{itemize}

Since $y^\star \neq y^\dagger$ and both worlds are consistent with the observed data, the target estimand $Y_t^\star$ is not identified.
\end{proof}

This proposition has a powerful corollary: under this agnostic model, observing \textit{any} output provides zero information about $Y_t^\star$.

\setcounter{corollary}{\numexpr\getrefnumber{cor:sharp-nonid}-1\relax}
\begin{corollary}[Sharp non-identification]\label{cor:sharp-nonid-appendix}
Under Assumption~\ref{ass:reach} and the operator model, for any observed output $Y_{\mathrm{constrained}}(P) = y$, the identified set for $Y_t^\star$ equals the entire label set:
\[
\mathcal{I}(y) \;=\; \mathcal{Y}.
\]
\end{corollary}

\begin{proof}
For any $\bar y\in\mathcal{Y}$, the construction in Proposition~\ref{prop:nonid} with $y^\star = \bar y$ shows that a world where $Y_t^\star = \bar y$ is observationally equivalent to the observed data $y$. Thus $\mathcal{I}(y) = \mathcal{Y}$.
\end{proof}

\subsection{Boundary Cases and Generalizations}

\paragraph{When the problem vanishes (future-invariant tasks).}
The non-identification problem arises because the answer to $Q_t$ is sensitive to post-$t$ information. If a task is ``future-invariant,'' the problem does not exist. Formally, if for the given task $Q_t$ and the information split at time $t$, we have $\delta_{\theta_t}(Q_t,\varnothing)=\delta_{\theta}(Q_t,\varnothing)$, then $Y_t^\star$ equals the factual decision $\delta_\theta(Q_t,\varnothing)$ and is trivially identified. However, as discussed in Section~\ref{sec:formalization}, tasks requiring judgment, selection, or prediction are typically \textit{not} future-invariant, limiting the practical applicability of this boundary case.

\paragraph{Stochastic-Decoding Generalization.}
The result is not an artifact of $\argmax$ (zero-temperature) decoding. With stochastic decoding (temperature $> 0$), the same non-identification holds whether we observe the full predictive distribution $G_\theta(\cdot\mid Q,P)$ or samples from it. Let $G_\phi(\cdot\mid Q,P)$ denote the predictive distribution over $\mathcal{Y}$ induced by parameters $\phi$. The proof follows the same construction: for any two counterfactual distributions $G^*, G^\dagger$ over $\mathcal{Y}$, we construct two worlds that produce identical observables but disagree on the target. This requires a distributional extension of Assumption~\ref{ass:reach}: for any distribution $G$ over $\mathcal{Y}$, there exist parameters $\phi_G$ such that $G_{\phi_G}(\cdot\mid Q,\varnothing) = G$. For neural networks with softmax output layers (the standard architecture that converts model scores into probabilities), this assumption is uncontroversial: the softmax function $G(y) = \frac{e^{z_y}}{\sum_{y'\in\mathcal{Y}} e^{z_{y'}}}$ maps vectors of scores (``logits'') $z \in \mathbb{R}^{|\mathcal{Y}|}$ continuously onto the probability simplex. Since $\mathcal{Y}$ is finite, any target distribution $G$ can be realized by choosing appropriate logit values (via solving $|\mathcal{Y}|-1$ independent equations with $|\mathcal{Y}|$ parameters). For the case of observed samples rather than the full distribution, non-identification follows by a simple informativeness argument: if observing the complete distribution $G_\theta(\cdot\mid Q,P)$ fails to identify $Y_t^\star$, then observing finite samples---which only partially reveal this distribution---certainly cannot identify it either.

\paragraph{Multiple or Adaptive Prompts.}
The impossibility result is not limited to single prompts. Observing outputs from any sequence of query-prompt pairs $\{(Q^{(k)},P^{(k)})\}_{k=1}^K$ (even chosen adaptively based on earlier responses) does not identify $Y_t^\star$, since the construction in Proposition~\ref{prop:nonid} applies independently to each observation in the sequence.

\paragraph{Masking procedures and valid masking.}
Definition~\ref{def:valid-masking} requires two conditions for valid masking. Condition (i) (future-invariance) requires both empirical and theoretical components.

\textbf{Empirical component: Reconstruction tests.} Let $R_\theta: \mathcal{Q} \to \mathcal{I}$ denote a reconstruction function that attempts to recover identifying information from a masked task, where $\mathcal{Q}$ is the space of tasks and $\mathcal{I}$ is the space of entity identifiers (e.g., company names, dates, industries). For a masked task $Q_t^{\text{mask}}$, if the model can successfully reconstruct the entity identifier with non-negligible probability---formally, if $\Pr[R_\theta(Q_t^{\text{mask}}) = I_{\text{true}}] > \epsilon$ for some threshold $\epsilon > 0$---this definitively proves condition (i) fails: the masked input contains sufficient information to trigger entity-specific memorization. However, reconstruction failure ($\Pr[R_\theta(Q_t^{\text{mask}}) = I_{\text{true}}] \leq \epsilon$) \textit{does not prove} condition (i) holds.

Formally, let $\mathcal{C}$ denote the set of all possible channels through which $Q_t^{\text{mask}}$ could trigger access to $I_{>t}$. Future-invariance requires that for every channel $c \in \mathcal{C}$, the decision $\delta_{\theta}(Q_t^{\text{mask}}, \varnothing)$ does not depend on $I_{>t}$ through that channel. Reconstruction tests probe only a finite subset $\{c_1, \ldots, c_k\} \subset \mathcal{C}$ of potential reconstruction pathways. By Remark~\ref{rem:lower-bound}, negative evidence does not prove memorization channels are absent.

\textbf{Theoretical component.} This requires arguing that no information pathway exists to trigger entity-specific memorized outcomes. For generic/statistical tasks where the underlying relationship is stable across time periods (e.g., market reactions to earnings surprises), researchers can argue that even if the model identified the entity, the answer would not change because the pattern is time-invariant.

Condition (ii) (detectable skill) addresses a distinct identification problem. Without observing that the model beats baseline, poor performance could reflect either (a) lack of model capability or (b) over-aggressive masking that removed information necessary for the task. These alternatives are observationally equivalent, creating a non-identification problem analogous to Proposition~\ref{prop:nonid}. Requiring $p^{\text{mask}} > p_0$ resolves this: beating baseline establishes both that the model has capability and that masking preserved sufficient information.

Valid masking enables two research uses. Task-preserving masking occurs when $\delta_{\theta}(Q_t^{\text{mask}}, \varnothing) = Y_t^\star = \delta_{\theta_t}(Q_t, \varnothing)$, recovering the original research objective. Capability-demonstrating masking occurs when the masked task represents a change in estimand; this validates model capability on the masked task class but does not enable entity-specific inference on the original task.

\subsection{Proof of Proposition~\ref{prop:lowpower}: Statistical Indistinguishability in Pre vs Post Comparisons}

\paragraph{Setup and assumptions.}
Let $n_{\text{pre}}$ and $n_{\text{post}}$ denote the pre-cutoff and post-cutoff sample sizes, respectively. We assume:
\begin{itemize}
    \item \textbf{Independence:} The pre-cutoff and post-cutoff samples are drawn independently.
    \item \textbf{Asymmetric sample sizes:} $n_{\text{pre}} \gg n_{\text{post}}$ (as is typical in practice).
\end{itemize}

Let $p_{\text{pre}} = \Pr[\delta_{\theta}(Q_t, \varnothing) = Y_{\text{true}} \mid t < t_{\text{cutoff}}]$ and $p_{\text{post}} = \Pr[\delta_{\theta}(Q_t, \varnothing) = Y_{\text{true}} \mid t \geq t_{\text{cutoff}}]$ denote the model's true accuracy on pre-cutoff and post-cutoff data, respectively, where $t_{\text{cutoff}}$ is the model's knowledge cutoff date and $Y_{\text{true}}$ is the realized outcome. Define the true memorization gap as $\Delta \coloneqq p_{\text{pre}} - p_{\text{post}}$. Let $\widehat{p}_{\text{pre}}$ and $\widehat{p}_{\text{post}}$ denote the corresponding sample estimates.

\noindent\textbf{Proposition~\ref{prop:lowpower} (Statistical indistinguishability in pre- vs post-cutoff comparisons).}
Consider testing $H_0: \Delta = 0$ where $\Delta \coloneqq p_{\text{pre}} - p_{\text{post}}$. When $n_{\text{pre}} \gg n_{\text{post}}$, the standard error of the estimated gap $\widehat{\Delta} = \widehat{p}_{\text{pre}} - \widehat{p}_{\text{post}}$ is approximately
\[
\text{SE}(\widehat{\Delta}) \;\approx\; \sqrt{\frac{p_{\text{post}}(1-p_{\text{post}})}{n_{\text{post}}}} \;=\; O(n_{\text{post}}^{-1/2}).
\]
There exist two data-generating processes that cannot be reliably distinguished by conventional hypothesis tests:
\begin{enumerate}[label=(\roman*)]
    \item \textbf{Time-Invariant Capability:} The true gap equals zero: $\Delta = 0$.
    \item \textbf{Undetected Memorization:} The true gap is positive but small: $\Delta >0$ with $\Delta/\text{SE}(\widehat{\Delta}) = c$ for some small constant $c = O(1)$.
\end{enumerate}
For sufficiently small $c$, hypothesis tests will have low power to detect such gaps. When $n_{\text{post}}$ is small, economically substantial memorization gaps remain statistically undetectable.

\begin{proof}[Proof of Proposition~\ref{prop:lowpower}]
We establish the result via a standard power analysis. Consider a one-sided hypothesis test of $H_0: \Delta = 0$ versus $H_1: \Delta > 0$ using the test statistic $\widehat{\Delta} = \widehat{p}_{\text{pre}} - \widehat{p}_{\text{post}}$. A one-sided test is appropriate because memorization is theoretically predicted to inflate pre-cutoff accuracy (since the model has access to outcome information for pre-cutoff observations), implying $\Delta \geq 0$; negative gaps ($\Delta < 0$) are not consistent with the memorization mechanism.

The true standard error of the difference is
\[
\text{SE}(\widehat{\Delta}) = \sqrt{\frac{p_{\text{pre}}(1-p_{\text{pre}})}{n_{\text{pre}}} + \frac{p_{\text{post}}(1-p_{\text{post}})}{n_{\text{post}}}}.
\]

When $n_{\text{pre}} \gg n_{\text{post}}$ (as is typical in practice), the first term is negligible relative to the second, so
\[
\text{SE}(\widehat{\Delta}) \approx \sqrt{\frac{p_{\text{post}}(1-p_{\text{post}})}{n_{\text{post}}}} \;=\; O(n_{\text{post}}^{-1/2}).
\]
Note that due to this sample size asymmetry, the variance of $\widehat{\Delta}$ is dominated by the post-cutoff sample variance, making the standard pooled-variance formulation unnecessary.

In practice, the true standard error must be estimated using sample proportions:
\[
\widehat{\text{SE}}(\widehat{\Delta}) = \sqrt{\frac{\widehat{p}_{\text{pre}}(1-\widehat{p}_{\text{pre}})}{n_{\text{pre}}} + \frac{\widehat{p}_{\text{post}}(1-\widehat{p}_{\text{post}})}{n_{\text{post}}}} \;\approx\; \sqrt{\frac{\widehat{p}_{\text{post}}(1-\widehat{p}_{\text{post}})}{n_{\text{post}}}}.
\]
By consistency, $\widehat{\text{SE}}(\widehat{\Delta}) \xrightarrow{p} \text{SE}(\widehat{\Delta})$ as $n_{\text{post}} \to \infty$. In what follows, we use the population standard error $\text{SE}(\widehat{\Delta})$ for theoretical power calculations (which characterize the limiting behavior of the test), while recognizing that practical hypothesis tests substitute the consistent estimator $\widehat{\text{SE}}(\widehat{\Delta})$.

\paragraph{Power analysis.}
By standard asymptotic theory for two-sample proportion tests, under the alternative hypothesis with true gap $\Delta > 0$, the test statistic is approximately
\[
\widehat{\Delta} \sim \mathcal{N}\left(\Delta, \,\text{SE}^2(\widehat{\Delta})\right).
\]
The power to reject $H_0$ at significance level $\alpha$ depends on the standardized effect size $\Delta / \text{SE}(\widehat{\Delta})$. As this ratio approaches zero, power approaches $\alpha$ (the test has no power beyond the false positive rate). Specifically, for a one-sided test at level $\alpha$, the power function is approximately
\[
\text{Power}(\Delta) \;=\; \Phi\left(\frac{\Delta}{\text{SE}(\widehat{\Delta})} - z_{1-\alpha}\right),
\]
where $\Phi$ is the standard normal CDF and $z_{1-\alpha}$ is the critical value. When $\Delta / \text{SE}(\widehat{\Delta}) = c$ for a small constant $c = O(1)$, the power is low.

\paragraph{Statistical indistinguishability.}
Consider two data-generating processes with different true parameters:
\begin{itemize}
    \item \textbf{DGP 1 (Time-Invariant Capability):} The true gap is zero: $\Delta = 0$. Any observed difference is purely sampling variation.

    \item \textbf{DGP 2 (Undetected Memorization):} The true gap is positive with standardized effect size $\Delta / \text{SE}(\widehat{\Delta}) = c$ for some small positive constant $c = O(1)$. Pre-cutoff accuracy is inflated by memorization.
\end{itemize}

For sufficiently small $c$ (depending on the significance level $\alpha$ and desired power), standard hypothesis tests cannot distinguish DGP 2 from DGP 1 with probability substantially exceeding the false positive rate $\alpha$. Both DGPs will produce $p > \alpha$ (``similar results'') with high probability.

When $n_{\text{post}}$ is small, $\text{SE}(\widehat{\Delta}) = O(n_{\text{post}}^{-1/2})$ is large. Any true gap $\Delta$ satisfying $\Delta = c \cdot O(n_{\text{post}}^{-1/2})$ for small $c$ will have low standardized effect size. Such memorization gaps, which may be substantial in absolute terms (e.g., 5--10 percentage points), remain statistically undetectable. Therefore, observing that pre-cutoff and post-cutoff accuracy appear similar provides no valid inference about whether the model has genuine forecasting skill ($\Delta = 0$) or undetected memorization ($\Delta > 0$).

\paragraph{Asymptotic identification.}
As $n_{\text{post}} \to \infty$, $\text{SE}(\widehat{\Delta}) \to 0$, and the set of undetectable gaps shrinks to zero. With sufficiently large $n_{\text{post}}$, any non-zero gap $\Delta > 0$ becomes detectable with power approaching 1.
\end{proof}

\subsection{Proof of Corollary~\ref{cor:finetuning}: Black-box Fine-tuning}

\begin{proof}[Proof of Corollary~\ref{cor:finetuning}]
Consider a black-box fine-tuning procedure $\mathcal{F}: \Phi \times \mathcal{D} \to \Phi$ that takes the original parameters $\theta = \theta(I_{\text{all}})$ and a ``forgetting'' dataset $D_{\text{forget}}$, producing fine-tuned parameters $\theta_{\text{ft}} = \mathcal{F}(\theta, D_{\text{forget}})$. We observe outputs from the fine-tuned model: $Y_{\text{ft}} = \delta_{\theta_{\text{ft}}}(Q_t, \varnothing)$.

The proof follows the same construction as Proposition~\ref{prop:nonid}. We impose no \textit{a priori} structure linking the black-box operator $\mathcal{F}$ to the target counterfactual $\theta_t$. This agnosticism is the source of non-identification.

Fix any observed output $y_{\text{ft}} \in \mathcal{Y}$ from the fine-tuned model. By Assumption~\ref{ass:reach}, there exists $\phi_{y_{\text{ft}}} \in \Phi$ such that $\delta_{\phi_{y_{\text{ft}}}}(Q_t, \varnothing) = y_{\text{ft}}$.

For any two distinct labels $y^\star, y^\dagger \in \mathcal{Y}$, we construct two observationally equivalent worlds:

\begin{itemize}
    \item \textbf{$W_\star$ (Genuine Forgetting):} Set the true counterfactual parameters $\theta_t^{(\star)} = \phi_{y^\star}$, so $Y_t^\star = \delta_{\theta_t^{(\star)}}(Q_t, \varnothing) = y^\star$. In this world, the fine-tuning procedure successfully removed post-$t$ information from the model's parameters.

    \item \textbf{$W_\dagger$ (Behavioral Suppression):} Set the true counterfactual parameters $\theta_t^{(\dagger)} = \phi_{y^\dagger}$, so $Y_t^\star = \delta_{\theta_t^{(\dagger)}}(Q_t, \varnothing) = y^\dagger$. In this world, the fine-tuning procedure only modified surface behavior: $\theta_{\text{ft}}$ still contains post-$t$ information embedded in its parameters, but learned to suppress outputs that reveal this knowledge.
\end{itemize}

In both worlds, the fine-tuning operator $\mathcal{F}$ produces parameters such that $\delta_{\theta_{\text{ft}}}(Q_t, \varnothing) = y_{\text{ft}}$. The observed output is identical, but the interpretations are contradictory: genuine forgetting ($W_\star$) versus learned suppression ($W_\dagger$).

Since $y^\star \neq y^\dagger$ and both worlds are consistent with the observed data, $Y_t^\star$ is not identified from black-box fine-tuned outputs. Without white-box verification (e.g., mechanistic interpretability techniques that identify and verify removal of specific computational pathways encoding the post-$t$ information), we cannot distinguish whether the model genuinely forgot or merely learned to hide its knowledge.
\end{proof}

\end{document}